\documentclass[preprint,
    10pt,
    3p,
    fleqn,
]{elsarticle}

\usepackage{main}

\usepackage[english]{babel}

\begin{document}

\begin{frontmatter}

  \title{Formally Verified Animation for RoboChart using Interaction Trees\tnoteref{t1,t2}}

  \tnotetext[t1]{This document presents results from the research project CyPhyAssure (\url{www.cs.york.ac.uk/circus/CyPhyAssure/}) funded by EPSRC.}

  \author[1]{Kangfeng Ye\corref{cor1}}

  \ead{kangfeng.ye@york.ac.uk}

  \author[1]{Simon Foster}

  \ead{simon.foster@york.ac.uk}

  \author[1]{Jim Woodcock}

  \ead{jim.woodcock@york.ac.uk}

  \cortext[cor1]{Corresponding author}


  \affiliation[1]{%
    organization={Department of Computer Science, University of York},
    addressline={Deramore Lane, Heslington},
    postcode={YO10 5GH},
    city={York},
    country={United Kingdom}
  }%
  
  \begin{abstract}
    RoboChart is a core notation in the RoboStar framework. It is a timed and probabilistic domain-specific and state machine-based language for robotics. RoboChart supports shared variables and communication across entities in its component model. It has formal denotational semantics given in CSP. The semantic technique of Interaction Trees (ITrees) represents behaviours of reactive and concurrent programs interacting with their environments. Recent mechanisation of ITrees, ITree-based CSP semantics and a Z mathematical toolkit in Isabelle/HOL bring new applications of verification and animation for state-rich process languages, such as RoboChart. In this paper, we use ITrees to give RoboChart novel operational semantics, implement it in Isabelle, and use Isabelle's code generator to generate verified and executable animations. We illustrate our approach using an autonomous chemical detector and patrol robot models, exhibiting nondeterminism and using shared variables. With animation, we show two concrete scenarios for the chemical detector when the robot encounters different environmental inputs and three for the patrol robot when its calibrated position is in other corridor sections. We also verify that the animated scenarios are trace refinements of the CSP denotational semantics of the RoboChart models using FDR, a refinement model checker for CSP. \changed[\C{31}]{This ensures that our approach to resolve nondeterminism using CSP operators with priority is sound and correct.}
  \end{abstract}
  
  \begin{keyword}
    Interaction trees\sep CSP\sep animation\sep theorem proving\sep RoboChart\sep code generation\sep robot software\sep operational semantics\sep nondeterminism.
  \end{keyword}
  
\end{frontmatter}


\section{Introduction}
\label{sec:intro}

The RoboStar%
\footnote{%
  \url{robostar.cs.york.ac.uk}.%
} framework~\cite{Cavalcanti2021} brings modern modelling and verification technologies into software engineering for robotics.  In this framework, models of the platform, environment, design, and simulations are given formal semantics in a unified semantic framework~\cite{Hoare1998}.  Correctness of simulation is guaranteed for particular models using a variety of analytical technologies, including model checking, theorem proving, and testing.  Additionally, modelling, semantics generation, verification, simulation, and testing are automated and integrated into an Eclipse-based tool, RoboTool.%
\footnote{%
  \url{robostar.cs.york.ac.uk/robotool/}.%
} The core of RoboStar is RoboChart~\cite{Miyazawa2019,Ye2021}, a timed and probabilistic domain-specific language \changed[\C{6}]{to model robotic software}, which provides UML-like architectural and state machine modelling notations.  RoboChart is \changed[\C{32}]{distinguished by} its formal semantics~\cite{Miyazawa2019,Woodcock2019,Ye2021}, which enables automated verification using model checking and theorem proving~\cite{Ye2021a}. \changed[\C{6}]{In RoboChart, physical robots are abstracted into robotic platforms through variables, events, and operations. RoboSim~\cite{Cavalcanti2019a} is a domain-specific language for simulation in robotics. It can be seen as a correct implementation or refinement of RoboChart into the simulation level in terms of control software, called d-model. RoboSim is also enriched to specify robotic platforms' physical and dynamic behaviours, provided through p-models. The d-model and p-model of RoboSim are related through a platform mapping, which describes how variables, events, and operations in the d-model are interpreted as continuous variables in the p-model. The hybrid models then can be verified using Differential Hoare logics~\cite{Foster2020a,Foster2021a}, or hardware/software co-verification~\cite{Murray2022}. Our work presented here targets the animation of RoboChart models for robotic control software, \changed[\C{34}]{and so users can interact with RoboChart models through provided command line interfaces. From this aspect, users play a role in the environment of the models. They inspect the behaviour of the models by choosing what the model is allowed to do and observing its response.}}

Previous work~\cite{Miyazawa2019} gives RoboChart a denotational semantics based on the CSP process algebra~\cite{Hoare1985,Roscoe2011}.  This paper defines direct operational semantics for RoboChart using Interaction Trees (ITrees)~\cite{Xia2019}. \changed[\C{5}]{ITrees support a coinductive encoding of labelled transition systems that can model infinite behaviours of a reactive system interacting with its environment. Crucially, ITrees provide abstract yet directly executable semantics, which means they can support animation to support model exploration, verification using coinductive proof techniques, and generation of correct-by-construction implementations~\cite{Foster2021}. An ITree-based semantics for RoboChart thus further empowers developers with techniques for prototyping and analysis of models}. ITrees have been mechanised in both Coq~\cite{Xia2019} and Isabelle/HOL~\cite{Foster2021}. ITrees also unify trace-based denotational failures-divergences semantics~\cite{Brookes1984,Roscoe2011} for CSP, transition-based operational semantics, and algebraic semantics, as demonstrated in our previous work~\cite{Foster2021}.


%

The existing implementation of RoboChart's semantics in RoboTool is restricted to machine-readable CSP (or CSP-M) for verification with FDR~\cite{GABR14}, a CSP refinement checker, and so only a subset of RoboChart's rich types and expressions (which is based on that of the Z notation~\cite{Toyn2002}) can be supported, and quantified predicates cannot be solved.  Our contribution here is a richer ITree-based CSP semantics for RoboChart in terms of types and expressions to address these restrictions, thanks to the mechanised Z toolkit\footnote{\url{https://github.com/isabelle-utp/Z_Toolkit}.}~\cite{Spivey1992} in Isabelle/HOL.
%
Our semantics also allow us to characterise systems with an infinite number of states symbolically, avoiding the need to generate an explicit transition system.
%
%

We mechanise the ITree-based CSP semantics in Isabelle/HOL, \changed[\C{7}]{which ensures that all our definitions and proofs are theoretically well-grounded, yet also practically applicable in development tasks. Isabelle/HOL provides us with an array of tools to support software engineering and verification, including a flexible syntax frontend, integration of automated theorem provers, and code generation}.  \changed[\C{12}]{Animation of a RoboChart model is realised in two stages. The first stage generates the ITree-based CSP semantics for the model, and the second stage utilises the code generator~\cite{Haftmann2010} in Isabelle/HOL to automatically produce Haskell code for animation. We note that the work presented in this paper results from manually generating CSP semantics for RoboChart models in the first stage. Supporting full automation for semantics generation is part of our future work. In the second stage, } Isabelle's code generator~\cite{Haftmann2010} translates executable definitions in the source HOL logic to target functional languages (such as SML and Haskell), and the translation preserves semantic correctness using higher-order rewrite systems~\cite{Mayr1998}.
As a result, the semantics of the source logic in Isabelle is preserved during code generation via translation to target functional languages. Our animation, therefore, is formally verified for RoboChart's semantics in ITrees. Thanks to the equational logic, functional algorithms and data refinement are supported in code generation, so less efficient algorithms and \changed[\C{33}]{data structures used for} verification in Isabelle can be replaced with more efficient ones for animation.



Our technical contributions are as follows: we %
\begin{inparaenum}[(1)]
\item implement extra CSP operators (generalised choice, interrupt, exception, and renaming), which are required to support the RoboChart semantics, and new CSP operators with priority (hiding with priority and renaming with priority) to resolve nondeterminism in RoboChart based on an order; %
\item implement a bounded list or sequence type (that is not defined in the Z toolkit) for code generation; %
\item use the new concepts to define an ITree-based operational semantics for RoboChart; %
\item implement the semantics of two RoboChart models for case studies; and %
\item apply our animator to explore the behaviour of the models. %
\end{inparaenum}
With our mechanisation and animation, we have detected several issues in one RoboChart model, explored the semantics of shared variables in RoboChart, and resolved nondeterminism in a particular way. Specifically, the prioritised renaming operator uses the order in which transitions are given in the renaming mappings to resolve nondeterminism. The benefit of resolving nondeterminism statically in semantics, instead of dynamically in the animator, is that animation becomes more automatic since the user only needs to resolve external choice and not nondeterministic choice. There is no need to overcome the big $\tau$ diamonds~\cite{Roscoe2011} as in the animator of ProB~\cite{Leuschel2003} and FDR. All definitions and theorems in this paper are mechanised, and accompanying icons (\isalogo) link to corresponding repository artefacts. \changed[\C{15}]{We assume basic knowledge of Isabelle/HOL from interested readers to understand these definitions and theorems.}

The remainder of this paper is organised as follows. Section~\ref{sec:robochart} gives a brief discussion of \changed[\C{4}]{CSP and then introduces} RoboChart through two examples: an autonomous chemical detector model and a patrol robot model.
Section~\ref{sec:itree} briefly describes the mechanisation of ITrees in Isabelle, \changed[\C{9}]{lists the semantics of previously defined CSP operators}, and presents the extra CSP operators in detail. Then we introduce the RoboChart semantics in ITrees in Sect.~\ref{sec:rc_to_itrees} \changed[\C{11}, \C{14}]{using general mathematical rules}, exemplified using the two models, and illustrate several scenarios for the two models in Sect.~\ref{sec:animation} using animation. We review related work in Sect.~\ref{sec:related} and conclude in Sect.~\ref{sec:concl}.

This paper is an extension of \cite{Ye2022}. It adds a new generalised choice operator, redefines the CSP external choice operator using the generalised choice, and introduces biased external choice operators in Sect.~\ref{ssec:itree_genchoice}. This work also gives the ITrees-based semantics to shared variables in RoboChart. It resolves nondeterminism in the semantics using the new CSP operators with priority defined in Sects.~\ref{ssec:itree_hiding_with_priority} and \ref{ssec:itree_renaming_with_priority}. The semantics for shared variables and nondeterminism are exemplified in a new case study introduced in Sect.~\ref{ssec:robochart_patrol} and illustrated in Sect.~\ref{ssec:animation_patrol} for its animation. We also add \changed[\C{11}, \C{14}]{semantic rules} and important details in Sect.~\ref{sec:rc_to_itrees} to give a clear understanding of our semantics for RoboChart with examples from the two models, including new subsections: Sect.~\ref{ssec:semantics_overview} to give an overview of RoboChart semantics, Sect.~\ref{ssec:semantics_channels} about channels and alphabet transformation, and Sect.~\ref{ssec:semantics_operation} to give semantics to operations in RoboChart. We give more details and examples to illustrate the semantics of the existing subsections. In particular, Sect.~\ref{ssec:semantics_stm} is substantially extended to give semantics to state machines, not only its sketch but also its detailed definitions and examples for memories, different kind of nodes, and their composition in parallel. Similarly, Sects.~\ref{ssec:semantics_ctrl} and \ref{ssec:semantics_module} are also extended with examples.

\section{CSP and RoboChart}
\label{sec:robochart}
\changed[\C{4}]{
\subsection{CSP}
\label{ssec:robochart_csp}

Communicating Sequential Processes or CSP~\cite{Hoare1987,Roscoe2011} is a well-established process algebra to model concurrent systems using communication (message-passing via channels) for the interaction between processes. It has primitives for specifying sequential behaviour and concurrent interaction using events. Table~\ref{table:csp_operators} summarises the CSP processes and operators used in our semantics for RoboChart.

\begin{table}[!htb]
  \setlength\extrarowheight{2pt} 
  \caption{\label{table:csp_operators} \changed[\C{4}]{A summary of CSP processes and operators.}}
  \resizebox{\textwidth}{!}{
  \begin{tabularx}{\textwidth}{l|p{3.5cm}|X}
    \hline
    \textbf{Symbol} & Name & Description \\
    \hline
    $\tau$ & internal event & Invisible event.\\
    $Skip$ & skip & Terminate immediately without change to the state.\\ 
    $Stop$ & deadlock & Refuse any interaction with the state unchanged. \\ 
    $c \then P$ & prefix & Synchronise on channel $c$ and then behave like $P$. \\
    $c?x \then P(x)$ & input & Accept an input of any value (of type $T$) on channel $c$ if $c$ has a type $T$, record the value on variable $x$, and then behave like $P(x)$. \\
    $c?x:S \then P(x)$ & restricted input & Similar to input, but it only accepts the values from set $S$.\\
    $c!v \then P$ & output & Synchronise on channel $c$ with value $v$, and then behave like $P$.\\
    $b \guard P$ & guarded process & Behave like $P$ if $b$ is true, and deadlock otherwise.\\
    $P \extchoice Q$ & external choice & Offer the environment a choice of the first events of $P$ and $Q$, and then behave accordingly.\\
    $P \intchoice Q$ & internal choice & Nondeterministic choice between $P$ and $Q$ without offering the environment a choice.\\
    $P ; Q$ & sequential composition & Behave like $P$ initially, and behave like $Q$ if $P$ terminates. \\
    $P \interrupt Q$ & interrupt & Behave like $P$, but offer the environment a choice of the initial events of $Q$ at any time until $P$ terminates. If one of these events is performed, $Q$ takes over and behaves accordingly.\\
    $ P \exception{E} Q$ & exception & Behave like $P$ until $P$ performs an event from set $E$, at that point, then behave like $Q$. \\
    $P \parallel_{E} Q$ & generalised parallel composition & $P$ and $Q$ run simultaneously, synchronise on the events in $E$, progress independently on the events not in $E$, and terminate if both terminate. \\
    $P \interleave Q$ & interleave & Equal to $P \parallel_{\emptyset} Q$ where $P$ and $Q$ always progress independently on any event.\\
    $ P \hide E$ & hiding & Behave like $P$ except that the events from $E$ become internal.\\
    $ \rename{P}{c \becomes d}$ & renaming & Rename the event $c$ in $P$ to $d$.\\
    $ \Extchoice i:I @ P(i)$ & replicated external choice & Offer a choice of the indexed processes $P(i)$ by $i$ from set $I$. The similar replicated internal choice, sequential composition, parallel composition, and interleaving are omitted here.\\
    \hline
  \end{tabularx}}
\end{table}

CSP has several rich semantic models~\cite{Roscoe2011}, including traces, stable failures, failures-divergences, and refusal testing models based on its denotational semantics and characterising the observable behaviours of processes. CSP also has consistent denotational and operational semantics~\cite{Roscoe2011}: all these semantic models have congruence theorems in the operational semantics. CSP also has been extended to specify time in concurrent systems like \emph{tock}-CSP~\cite{Roscoe2011,Baxter2021} and Timed CSP~\cite{Roscoe2011,Schneider1999}, and to specify shared variables, priority, and mobility~\cite{Roscoe2011} in concurrent systems. Several languages are also built on CSP, such as Circus~\cite{Woodcock2002}, CML~\cite{Woodcock2012}, and RoboChart presented here.

FDR is a widely used model-checking tool for verifying CSP processes through refinement: $P \refinedby_\mathcal{M} Q$ represents a specification $P$ is refined by an implementation $Q$ in terms of a semantic model $\mathcal{M}$. FDR supports hierarchical compression methods\footnote{\url{https://cocotec.io/fdr/manual/cspm/prelude.html\#compressions}} to tackle the state space explosion problem in model checking. According to~\cite{GABR14}, the cluster version of FDR3 is able to analyse problems with $10^{12}$ compressed states (approximately $10^{20}$ uncompressed states). FDR also has a built-in CSP process animator called ProBE~\cite{GABR14}, which can be used to manually explore the full transition tree of a CSP process. 

This paper applies an ITree-based semantics for CSP, which we developed previously~\cite{Foster2021}. It is consistent with the failures-divergences semantics~\cite{Roscoe2011} but focuses on deterministic CSP processes, which are directly executable. Since our encoding of CSP is symbolic and implicit, we can avoid the state explosion problem.
}

\subsection{RoboChart}
{
\changed[\C{8}]{RoboChart is motivated by problems in the current practice of programming robotic applications using standard state machines:
\begin{enumerate*}[label={(\alph*)}]
\item no precise syntax and formal semantics, 
\item informally discussed time and uncertainty requirements, 
\item loosely connected artefacts, 
\item no tool support, and 
\item no assurance. 
\end{enumerate*}
RoboChart is a state machine-based notation designed for roboticists to model robotic control software and apply modern verification techniques. RoboChart has precise syntax and formal semantics and allows users to specify time and probability features in state machines formally. Modelling using RoboChart and verification is supported by its accompanying tool, RoboTool. Roboticists can use RoboTool to create RoboChart models using consistent diagrammatic and textual modelling. They can either draw diagrams (and RoboTool automatically generates the corresponding textual model) or program models (RoboTool can also automatically render its diagrammatic representation).

RoboChart has an unambiguous mathematical semantics. RoboTool automatically generates semantics for models and applies verification techniques like model checking and theorem proving to analyse the semantics based on the properties specified by users. RoboChart models and properties use languages familiar to roboticists. Other artefacts, like semantics, are automatically linked to the models and properties, making modern software engineering and verification techniques more accessible for roboticists.
}

RoboChart also has a component model with notions of controller, module, and state machines to foster reuse. 
In a RoboChart model, physical robots are abstracted into robotic platforms through variables, events, and operations. 
We describe \changed[\C{35}]{some of the features} of RoboChart for modelling controllers of robots using two examples: an autonomous chemical detector~\cite{Hilder2012,Miyazawa2019,RoboChartRef} in Sect.~\ref{ssec:robochart_chemical} and a patrol robot in Sect.~\ref{ssec:robochart_patrol}.
We refer to the RoboChart reference manual~\cite{RoboChartRef} for a complete account of the notation and its semantics.
}

\subsubsection{Autonomous chemical detector}
\label{ssec:robochart_chemical}
The robot is equipped with sensors to 
\begin{inparaenum}[(1)]
\item analyse the air to detect dangerous gases;  
\item detect obstacles; and
\item estimate change in position (using an odometer).
\end{inparaenum}
The controller of the robot performs a random walk with obstacle avoidance. Upon detecting a chemical with its intensity above a threshold, the robot drops a flag and stops there.
This model\footnote{\url{robostar.cs.york.ac.uk/case_studies/autonomous-chemical-detector/autonomous-chemical-detector.html}}~\cite{Miyazawa2019} has been studied and analysed in RoboTool, using \changed[\C{36}]{FDR}.

\begin{figure}[t]
  \centering%
  \includegraphics[width=.90\textwidth]{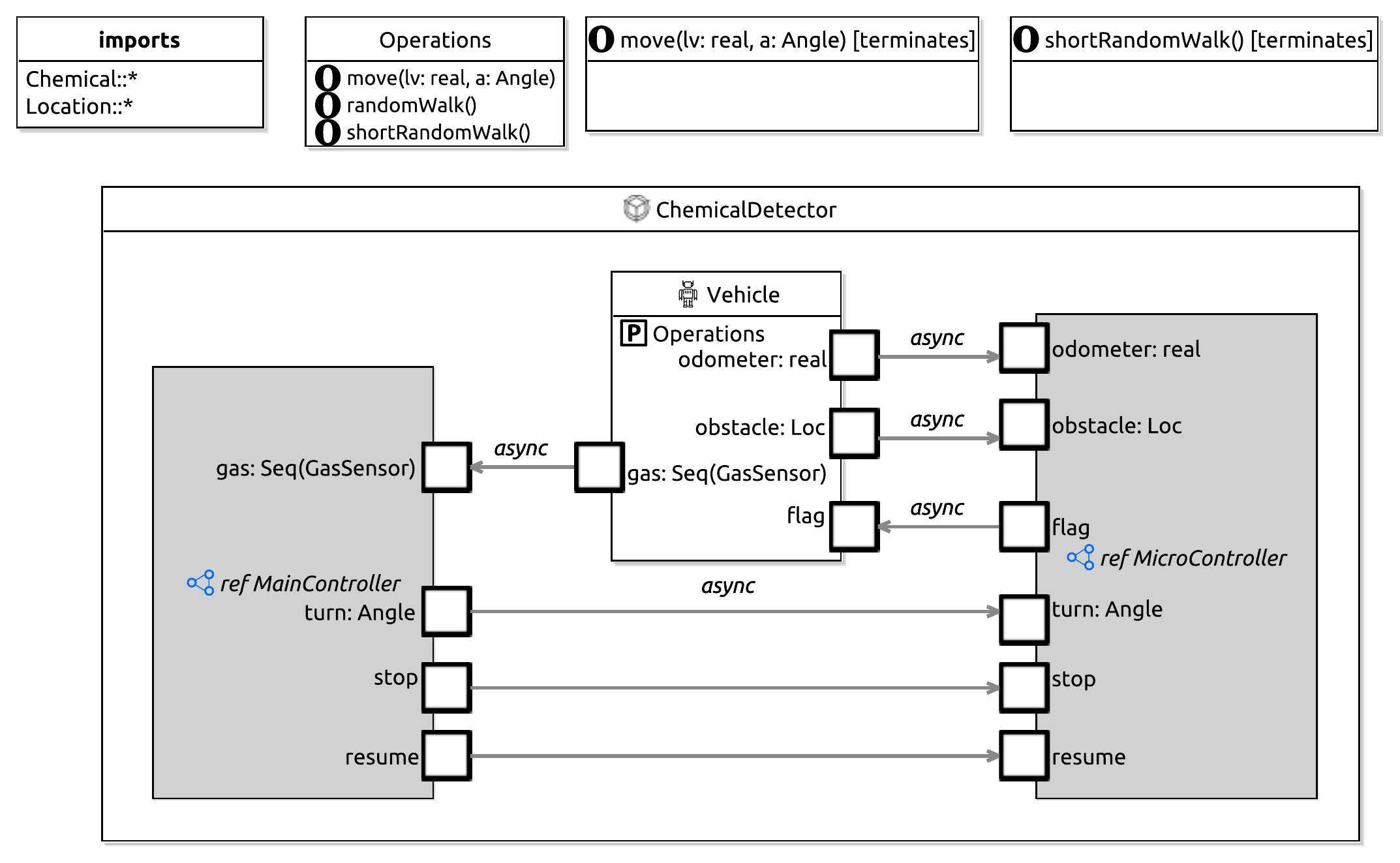}%
  \vspace{-2ex}
  \caption{The module of the autonomous chemical detector model.}%
\vspace{-4ex}
  \label{fig:robochart_acd_module}%
\end{figure}

The top-level structure of a RoboChart model, a module, is shown in Fig.~\ref{fig:robochart_acd_module}. 
The module \rcitem{ChemicalDetector} contains a robotic platform \rcitem{Vehicle} and two controller references \rcitem{MainController} and 
\rcitem{MicroController}. The physical robot is abstracted into the robotic platform through variables, events, and operations. The platform provides the controllers with services  
\begin{inparaenum}[(1)]
\item to read its sensor data through three events: \rcitem{gas}, \rcitem{obstacle}, and \rcitem{odometer};
\item for movement through three operations: \rcitem{move}, \rcitem{randomWalk}, and \rcitem{shortRandomWalk} as grouped in an interface \rcitem{Operations}; and 
\item to drop a flag through receiving a \rcitem{flag} event.
\end{inparaenum}
{These services represent the controllers' observable behaviour or external interaction with the physical robot. The controllers \rcitem{MainController} is responsible for gas analysis, and \rcitem{MicroController} accounts for the robot movement with obstacle avoidance. 
}

A platform and controllers communicate using directional connections. {For example, the platform is linked to \rcitem{MainController} through an asynchronous connection on event \rcitem{gas} of type \rcitem{seq(GasSensor)}, sequences of type \rcitem{GasSensor}.} Furthermore, the \rcitem{MainController} and \rcitem{MicroController} interact using the events \rcitem{turn}, \rcitem{stop}, and \rcitem{resume} {to allow \rcitem{MainController} to instruct \rcitem{MicroController} to \emph{turn} towards the location where the gas is detected, \emph{stop} the robot if the gas is dangerous, or \emph{resume} its movement behaviour (by ignoring the current movement operation) if no gas is detected. These interactions are the internal behaviour of controllers and are not observable.}


{
\begin{figure}
  \centering%
  \includegraphics[width=.95\textwidth]{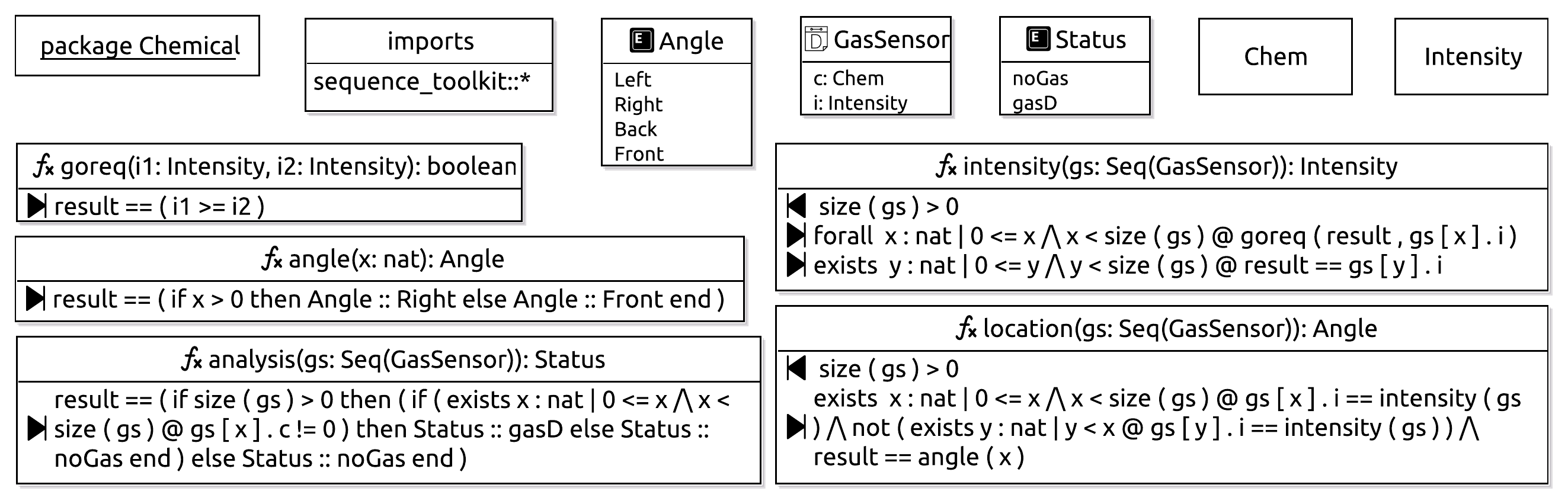}%
  \caption{The Chemical package of the autonomous chemical detector model.}%
  \label{fig:robochart_acd_chemical}%
\end{figure}
}

The types used in the module are defined in the two imported packages: \rcitem{Chemical} and \rcitem{Location} 
{shown in Figures~\ref{fig:robochart_acd_chemical} and~\ref{fig:robochart_acd_location}. 
The \rcitem{Chemical} package declares primitive types \rcitem{Chem} and \rcitem{Intensity}, enumerations \rcitem{Status} and \rcitem{Angle}, a record \rcitem{GasSensor} containing two fields (\rcitem{c} of type \rcitem{Chem} and \rcitem{i} of type \rcitem{Intensity}), and five functions specified using preconditions and postconditions (in the original model, two are specified and three are unspecified).
%
\begin{figure}
  \centering%
  \includegraphics[width=.65\textwidth]{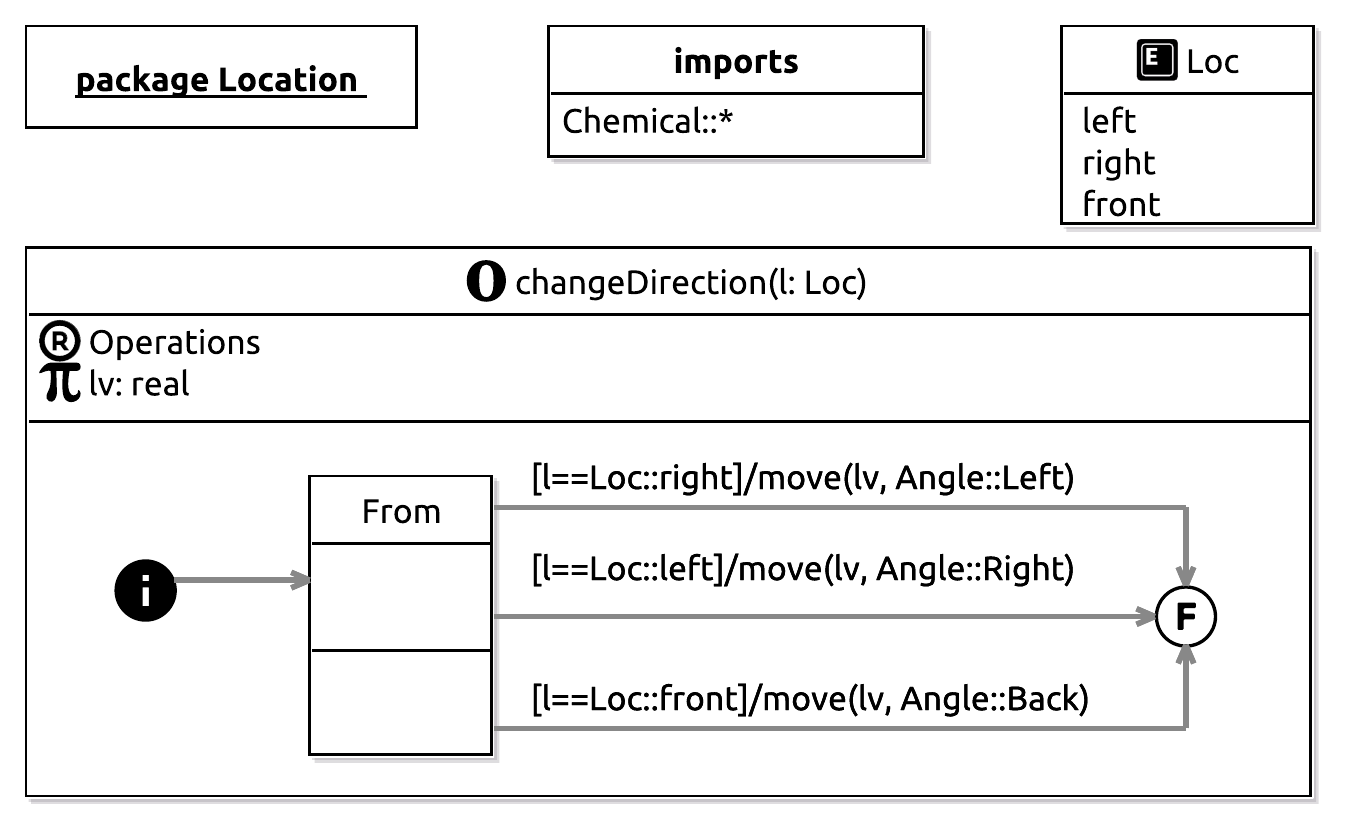}%
  \caption{The Location package of the autonomous chemical detector model.}%
  \label{fig:robochart_acd_location}%
\end{figure}
The \rcitem{Location} package declares an enumeration \rcitem{Loc} and defines an operation \changed[\C{37}]{\rcitem{changeDirection}} using a state machine. The operation has a parameter \rcitem{l} of type \rcitem{Loc} and a constant \rcitem{lv}.
This operation aims to move the robot in the opposite direction of the currently detected gas location \rcitem{l} using a constant linear velocity \rcitem{lv}.

\rcitem{MainController}, defined in the left diagram of Fig.~\ref{fig:robochart_acd_gasanalysis},  is implemented using a state machine \rcitem{GasAnalysis} (by a contained reference to the machine), presented in the right diagram of Fig.~\ref{fig:robochart_acd_gasanalysis}.
}
{
\begin{figure}[t]
  \centering%
  \includegraphics[width=0.4\textwidth]{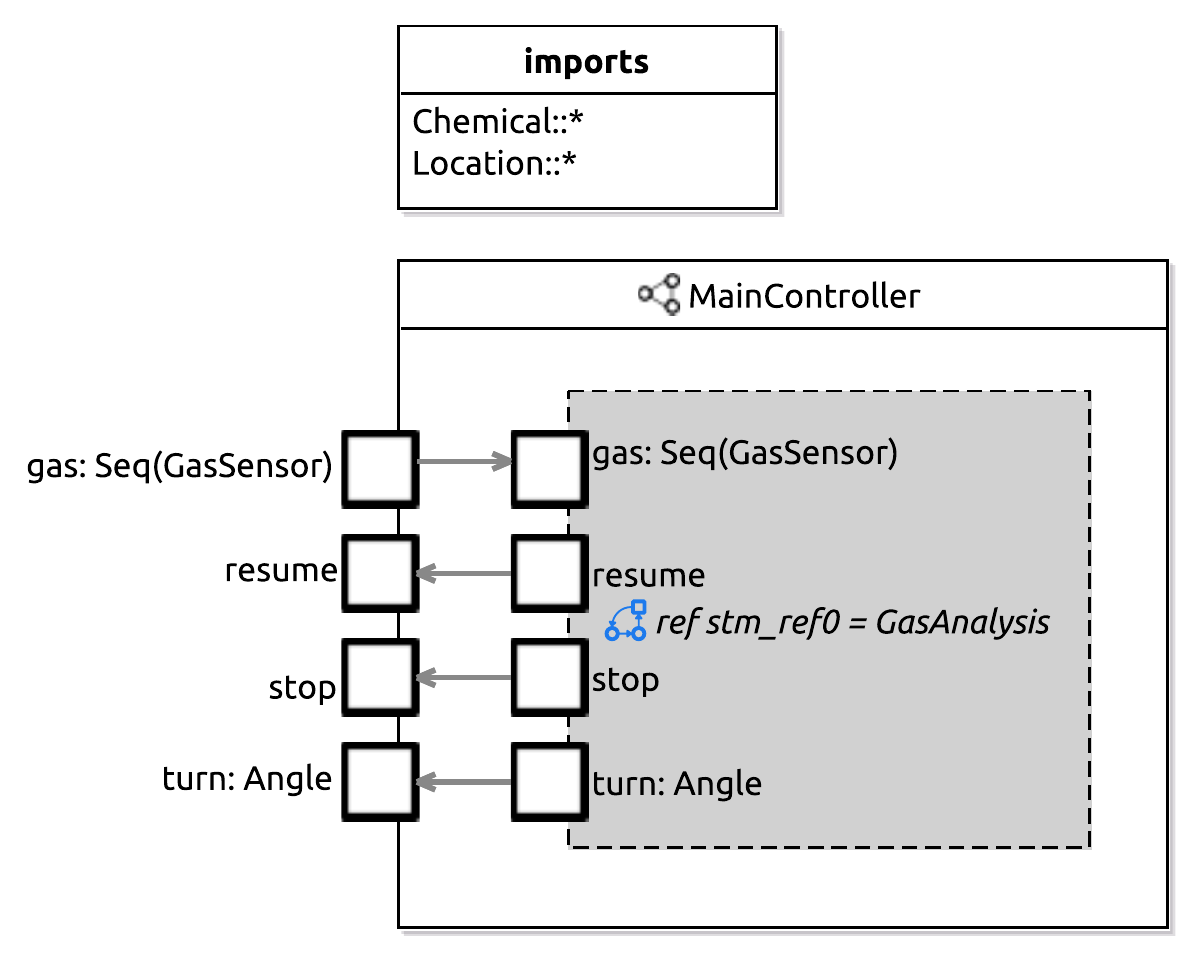}%
  \includegraphics[width=0.6\textwidth]{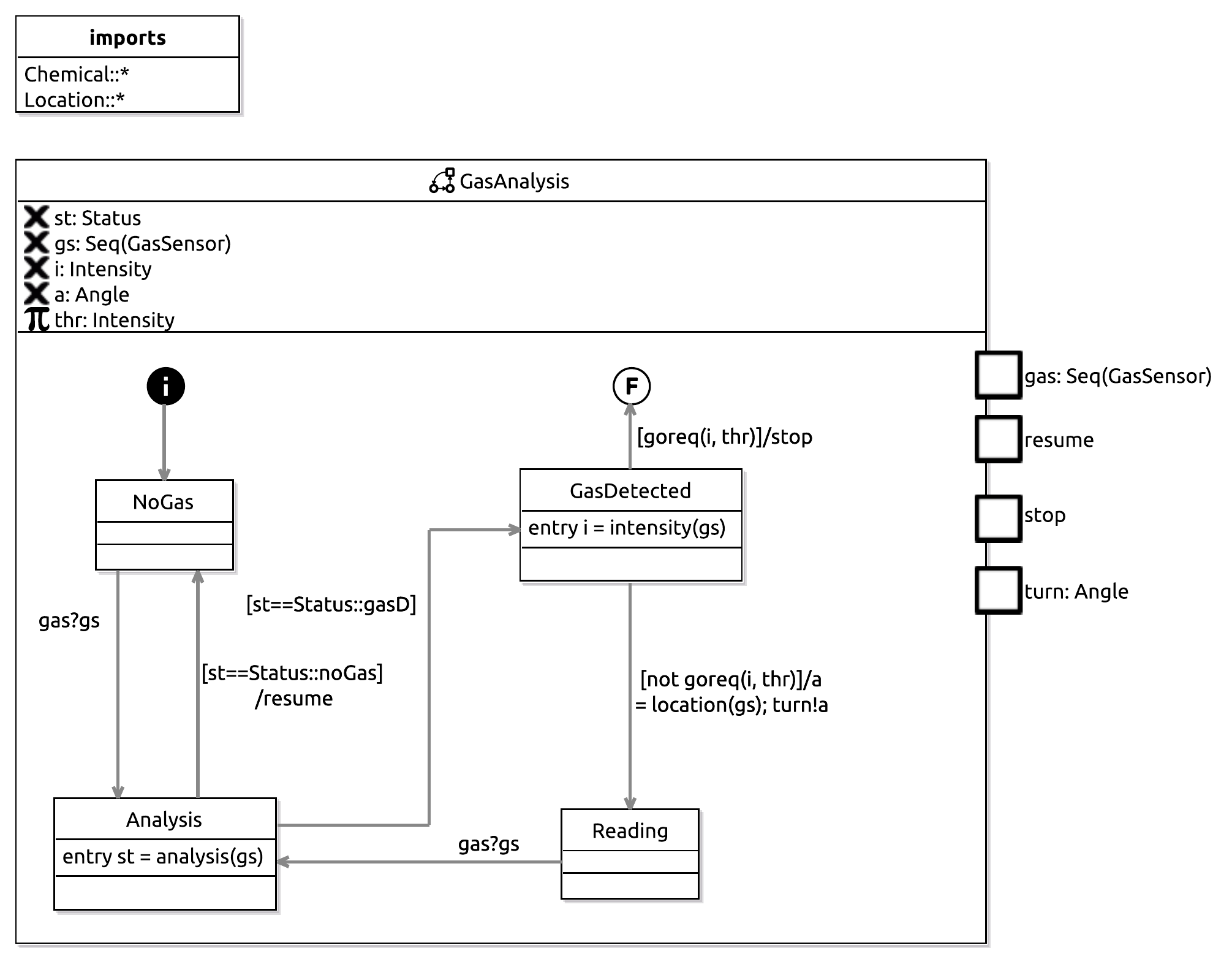}%
  \vspace{-2ex}
  \caption{\rcitem{MainController} and GasAnalysis state machine of the autonomous chemical detector model.}%
  \vspace{-4ex}
  \label{fig:robochart_acd_gasanalysis}%
\end{figure}
}
{
\noindent
The machine \rcitem{GasAnalysis} declares one constant \rcitem{thr} of type \rcitem{Intensity} for the intensity threshold, and four variables (\rcitem{gas} of \rcitem{Seq(GasSensor)}, \rcitem{st} of \rcitem{Status}, \rcitem{i} of \rcitem{Intensity}, and \rcitem{a} of \rcitem{Angle}) for a sequence of gas sensor readings (from the platform), and the gas analysis results including its status (either no gas \rcitem{NoGas} or a gas \rcitem{gasD} detected), intensity and angle. 
The machine also contains a variety of nodes: one initial junction (\includegraphics[align=c,height=8pt]{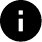}), seven normal states such as \rcitem{NoGas} and \rcitem{Analysis}, and a final state (\includegraphics[align=c,height=8pt]{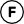}). A state may have an \rcitem{entry} action such as an assignment of \rcitem{st} from an application of function \rcitem{analysis} to \rcitem{gs} in the state \rcitem{Analysis}, an exit action, or a \rcitem{during} action.

In state machines, transitions connect states and junctions. Transitions have a label with optional features: a trigger event, a clock reset, a guard condition, and an action. For example, 
the transition of \rcitem{GasAnalysis} from \rcitem{NoGas} to \rcitem{Analysis} has an input trigger \rcitem{gas?gs} enabling the machine to receive sensor readings from the channel \changed[\C{38}]{\rcitem{gas}} and store the value in the variable \rcitem{gas}, and the transition from \rcitem{GasDetected} to \rcitem{Reading} has a guard (\rcitem{not goreq(i, thr)}) and an action (\rcitem{a=location(gs); turn!a}) which is a sequential composition of an assignment and an output communication (\rcitem{turn!a}) enabling the machine to send the angle \rcitem{a} of the detected gas over the channel \rcitem{turn}.

This machine gives the behaviour of the robot's gas analysis: 
\begin{inparaenum}[(1)]
\item enter the state \rcitem{NoGas} after the transition from the initial junction is taken;
\item wait for the gas sensor to be ready on channel \rcitem{gs}, then receive readings, recorded in \rcitem{gs}, on the channel from the platform (via connections from \rcitem{MainController}), and at the same time the transition to state \rcitem{Analysis} is taken;
\item upon entering the state \rcitem{Analysis} whose entry action is executed first to analyse the sensor readings by the function \rcitem{analysis} and to record the result in variable \rcitem{st}; 
\item signal the event \rcitem{resume} if no gas is detected (guard [\rcitem{st == noGas]}) and return to state \rcitem{NoGas};
\item go to state \rcitem{GasDetected} otherwise (guard [\rcitem{st == gasD]});
\item upon entering the state \rcitem{GasDetected} whose entry action is executed to determine the intensity by the function \rcitem{intensity} and to record the result in variable \rcitem{i}; 
\item {signal \rcitem{stop} if the intensity \rcitem{i} is larger than or equal to (implemented in function \rcitem{goreq}) the intensity threshold \rcitem{thr}, and terminate by going to the final state;}
\item {take the transition to state \rcitem{Reading} otherwise with the action of the transition being executed to determine the angle \rcitem{a} of the detected gas and signal \rcitem{turn} towards the angle;}
\item {try to read the gas sensor again at state \rcitem{Reading} with its only outgoing transition, and if the \changed[\C{39}]{transition} is taken, the machine is back to state \rcitem{Analysis}.}
\end{inparaenum}
}

\begin{figure}[t]
  \centering%
  \includegraphics[width=0.6\textwidth]{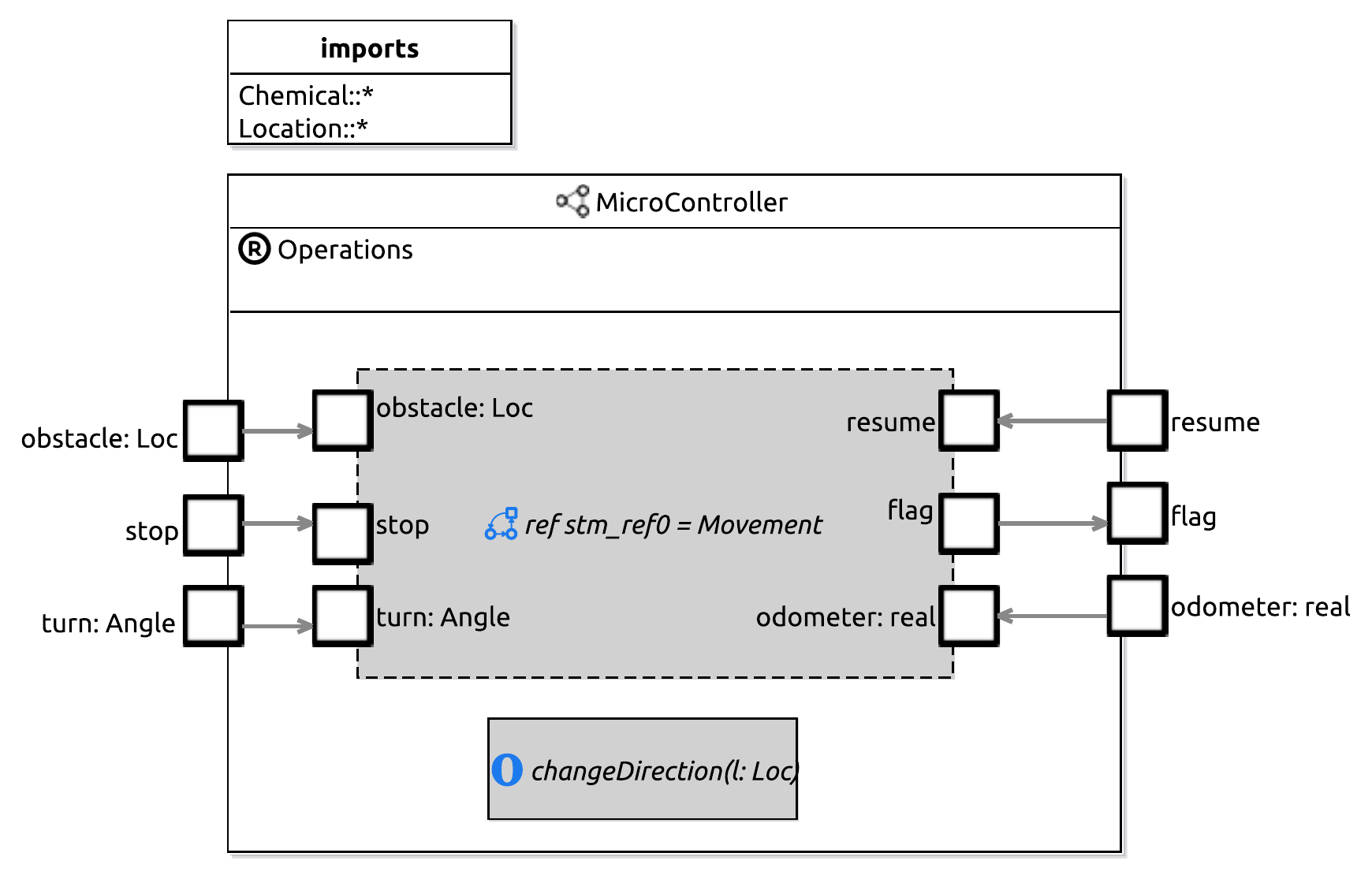}%
  \caption{\rcitem{MicroController} of the autonomous chemical detector model.}%
  \label{fig:robochart_acd_microcontroller}%
\end{figure}

\begin{figure}[t]
  \centering%
  \includegraphics[width=1.0\textwidth]{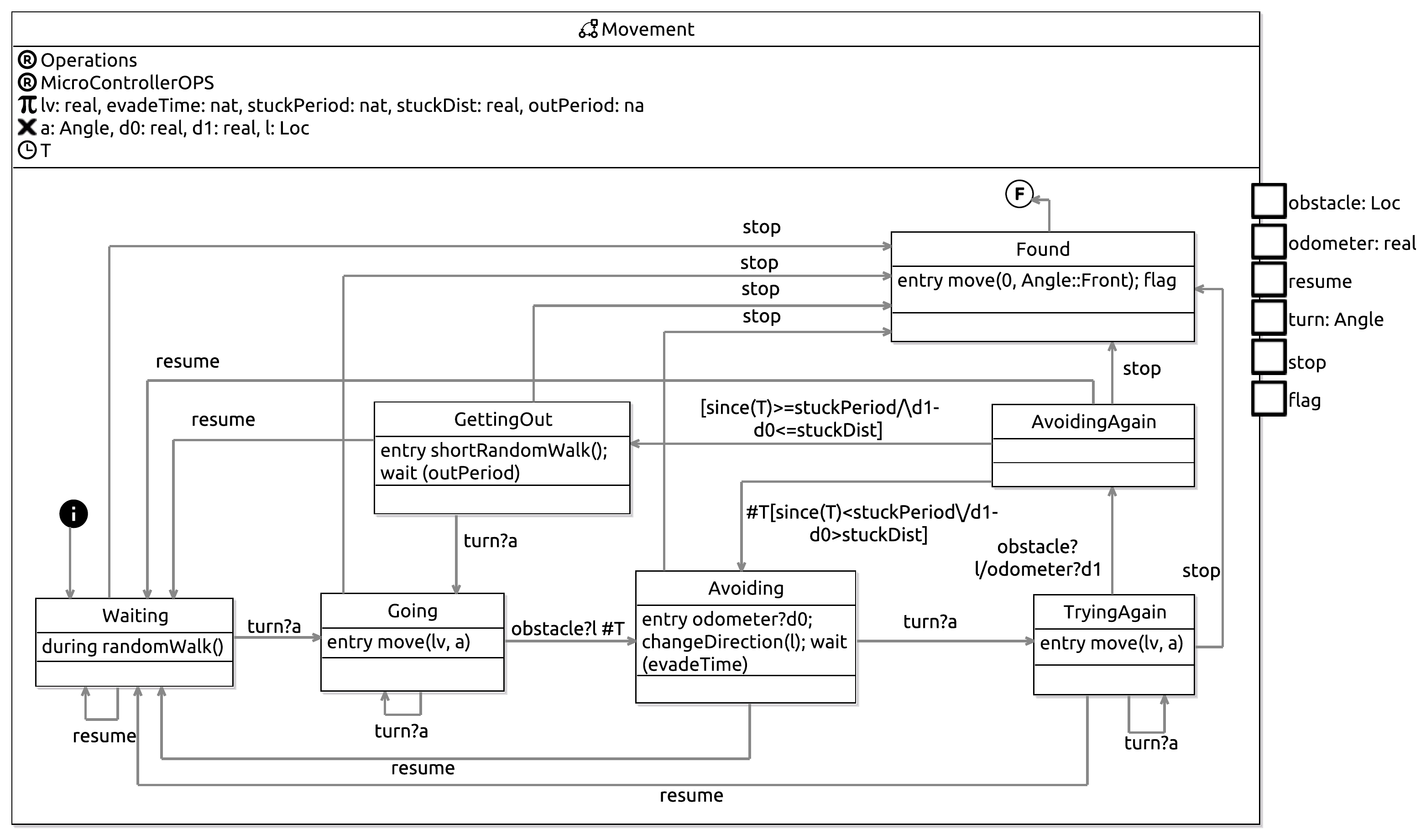}%
  \caption{Movement state machine of the autonomous chemical detector model.}%
  \label{fig:robochart_acd_movement}%
\end{figure}

\rcitem{MicroController}, defined in Fig.~\ref{fig:robochart_acd_microcontroller},  is implemented using a state machine \rcitem{Movement} (by a contained reference), presented in Fig.~\ref{fig:robochart_acd_movement}, and a reference to the operation \changed[\C{40}]{\rcitem{changeDirection}}.
\noindent
The machine \rcitem{Movement} declares various constants such as \rcitem{lv} for linear velocity, four variables (\rcitem{a}, \rcitem{d0}, \rcitem{d1}, and \rcitem{l}) for the preservation of values (angle, odometer readings, and location) carried on communication, and a clock \rcitem{T}. The machine also contains one initial junction, seven normal states such as \rcitem{Waiting} and \rcitem{Going}, and a final state. Notably, the state \rcitem{Waiting} has a \rcitem{during} action, an operation call \rcitem{randomWalk()}, which provides parallelism in a machine and means the robot is doing a random walk \changed[\C{41}]{when waiting for further instructions from \rcitem{MainController} based on the gas analysis result}. This operation can be interrupted at any time as long as a transition from the state is taken.  
The transitions of this machine have labels with various features. The transition from \rcitem{Going} to \rcitem{Avoiding} has an input trigger \rcitem{obstable?l} and a clock reset \rcitem{\rcreset T}, and the transition from \rcitem{TryingAgain} to \rcitem{AvoidingAgain} has an input trigger and an action \rcitem{odometer?d1} (an input communication). The transition from \rcitem{AvoidingAgain} to \rcitem{Avoiding} has a clock reset \rcitem{\rcreset T} and a {disjunctive} guard in which \rcitem{since(T)} counts the elapsed time since the last reset of \rcitem{T}.

This machine gives the behaviour of the robot's response to outcomes of the chemical analysis: 
\begin{inparaenum}[(1)]
\item \rcitem{resume} to state \rcitem{Waiting} {if no gas is detected (implemented in \rcitem{GasAnalysis})};
\item \rcitem{stop} to state \rcitem{Found} and then terminate {if a gas above the threshold is detected};
\item \rcitem{turn} {to the direction, where a gas is detected but not above the threshold, with obstacle avoidance in state \rcitem{Going}};
\item {upon the first detection of an \rcitem{obstacle}, reset \rcitem{T} and start \rcitem{Avoiding} with an initial \rcitem{odometer} reading and the movement direction changed (software \rcitem{wait}s for \rcitem{evadeTime} for the effect of that change);}
\item {if a gas is still detected after the changed direction, \rcitem{TryingAgain} to \rcitem{turn} and \rcitem{move} to the gas direction;}
\item {if another obstacle is detected during avoidance, \rcitem{AvoidingAgain} by reading the \rcitem{odometer} to check the distance of two obstacles;}
\item {if the robot has moved far enough between the two obstacles or not got stuck long enough, go back to continue \rcitem{Avoiding};}
\item {otherwise, the robot has got stuck in a corner, use a \rcitem{shortRandomWalk} for \rcitem{GettingOut} of the area, then resume normal activities.}
\end{inparaenum}

{
\subsubsection{One-dimensional patrol robot}
\label{ssec:robochart_patrol}

\begin{figure}[t]
  \centering%
  \includegraphics[width=0.7\textwidth]{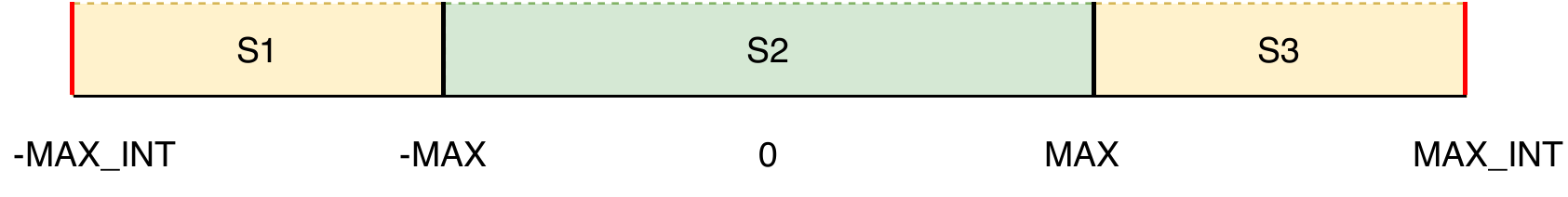}%
  \vspace{0ex}
  \caption{Sections for the patrol robot: S1 - adjacent to the left boundary; S2 - central area; S3 - adjacent to the right boundary.}%
  \vspace{0ex}
  \label{fig:robochart_patrol_sections}%
\end{figure}
In addition to the features introduced in the chemical detector example, RoboChart supports abstraction via nondeterministic choice and interaction via shared variables, exemplified in the model for a one-dimensional patrol robot on a corridor. 
The corridor, as shown in Fig.~\ref{fig:robochart_patrol_sections}, is split into three sections by four boundaries, denoted by their coordinates: \rcitem{-MAX\_INT}, \rcitem{-MAX}, \rcitem{MAX}, and \rcitem{MAX\_INT}. The left and right boundaries are hard and represent the walls, and the other two are soft and represent controlled limits. Section S2, soft boundaries exclusive, is the central working area. The sections S1 and S3, soft boundaries inclusive, are limited, and the robot will tend to move to S2 if it is in these sections. 

The control software of this patrol robot behaves as follows: 
\begin{inparaenum}[(1)]
\item it maintains a belief state ($x$) of the robot, default at 0 (denoting the centre of the corridor) and able to reset to 0 by an event \rcitem{reset};
\item when $x$ is 0, it can be calibrated to the actual position of the robot from its sensor by a \rcitem{cal} event; 
\item when $x$ is within S2, it can be either increased or decreased by 1, denoting the robot moves either to the left or to the right nondeterministically;  
\item when $x$ is within S1, it can only be increased by 1, denoting the robot moves to the right; and 
\item when $x$ is within S3, it can only be decreased by 1, denoting the robot moves to the left.
\end{inparaenum}

\begin{figure}[t]
  \centering%
  \includegraphics[width=1.0\textwidth]{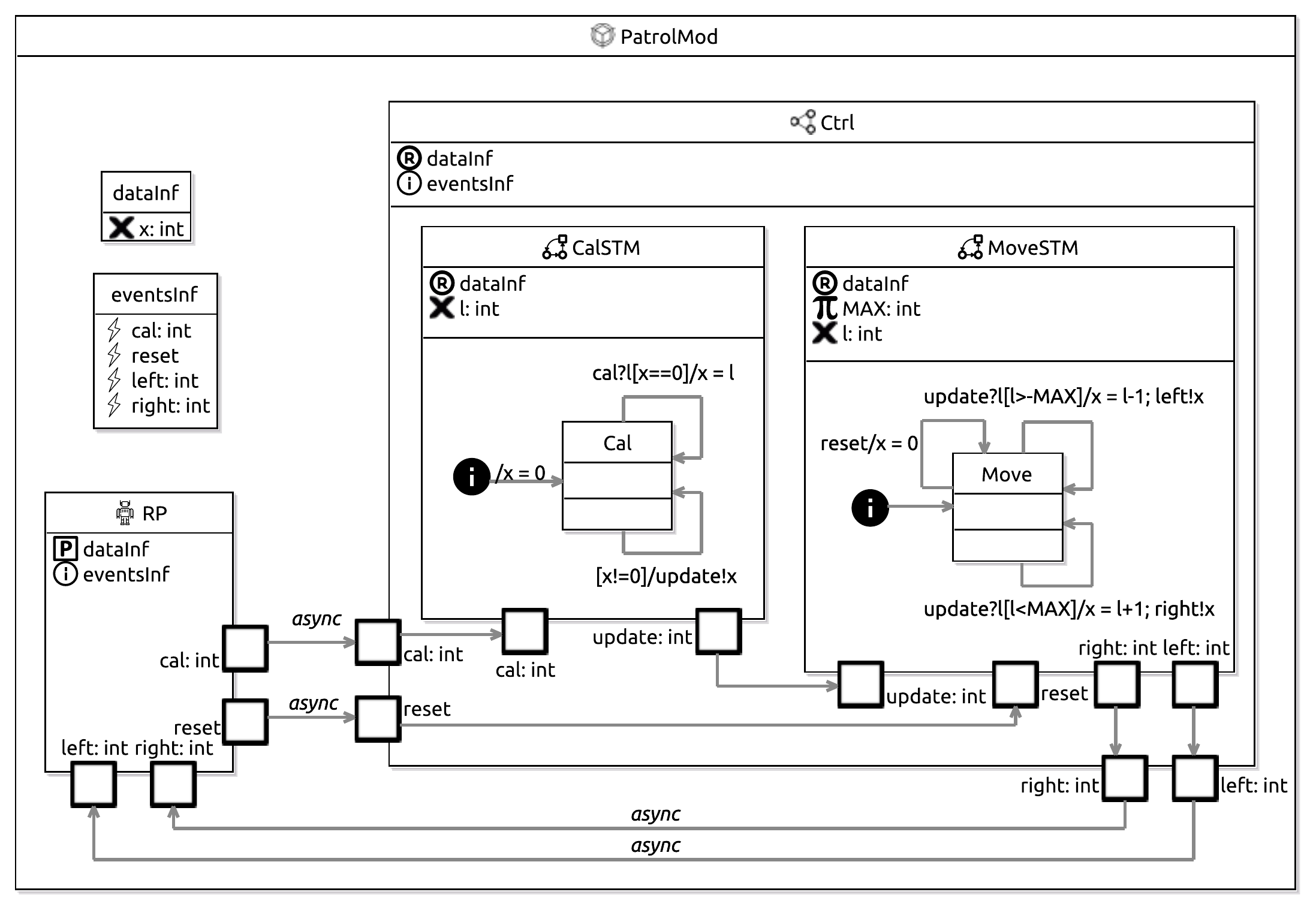}%
  \vspace{0ex}
  \caption{A RoboChart model for a patrol robot with nondeterministic behaviour and interaction using shared variables.}%
  \vspace{0ex}
  \label{fig:robochart_patrol}%
\end{figure}
We illustrate in Fig.~\ref{fig:robochart_patrol} the RoboChart model for this patrol robot. The module \rcitem{PatrolMod} contains a robotic platform \rcitem{RP} and a controller \rcitem{Ctrl} composed of two state machines \rcitem{CalSTM} and \rcitem{MoveSTM}. \rcitem{RP} declares two output events \rcitem{cal} of type \rcitem{int} and \rcitem{reset}, and two inputs events \rcitem{left} and \rcitem{right} (to indicate the direction of moving and its new position) of type \rcitem{int} through the interface \rcitem{eventsInf}, and provides a variable \rcitem{x} of type \rcitem{int} through the interface \rcitem{dataInf}. This variable is shared in \rcitem{Ctrl} and to the two machines by requesting \rcitem{dataInf}.

The machine \rcitem{CalSTM} sets \rcitem{x} to 0 in action (\rcitem{x=0}) of its default transition \rcitem{t0} to state \rcitem{cal}, and then it is ready for calibration from the input trigger \rcitem{cal?l} of the transition \rcitem{t1} if \rcitem{[x==0]}, or to update $x$ (the action \rcitem{update!x} of the transition \rcitem{t2} ) to \rcitem{MoveSTM} through a connection on event \rcitem{update} of type \rcitem{int} otherwise \rcitem{[x!=0]}. We note that it is not mandatory to use this communication mechanism to update \rcitem{x} to \rcitem{MoveSTM} because \rcitem{x} is shared in \rcitem{MoveSTM} too. This illustrates how nondeterminism is introduced in \rcitem{MoveSTM} and how it interleaves with the shared variable. 

The machine \rcitem{MoveSTM} is responsible for resetting \rcitem{x} (\rcitem{x=0}) by the self-transition \rcitem{t2} of state \rcitem{move} with trigger \rcitem{reset}. The robot movement around the corridor is based on the value (of variable \rcitem{x}), which passes on \rcitem{update} and is stored in a local variable \rcitem{l} by the other two self-transitions (\rcitem{t1} and \rcitem{t3}) from state \rcitem{move}. If \rcitem{l} is larger than \rcitem{-MAX} (guard \rcitem{[l>-MAX]} of \rcitem{t3} where \rcitem{MAX} is a constant variable), denoting sections S2 and S3, \rcitem{x} is updated to \rcitem{l-1}, followed by a \rcitem{left!x} event. If \rcitem{l} is less than \rcitem{MAX} (guard \rcitem{[l<MAX]} of \rcitem{t1}), denoting sections S2 and S1, \rcitem{x} is updated to \rcitem{l+1}, followed by a \rcitem{right!x} event. Consequently, if \rcitem{l} is in section S2, the choice between the two transitions is nondeterministic. We note that \rcitem{MAX\_INT} is not explicitly declared in the model and will be given in verification or animation where \rcitem{int} is bounded. 

Next, we describe the extra ITree-based CSP operators for the RoboChart semantics and return to the two models in Sect.~\ref{sec:rc_to_itrees}.
}

\section{Interaction trees}
\label{sec:itree}
This section briefly introduces interaction trees and extends our existing CSP semantics with additional operators to support the RoboChart semantics. These include three operators (interrupt,  exception,  and renaming) introduced in the previous work~\cite{Ye2022}, one generalised choice operator, and two prioritised hiding and renaming operators presented in this extension paper.

Interaction trees (ITrees)~\cite{Xia2019} are a data structure for modelling reactive systems interacting with their environment through events. They are potentially infinite and defined as coinductive trees
~\cite{Blanchette2014} 
in Isabelle/HOL.
\newcommand{\itreedef}{\isalink{https://github.com/isabelle-utp/interaction-trees/blob/418b37554f808828610f10b40c051a562fe0c716/Interaction_Trees.thy\#L22}}
\begin{alltt}
\isakwmaj{codatatype} (\textquotesingle{e}, \textquotesingle{r}) itree = \(\itreedef\) 
  Ret \textquotesingle{r} | Sil "(\textquotesingle{e}, \textquotesingle{r}) itree" | Vis "\textquotesingle{e} \(\pfun\) (\textquotesingle{e}, \textquotesingle{r}) itree"
\end{alltt}
\noindent ITrees are parameterised over two types: \texttt{\textquotesingle e} for events ($E$), and \texttt{\textquotesingle r} for return values or states ($R$) . Three possible interactions are provided: 
\begin{inparaenum}[(1)]
	\item $\Ret~x$: termination with a value $x$ of type $R$ returned, denoted as $\tick_x$;\footnote{\changed[\C{27}]{We note that $\tick_x$ is just a shorthand for $\Ret~x$ and $\tick$ cannot be used alone. In some CSP literature, such as ~\cite{Roscoe2011}, $\tick$ means a special event representing termination. We do not refer to the single $\tick$ in this paper.}}
	\item $\Sil~P$: an internal silent event, denoted as $\tau P$ for a successor ITree $P$; or
	\item $\Vis~F$: a choice among several visible events represented by a partial function $F$ of type $E \pfun (E,R) \cspkey{itree}$.
\end{inparaenum}
Partial functions are part of the Z toolkit\footnote{\url{https://github.com/isabelle-utp/Z_Toolkit}.}~\cite{Spivey1992}, which is also mechanised in Isabelle/HOL. We also use a notation $\vbar\, e\!\in\!E \then P(e)$ for $\Vis~(\lambda e \in E @ P(e))$.

\changed[\C{7}]{Though ITrees are elementary structures, they can be used to encode, animate, and symbolically reason about the complex behaviour of systems of interacting concurrent processes. ITrees is a model that gives semantics to various languages and potentially unifies them.} In particular, deterministic CSP processes can be given executable semantics using ITrees. Determinism is inherent since we use a partial function to model events and their continuations. Precisely, each unique event must map to at most one continuation. The benefit of this approach is that ITrees are easy to implement and animate since the behaviour depends solely on provided input events. Therefore, our CSP operators cannot introduce nondeterminism, which must be statically resolved.

Our animator takes a model with ITree-based semantics in Isabelle/HOL and generates Haskell code for the underlying ITree. Though this is often an infinite structure, Haskell's intrinsic use of lazy evaluation allows us to model and partially evaluate infinite objects. We can, therefore, step through an ITree's behaviour by unfolding the various constructors of the underlying algebraic (co)datatype. When a $\Vis$ constructor is encountered, the user is presented with a choice for one of the enabled events. When a $\Sil$ is encountered, it is removed, and the successor ITree is animated. This allows us to compress long sequences of $\tau$ events and simplified animation. Finally, $\Ret$ leads to the termination of the animation.

Previously\changed[\C{17}]{~\cite{Foster2021}}, the following CSP processes and operators have been defined: 
\begin{inparaenum}[(1)]
	\item basic processes: \cspkey{skip}, \cspkey{stop}, and \cspkey{div}; 
    \item input prefixing: $\cspkey{inp}~c~V$ (communicate any value from \changed[\C{42}]{$V$} over the channel $c$);
	\item output prefixing $c!v$ \changed[\C{43}]{(send a value $v$ over the channel $c$)};
	\item $\cspkey{guard}~b$ \changed[\C{43}]{(guarded based on a Boolean value $b$)};
	\item external choice $P \extchoice Q$ \changed[\C{43}]{between two processes $P$ and $Q$}; 
	\item parallel composition $P \parallel_A Q$;
	\item hiding $P \hide A$; 
	\item sequential composition $P \fatsemi Q$; 
	\item \cspkey{loop} and \cspkey{iterate}.
\end{inparaenum} 
\changed[\C{9},\C{10}]{
We summarise their definitions in Table~\ref{table:existing_csp_itrees} and show some definitions omitted in the table (due to the large space they will take) as follows.

\newcommand{\skippdef}{\isalink{https://github.com/isabelle-utp/interaction-trees/blob/418b37554f808828610f10b40c051a562fe0c716/UTP/ITree_CSP.thy\#L50}}
\newcommand{\deadlockdef}{\isalink{https://github.com/isabelle-utp/interaction-trees/blob/418b37554f808828610f10b40c051a562fe0c716/ITree_Deadlock.thy\#L9}}
\newcommand{\stopdef}{\isalink{https://github.com/isabelle-utp/interaction-trees/blob/418b37554f808828610f10b40c051a562fe0c716/RoboChart/ITree_RoboChart.thy\#L11}}
\newcommand{\divdef}{\isalink{https://github.com/isabelle-utp/interaction-trees/blob/418b37554f808828610f10b40c051a562fe0c716/ITree_Divergence.thy\#L77}}
\newcommand{\rundef}{\isalink{https://github.com/isabelle-utp/interaction-trees/blob/418b37554f808828610f10b40c051a562fe0c716/Interaction_Trees.thy\#L240}}
\newcommand{\runaltdef}{\isalink{https://github.com/isabelle-utp/interaction-trees/blob/418b37554f808828610f10b40c051a562fe0c716/Interaction_Trees.thy\#L243}}
\newcommand{\inpdef}{\isalink{https://github.com/isabelle-utp/interaction-trees/blob/418b37554f808828610f10b40c051a562fe0c716/UTP/ITree_CSP.thy\#L59}}
\newcommand{\outpdef}{\isalink{https://github.com/isabelle-utp/interaction-trees/blob/418b37554f808828610f10b40c051a562fe0c716/UTP/ITree_CSP.thy\#L155}}
\newcommand{\guarddef}{\isalink{https://github.com/isabelle-utp/interaction-trees/blob/418b37554f808828610f10b40c051a562fe0c716/UTP/ITree_CSP.thy\#L161}}
\newcommand{\extchoicedef}{\isalink{https://github.com/isabelle-utp/interaction-trees/blob/418b37554f808828610f10b40c051a562fe0c716/UTP/ITree_CSP.thy\#L424}}
\newcommand{\paralleldef}{\isalink{https://github.com/isabelle-utp/interaction-trees/blob/418b37554f808828610f10b40c051a562fe0c716/UTP/ITree_CSP.thy\#L844}}
\newcommand{\interleavedef}{\isalink{https://github.com/isabelle-utp/interaction-trees/blob/418b37554f808828610f10b40c051a562fe0c716/UTP/ITree_CSP.thy\#L846}}
\newcommand{\hidedef}{\isalink{https://github.com/isabelle-utp/interaction-trees/blob/418b37554f808828610f10b40c051a562fe0c716/UTP/ITree_CSP.thy\#L898}}
\newcommand{\seqdef}{\isalink{https://github.com/isabelle-utp/interaction-trees/blob/418b37554f808828610f10b40c051a562fe0c716/Interaction_Trees.thy\#L142}}
\newcommand{\iteratedef}{\isalink{https://github.com/isabelle-utp/interaction-trees/blob/418b37554f808828610f10b40c051a562fe0c716/ITree_Iteration.thy\#L9}}
\newcommand{\loopdef}{\isalink{https://github.com/isabelle-utp/interaction-trees/blob/418b37554f808828610f10b40c051a562fe0c716/ITree_Iteration.thy\#L12}}

\begin{table}[!htb]
  \setlength\extrarowheight{2pt} 
  \caption{\label{table:existing_csp_itrees} \changed[\C{9}]{A summary of the existing basic CSP operators defined in the previous work~\cite{Foster2021}. The definitions for $\extchoice$, $\Vert_A$, 
  $\interleave$, $\hide$, and $\fatsemi$ are omitted in the table (because their definitions take a large space) and are presented and discussed in the overview of Sect.~\ref{sec:itree}.}}
  \resizebox{\textwidth}{!}{
  \begin{tabularx}{\textwidth}{c|c|X}
    \hline
    \textbf{Operator} & Definition & Description \\
    \hline
    \cspkey{skip} & $\Ret ()$ \skippdef & Terminate immediately, and return a unit type $()$, a degenerate form of $\Ret$.\\ 
    \hline
    \cspkey{stop} & $\Vis~ \{\mapsto \}$ \deadlockdef\stopdef & Deadlock, or no event is possible. \\ 
    \hline
    \cspkey{div} & $\isakwmaj{primcorec}\ \cspkey{div} = \tau \cspkey{div} $ \divdef & ITree of infinite depth, a divergent ITree that does not terminate and only performs internal activity. \\ 
    \hline
    $\cspkey{run}~E$ & $\vbar\, e\!\in\!E \then \cspkey{run}~E$ \runaltdef & Repeatedly perform any event from $E$. We give an alternative definition instead of the original definition for simplicity.\\
    \hline
    $\cspkey{inp}~c~V$ & $\Vis~\left(
        \begin{array}{l}
        \lambda e \in \left(
        \begin{array}{l}
            \dom(\pmatch_c) \\
            \cap \pbuild_c \limg A \rimg
        \end{array}\right) \\
        @ \Ret~(\pmatch_c~e)
        \end{array}\right)$ \inpdef & Accept a constrained input (only the values from $V$) over channel $c$ and return the value. \\
    \hline
    $\cspkey{outp}~c~v$ & $\Vis~\{\pbuild_c~v \mapsto \Ret~()\}$ \outpdef & Send a value $v$ over channel $c$, or denoted as $c!v$.\\
    \hline
    $\cspkey{guard}~b$ & $\IF b \THEN skip \ELSE stop$ \guarddef & Behave as $\cspkey{skip}$ if $b$ is true and otherwise $\cspkey{stop}$. \\
    \hline
    $b \& P$ & $\isakwmin{do} \{ \cspkey{guard}~b; P\}$ & Guarded process. \\
    \hline
    $P \fatsemi Q$ & \seqdef & Sequential composition of $P$ and $Q$.\\
    \hline
    $P \extchoice Q$ & \extchoicedef & External choice, redefined in Sect.~\ref{ssec:itree_genchoice}. \\
    \hline
    $P \parallel_A Q$ & \paralleldef & Parallel composition of $P$ and $Q$ over $A$.\\
    \hline
    $P \interleave Q$ & \interleavedef & Interleave of $P$ and $Q$.\\
    \hline
    $P \hide A$ & 
    \hidedef & Hide the events in $A$ from $P$.\\
    \hline
    $\cspkey{iterate}~b~P~s$ & $\begin{array}{l}
        \IF (b~s) \THEN \\
        \quad\Sil (P~s \mbind \cspkey{iterate}~b~P) \\
        \ELSE \tick_s
    \end{array}$ \iteratedef & Continue to execute $P$ 
    while the condition $b~s$ holds and otherwise terminates and returns the current state $s$. Also called $\cspkey{while}$ in~\cite{Foster2021}.\\
    \hline
    $\cspkey{loop}~P$ & $\cspkey{iterate}~(\lambda s. True)~P$ \loopdef& Infinite loop.\\
    \hline
  \end{tabularx}}
\end{table}

\changed[\C{21}]{External choice $P \extchoice Q$ is defined corecursively. \changed[\C{23}]{A corecursive definition can have several equations ordered by priority.} \isalink{https://github.com/isabelle-utp/interaction-trees/blob/ff9f73f98c653b265bd9da55689715cf973499c1/ITree_CSP.thy\#L75}
\begin{align*}
    (\Vis~F) \extchoice (\Vis~G) &= \Vis~(F \odot G) \\
    (\Sil~P') \extchoice Q &= \Sil~(P' \extchoice Q) \\
    P \extchoice (\Sil~Q') &= \Sil~(P \extchoice Q') \\
    (\Ret~x) \extchoice (\Vis~G) &= \Ret~x \\
    (\Vis~F) \extchoice (\Ret~y) &= \Ret~y \\
    (\Ret~x) \extchoice (\Ret~y) &= (\textit{if}~x = y~\textit{then}~(\Ret~x)~\textit{else}~\skey{stop})
\end{align*}
The merge function $F \odot G \defs (\dom(G) \ndres F) \oplus (\dom(F) \ndres G)$ is used to define the \Vis~case. \changed[\C{19}]{The $\ndres$ is called the domain anti-restriction, and $A \ndres R$ denotes the domain restriction of relation $R$ to the complement of set $A$.} \changed[\C{18}]{The $\oplus$ is a relational overriding operator.} For example, $A \oplus B$ agrees with the relation $B$ and with the relation in $A$ outside the domain of $B$. The relation in $A$ inside the domain of $B$ is overridden by $B$. This function $F \odot G$ combines all event maplets from $F$ and $G$, ignoring any maplets whose events occur in the intersection $\dom(F) \cap \dom(G)$. For example $\{e_1 \mapsto P_1, e_2 \mapsto P_2\} \odot \{e_3 \mapsto P_3, e_2 \mapsto P_4\} = \{e_1 \mapsto P_1, e_3 \mapsto P_3\}$, since $e_2$ is ignored. This avoids nondeterminism; when the two domains are disjointed, the operator can be considered a union.}

Sequential composition ($P \fatsemi Q$) of ITrees is supported through a monadic bind operator ($\mbind$): $P \fatsemi Q \defs \left(\lambda x. P(x) \mbind Q\right)$ which executes $P$ first, and then passes the value of $x$ to its continuation $Q$ upon the termination of $P$. The $P \mbind Q$ is defined corecursively, including several equations ordered by priority.  \isalink{https://github.com/isabelle-utp/interaction-trees/blob/418b37554f808828610f10b40c051a562fe0c716/Interaction_Trees.thy\#L142}
\begin{align*}
    \tick_r \mbind Q &= Q(r) \\
    \tau P' \mbind Q &= \tau(P' \mbind Q) \\
    \Vis~F \mbind Q &= \Vis~(\lambda e \in \dom(F) @ F(e) \mbind Q) 
\end{align*}
With the monadic \isakwmin{do} notation, we can write a sequential composition like the one shown below. 
$$\isakwmin{do}\{x \gets \cspkey{inp}~c~V; outp~d~(x+1) ; \tick_x \}$$ 
This is equivalent to the composition:
$$inp~c~V \mbind \left(\lambda x. outp~d~(x+1) \mbind (\lambda y. \tick_x)\right)$$
This process accepts input from set $V$ on channel $c$, records the value in $x$, passes $x$ to its continuation, which sends $x+1$ on channel $d$, and afterwards terminates and returns the value of $x$.

Parallel composition $P \parallel_E Q$ over a set $E$ of events is defined corecursively below. \isalink{https://github.com/isabelle-utp/interaction-trees/blob/418b37554f808828610f10b40c051a562fe0c716/UTP/ITree_CSP.thy\#L844}
\begin{align*}
    (\Sil~P') \parallel_E Q &= \Sil~(P' \parallel_E Q) \\
    P \parallel_E (\Sil~Q') & = \Sil~(P \parallel_E Q') \\
    (\Vis~F) \parallel_E (\Vis~G) &= 
        \Vis\left(\begin{array}{l}
            \{e \mapsto (P' \parallel_E (\Vis~G)) | (e \mapsto \skey{Left}(P')) \in merge_E(F, G)\} \\
            \oplus~ \{e \mapsto ((\Vis~F) \parallel_E Q') | (e \mapsto \skey{Right}(Q')) \in merge_E(F, G)\} \\
            \oplus~ \{e \mapsto (P' \parallel_E Q') | (e \mapsto \skey{Both}(P', Q')) \in merge_E(F, G)\}
        \end{array}\right) \\
    (\Ret~x) \parallel_E (\Ret~y) &= \Ret~(x, y) \\
    (\Ret~x) \parallel_E (\Vis~G) &= \Vis~\{e \mapsto \left(\Ret~x \parallel_E Q'\right) | (e \mapsto Q') \in G\} \\
    (\Vis~F) \parallel_E (\Ret~y) &= \Vis~\{e \mapsto \left(P' \parallel_E \Ret~y\right) | (e \mapsto P') \in F\}
\end{align*}
The definition of the \Vis\ case uses an operator $merge_E(F, G)$ to merge two event functions. We omit its definition here for simplicity and refer to~\cite{Foster2021} for more details. \changed[\C{44}]{For the sake of presentation, we present partial functions as sets and use set comprehensions for construction.} \changed[\C{18}]{A set comprehension $\{e | P\}$ is a shorthand for $\{e | x~y. P\}$ if $x$ and $y$ are free variables of $e$ and occur in $P$. For example, $\{e \mapsto \left(\Ret~x \parallel_E Q'\right) | (e \mapsto Q') \in G\}$ in the above definition means $\{e \mapsto \left(\Ret~x \parallel_E Q'\right) | e~Q'. (e \mapsto Q') \in G\}$, that is, any maplet $e \mapsto Q'$ (a visible event $e$ and its continuation $Q'$) in the partial function $G$ becomes $e \mapsto \left(\Ret~x \parallel_E Q'\right)$. }

Interleave $P \interleave Q$ is simply the parallel composition over the empty set: $P \parallel_{\emptyset} Q$. \isalink{https://github.com/isabelle-utp/interaction-trees/blob/418b37554f808828610f10b40c051a562fe0c716/UTP/ITree_CSP.thy\#L846}

Hiding $P \hide A$ is defined corecursively below. \isalink{https://github.com/isabelle-utp/interaction-trees/blob/418b37554f808828610f10b40c051a562fe0c716/UTP/ITree_CSP.thy\#L898}
\begin{align*}
    & \Vis~F = \begin{cases}
        \Sil~\left(F(e) \hide A\right) & \IF A \cap \dom(F)=\{e\} \\
        \Vis~\{(e, P \hide A \mid (e, P) \in F\} & \IF A \cap \dom(F)= \emptyset\\
        \cspkey{stop} & \text{Otherwise}
    \end{cases} \\
    & \Sil(P) \hide A = \Sil(P\hide A)\\
    & \Ret x \hide A = \Ret~x
\end{align*}

The hiding operator is restricted to at most one event once to avoid nondeterminism introduced by hiding multiple events. However, hiding more than one event can be achieved through hiding with priority described in Sect.~\ref{ssec:itree_hiding_with_priority}. 

Though operators like external choice and parallelism can introduce nondeterminism, we restrict this by construction. Nevertheless, different strategies \changed[\C{20}]{such as biased operators, which give priority to the left-hand side or right-hand side process of the operators, and priority based on an order of events for hiding or renaming,} can be employed for statically resolving nondeterminism, which we explore further in this section. \changed[\C{20}]{We use these strategies to define alternative operators for different purposes.}
}

Next, we give an ITree semantics to extra CSP operators to allow us to give an ITree-based semantics to RoboChart. We 
\begin{inparaenum}[(1)]
    \item generalise external choice; 
    \item introduce three new operators---interrupt, exception, and renaming operators---used in the RoboChart's semantics to allow interruption of a during-action, termination of a state machine, a controller, or a module, and alphabet transformation of processes; and 
    \item add prioritised variants of renaming and hiding to resolve nondeterminism based on an order statically.
\end{inparaenum}
We restrict ourselves to deterministic operators as it makes the animation of large models more efficient. 

\subsection{Generalised choice and external choice}
\label{ssec:itree_genchoice}

Previously~\cite{Foster2021}, we have given semantics to CSP's external choice operator as a corecursive definition. Our mechanisation of ITrees intrinsically supports external choice through the use of partial functions to model visible events. However, our definition of choice can be generalised to support more flexible choice schemes. For example, priority can be given to one branch of the choice to resolve any nondeterminism statically. We achieve this through a novel generalised choice operator, $\genchoice{P}{\mathcal{M}}{Q}$. For this, we use a merge function $\mathcal{M}$, which merges two choice functions of type $E \pfun (E, R)\textit{itree}$. The operator is defined as a corecursive function using the equations listed below.

\begin{definition}[Generalised choice]\label{def:genchoice}
$ $\isalink{https://github.com/isabelle-utp/interaction-trees/blob/418b37554f808828610f10b40c051a562fe0c716/UTP/ITree_CSP.thy\#L169}
\begin{align*}
\arraycolsep=1.4pt
\begin{array}{rlrl}
    \genchoice{\left(\Vis~F\right)}{\mathcal{M}}{\left(\Vis~G\right)} &= \Vis~\left(\mathcal{M}~F~G\right)
    &\qquad \genchoice{\left(\Ret~x\right)}{\mathcal{M}}{\left(\Ret~y\right)} &= \left(\IF x = y \THEN \Ret~x \ELSE \cspkey{stop}\right)  \\ [3pt]
 \genchoice{\left(\Sil~P'\right)}{\mathcal{M}}{Q} &= \Sil~\left(\genchoice{P'}{\mathcal{M}}{Q}\right) 
 &\qquad \genchoice{P}{\mathcal{M}}{\left(\Sil~Q'\right)} &= \Sil~\left(\genchoice{P}{\mathcal{M}}{Q'}\right) \\ [3pt]
 \genchoice{\left(\Ret~v\right)}{\mathcal{M}}{\left(\Vis~G\right)} &= \Ret~v 
 &\qquad \genchoice{\left(\Vis~F\right)}{\mathcal{M}}{\left(\Ret~v\right)} &= \Ret~v\\ 
\end{array}
\end{align*}
\end{definition}

\noindent When choosing between two visible event functions, $\Vis~F$ and $\Vis~G$, the merge function is applied to combine the two. When combining two return value ITrees, $\Ret~x$ and $\Ret~y$, we require the two possible values to be identical and otherwise deadlocked to avoid nondeterminism. For silent events ($\tau$) and returns, we also prioritise their occurrence before any visible activity can occur. In particular, any $\tau$ events are greedily consumed before any visible event or return can occur.

With generalised choice, we can redefine external choice $P \extchoice Q \defs \genchoice{P}{\odot}{Q}$.

Another benefit of the generalised choice operator is that, as with Hoare and He's parallel-by-merge operator~\cite{Hoare1998}, its properties reduce to the merge function itself. This simplifies proof of algebraic properties for choice functions. The most basic property of a merge function is well-formedness:

\begin{definition}[Wellformed merge function]
A merge function $\mathcal{M}$ is well-formed provided that for any choice function $F$, $\mathcal{M}~\emptyset~F = \mathcal{M}~F~\emptyset = F$.
\end{definition}

\noindent A wellformed merge function has the empty choice function $\emptyset$ as a left and right identity. For example, it is clear that $\odot$ is well-formed because $\dom(\emptyset) = \emptyset$. From this definition of well-formedness, we obtain the following properties.

\begin{thm}[Generalised choice] If $\mathcal{M}$ is well-formed, then 
\isalink{https://github.com/isabelle-utp/interaction-trees/blob/418b37554f808828610f10b40c051a562fe0c716/UTP/ITree_CSP.thy\#L238}
\begin{align*}
    \genchoice{P}{\mathcal{M}}{\cspkey{stop}} =\,& \genchoice{\cspkey{stop}}{\mathcal{M}}{P} = P \\ 
    \genchoice{P}{\mathcal{M}}{\cspkey{div}} =\,& \genchoice{\cspkey{div}}{\mathcal{M}}{P} = \cspkey{div} \\
    \genchoice{P}{\mathcal{M}}{Q} =& \genchoice{Q}{\mathcal{M}^{\sim}}{P}
\end{align*}
\end{thm}

\noindent Generalised choice has $\cspkey{stop}$ as a unit since it can add no further behaviour. Moreover, $\cspkey{div}$ is a zero since $\tau$ events always take priority, so a divergent process prevents choices. We can commute a choice by taking the converse of the merge function, where $\mathcal{M}^{\sim} = (\lambda F~G.\, \mathcal{M}~G~F)$. As a result of the final law, if $\mathcal{M}$ is symmetric, that is $\mathcal{M} = \mathcal{M}^{\sim}$, then choice is commutative: $\genchoice{P}{\mathcal{M}}{Q} = \genchoice{Q}{\mathcal{M}}{P}$. These properties show that external choice is commutative and has $\cspkey{stop}$ as a unit. The former follows because $f \oplus g = g \oplus f$ whenever $\dom{f} \cap \dom{g} = \emptyset$, a property that is ensured by the construction of $\odot$.

We now consider how we can derive alternative choice schemes. In some circumstances, it may be undesirable that possible events are lost by external choice. For example, $a \then P \extchoice a \then Q$ ends as $\cspkey{stop}$ since both processes have $a$ as an initial event. Instead, We can resolve any nondeterminism by prioritising the addition of events from either the left or right branches. We introduce a biased choice operator, $P \extchoicel Q \defs \genchoice{Q}{\oplus}{P}$, which chooses events from $P$ whenever initial events are present in both $P$ and $Q$. For example, $a \then P \extchoicel a \then Q = a \then P$, since the event from the left branch is prioritised.



\subsection{Interrupt}
The second operator we introduce is interrupt~\cite{Hoare1985,Roscoe2011}, $P \interrupt Q$, which behaves like $P$ except that if at any time $Q$ performs one of its initial events, it takes over. This operator, along with the other two, is defined corecursively, which allows them to operate on the infinite structure of an ITree. In corecursive definitions, every corecursive call on the right-hand side of each equation must be guarded by an ITree constructor. 
\begin{definition}[Interrupt]\label{def:interrupt}
$ $\isalink{https://github.com/isabelle-utp/interaction-trees/blob/418b37554f808828610f10b40c051a562fe0c716/UTP/ITree_CSP.thy\#L952}
\begin{align*}
\begin{array}{l}
\left(\Sil~P'\right) \interrupt Q = \Sil~\left(P' \interrupt Q\right)
\qquad P \interrupt \left(\Sil~Q'\right) = \Sil~\left(P \interrupt Q'\right) \\ [1pt]
\left(\Ret~x\right) \interrupt Q = \Ret~x 
\qquad\qquad\quad\  P \interrupt \left(\Ret~x\right) = \Ret~x \\[1pt]
\left(\Vis~F\right) \interrupt \left(\Vis~G\right)  = \Vis \left(
\begin{array}{l}
\left\{ e \mapsto \left(P' \interrupt Q\right) | \left(e \mapsto P' \right) \in \left(\ZKey{$\dom$}(G) \ndres F\right) \right\} 
\oplus G
\end{array}
\right) 
\end{array}
\end{align*}
\end{definition}
%
The $\Sil$ cases allow $\tau$ events to happen independently with priority and without resolving $\interrupt$. The $\Ret$ cases terminate with $x$ returned from $ \ interrupt$'s left or right side.

\changed[\C{45}]{The $\Vis$ case also results in a $\Vis$ process} constructed from an overriding $\oplus$ of the further two sets, representing two partial functions.  In the partial function,  $(\ZKey{$\dom$}(G) \ndres F)$ restricts the domain of $F$ to the complement of the domain of $G$.  
The first partial function denotes that an initial event $e$ of $P$, not the initial event of $Q$, can occur independently (without resolving the interrupt), and its continuation is a corecursive call $P' \interrupt Q$.  
The second function is just $G$,  which 
denotes that the initial events of $Q$ can happen whether they are in $F$ or not. 
If $P$ and $Q$ share events, $Q$ has priority.  This prevents nondeterminism.

\subsection{Exception}
Next, we present the exception operator, $\except{P}{A}{Q}$, which behaves like $P$ initially, but if $P$ ever performs an event from the set $A$, then $Q$ takes over. 
\begin{definition}[Exception]\label{def:exception}
$ $\isalink{https://github.com/isabelle-utp/interaction-trees/blob/418b37554f808828610f10b40c051a562fe0c716/UTP/ITree_CSP.thy\#L1012}
\begin{align*}
\begin{array}{l}
\except{\left(\Ret~x\right)}{A}{Q} = \Ret~x \qquad 
\except{\left(\Sil~P'\right)}{A}{Q} = \Sil~\left( \except{P'}{A}{Q} \right) \\[1pt]
\except{\left(\Vis~F\right)}{A}{Q}
 = 
\Vis \left(
\begin{array}{l}
\left\{ e \mapsto \left(\except{P'}{A}{Q}\right) | \left(e \mapsto P' \right) \in (A \ndres F) \right\} 
\oplus \\
\left\{ e \mapsto Q | e \in (A \cap \ZKey{$\dom$}(F)) \right\}
\end{array}
\right) 
\end{array}
\end{align*}
\end{definition}
The $\Ret$ case terminates immediately with the value $x$ returned, and $Q$ will not be performed. The $\Sil$ case consumes the $\tau$ event.


Similar to Definition~\ref{def:interrupt}, the $\Vis$ case is also represented by the overriding of two partial functions. The first partial function means that an initial event $e$ of $P$ that is not in $A$ (that is, $e \in \ZKey{$\dom$}(A \ndres F)$)  can occur independently. Its continuation is a corecursive call $\except{P'}{A}{Q}$. Following the execution of an initial event $e$ of $P$ that is in $A$ (that is,  $e \in (A \cap \ZKey{$\dom$}(F)$),  the exception behaves like $Q$, which is expressed by the second partial function.

\subsection{Renaming}
\label{ssec:itree_renaming}
The other new operator we define for this work is renaming, $\rename{P}{\rho}$, which renames events of $P$ according to the renaming relation $\rho:E_1 \rel E_2$, which is equivalent to $\power (E_1 \cross E_2)$. 
This relation is possibly heterogeneous, so $E_1$ and $E_2$ are different types of events.
First, we define an auxiliary function for making a relation functional by removing any pairs with duplicate distinct values. This is the case when the renaming relation is functional, restricted to the initial events of $P$.
\begin{align*}
mk\_functional(R) = \{(x, y) \in R. \forall y'. (x, y') \in R \implies y=y' \}
\end{align*}
This produces the minimal functional relation that is consistent with $R$. For example, 
\begin{align*}
mk\_functional \left(\{e_1 \mapsto e_2, e_1 \mapsto e_3, e_2 \mapsto e_3 \}\right) = \{e_2 \mapsto e_3\}
\end{align*}
This function avoids nondeterminism introduced by renaming multiple events to the same event. We use this function to define the renaming operator.
\begin{definition}[Renaming]\label{def:rename}
$ $\isalink{https://github.com/isabelle-utp/interaction-trees/blob/418b37554f808828610f10b40c051a562fe0c716/UTP/ITree_CSP.thy\#L1048}
\begin{align*}
\begin{array}{l}
\rename{\left(\Ret~x\right)}{\rho} = \Ret~x \\[1pt] 
\rename{\left(\Sil~P'\right)}{\rho} = \Sil~\left( \rename{P'}{\rho} \right) \\[1pt]
\rename{\left(\Vis~F\right)}{\rho}
 = \left(
\begin{array}{l}
\LET 
G = F \circ mk\_functional\left( (\ZKey{$\dom$}(F) \dres \rho)^{\inv}\right) \\
@ \Vis \left(
\begin{array}{l}
\left\{ e_2 \mapsto \left(\rename{P'}{\rho}\right) | \left(e_2 \mapsto P' \right) \in G \right\}
\end{array}
\right) 
\end{array}
\right)
\end{array}
\end{align*}
\end{definition}
The $\Ret$ case behaves like $P$, and the renaming does not affect it. The $\Sil$ case allows $\tau$ events to be consumed since they are not subject to renaming. 

In the $\Vis$ case, $G$ is a partial function ($E_2 \pfun (E_1, R)\cspkey{itree}$) that is the backward partial function composition $\circ$ of $F$ and a partial function made using $mk\_functional$ from the inverse $\inv$ of the relation $(\ZKey{$\dom$}(F) \dres \rho)$ which is the domain restriction $\dres$ of $\rho$ to the domain $\ZKey{$\dom$}(F)$ of $F$. 
The multiple events of $E_1$ that are mapped to the same event of $E_2$ in $\rho$ and also are the initial events of $P$, or in $\ZKey{$\dom$}(F)$, are removed in $G$. 
The renaming result is a partial function in which each event $e_2$ in the domain of $G$ is mapped to a renamed process by a corecursive call $\rename{P'}{\rho}$ where $\left(e_2 \mapsto P'\right) \in G$. 

{
 The potential nondeterminism is excluded because many-to-one mappings in $\rho$ are removed by $mk\_functional$ in $G$. For example, 
 \begin{align*}
 &\rename{\left(
 e_1 \then P \extchoice 
 e_2 \then Q \extchoice 
 e_3 \then R
 \right)}{
 \{
 e_1 \mapsto e, 
 e_2 \mapsto e, 
 e_3 \mapsto ea,
 e_4 \mapsto eb
 \}} \\
 =& \left(ea \then \rename{R}{\{
 e_1 \mapsto e, 
 e_2 \mapsto e, 
 e_3 \mapsto ea,
 e_4 \mapsto eb
 \}}\right) \tag*{(renaming example 1)} \label{eqn:renaming_example}
 \end{align*}
 Here, the only available initial event after renaming is $ea$ because the relation $(\ZKey{$\dom$}(F) \dres \rho)^{\inv}$ is equal to $\{e \mapsto e_1, e \mapsto e_2, ea \mapsto e_3\}$ and $mk\_functional$ removes the first two pairs (because of duplicate distinct values), and so results in $\{ea \mapsto e_3\}$.

\begin{rmk}
Roscoe~\cite{Roscoe2011} defines three ways for renaming in CSP. Injective functional renaming will not change the behaviour of a CSP process and is an alternative to parametrised CSP processes on channels. Non-injective functional renaming may change the behaviour of a process by ignoring some level of detail or introducing nondeterminism. Relational renaming, a more general and powerful operator than the other two functional renamings, allows many-to-one (may introduce nondeterminism) or one-to-many (to offer more choice) mappings. What machine-readable CSP (CSP-M) supports and FDR implements is relational renaming. We investigated all these approaches and chose to implement the relational renaming with many-to-one and one-to-many mappings. This is mainly because RoboChart's semantics is defined using CSP-M and verified using FDR. For many-to-one mappings, our definition here, however, blocks these many events, and the definition of renaming with priority in Sect.~\ref{ssec:itree_renaming_with_priority}, as follows, chooses one of these many events according to their priority.
\end{rmk}

\subsection{Hiding with priority}
\label{ssec:itree_hiding_with_priority}
The current semantics~\cite{Foster2021} of hiding $P \hide A$ is deadlock if more than one initial event of $P$ is in $A$. This is to avoid nondeterminism caused by the hiding of two possible events. This restriction can be relaxed by hiding events in an order. For example, $\left(a \then P \extchoice b \then Q\right) \hide \{a, b\} = stop$, but $\left(\left(a \then P \extchoice b \then Q\right) \hide \{a\} \right) \hide \{b\} = \tau \left(\left(P\hide \{a\}\right) \hide \{b\}\right)$, and $\left(\left(a \then P \extchoice b \then Q\right) \hide \{b\} \right) \hide \{a\} = \tau \left(\left(Q\hide \{b\}\right) \hide \{a\}\right)$. Hiding events in a different order resolves the external choice differently without deadlock. This difference is due to the maximal progress assumption of hiding: $\left(a \then P \extchoice b \then Q\right) \hide \{a\}$ is equal to $\tau \left(P\hide \{a\}\right)$.

We define hiding with priority, $P \hidep el$ (\isaref{https://github.com/isabelle-utp/interaction-trees/blob/418b37554f808828610f10b40c051a562fe0c716/RoboChart/ITree_RoboChart.thy\#L17}), to put these events to be hidden in an order based on their order in a list $el$.
\begin{align*}
    & P \hidep el = foldl\left(\left(\lambda Q~e.~Q \hide \{e\}\right), P, el\right) &
\end{align*}
The $foldl$ builds a return value by applying the function $\left(\lambda Q~e.~Q \hide \{e\}\right)$ (say $f$) to the combined result (initially $P$) and elements in $el$ based on their orders. For example, $foldl\left(f, P, [a,b]\right)$ will be expanded to $f(f(P, a), b)$, which is just $\left(P \hide \{a\}\right) \hide \{b\}$.
The above examples now can be expressed as $\left(a \then P \extchoice b \then Q\right) \hidep [a,b]$ and $\left(a \then P \extchoice b \then Q\right) \hidep [b,a]$.

\subsection{Renaming with priority}
\label{ssec:itree_renaming_with_priority}
Because the relation $\rho$ of type $E_1 \rel E_2$ in the renaming definition~\ref{def:rename} is possibly heterogeneous (so $E_1$ and $E_2$ are different types), renaming cannot be like hiding with priority to place the events to be renamed in an order to rename events one by one. This is because renaming a process changes its event type from $E_1$ to $E_2$, so we can no longer rename other events of type $E_1$. For this reason, we define renaming with priority, $\renamep{P}{\varrho}$, which renames events of $P$ according to a finite sequence $\varrho$ of type $\seq\left(E_1 \cross E_2\right)$. We use the finite sequence type here to describe the mathematical definition of this operator, and its representation in Isabelle is \changed[\C{47}]{a list}. \changed[\C{48}]{A finite sequence of type, $\seq X$, is a finite partial function $\nat \ffun X$ from natural numbers $\nat$ to $X$.} With $\varrho$, a priority is given based on the indices of pairs in the sequence to resolve potential nondeterminism in a particular way. For pairs with the same second element (in other words, many-to-one mappings), the pair with the smallest index has the highest priority. This renaming with a priority operator will only rename the event with the highest priority and block other events with lower priority. 

Before defining $\renamep{P}{\varrho}$, we need to define another two functions. 
The first function is the domain restriction $\dresl$ of $\varrho$ to a set $A$. 
\begin{align*}
    A \dresl \varrho = squash~\left\{s : \seq\left(E_1 \cross E_2\right) | s \in \varrho \bullet (s.2).1 \in A\right\} \tag*{\isalink{https://github.com/isabelle-utp/interaction-trees/blob/418b37554f808828610f10b40c051a562fe0c716/RoboChart/ITree_RoboChart.thy\#L31}}
\end{align*}
This function produces a new sequence, compacted from a function, or a set of ordered pairs, in which each member $s$ is in $\varrho$ and the first element $(s.2).1$ (of type $E_1$) of the second element $s.2$ (of type $E_1 \cross E_2$) of $s$ is in $A$, by the $squash$ function~\cite{Spivey1992}. \changed[\C{48}]{Here, we use the selection operator ($s.i$) to select the \emph{i}th element in a tuple $s$.} When a type is obvious, we use a short form $x \in A \bullet P(x)$ for $x : T | x \in A \bullet P(x)$. This could be used in set comprehension, quantification, etc.
%

We give an example below to illustrate how $\dresl$ works.
\begin{align*}
    \{e_1, e_2, e_4\} \dresl \langle (e_1 , e), (e_2 , e), (e_3 , ea), (e_4 , eb) \rangle = \langle (e_1 , e), (e_2 , e), (e_4 , eb) \rangle
\end{align*}
For the second function $drop\_dup(\varrho)$, we need to drop the pairs with lower priority (and bigger indices). We define an auxiliary $least$ function first. 
\begin{align*}
    & least(\varrho, p) \defs \left(\nexists q \in \varrho \bullet q.1 < p.1 \land (q.2).2 = (p.2).2\right) \tag*{($least$ definition)} \label{def:drop_least}
\end{align*}
This function characterises if $p$ (of type $\nat \cross (E_1 \cross E_2)$) has the least index number $p.1$ in the members $q$ of $\varrho$ that have the same target event $(q.2).2$ (of type $E_2$) as that $(p.2).2$ of $p$. In other words, $p$ has the least index number in a set of renaming pairs with multiple events renamed to the same event as in $p$. Now $drop\_dup(\varrho)$ is defined below.
%
\begin{align*}
    & \left(\forall p \in drop\_dup(\varrho) \bullet least(\rho, p)\right) \land \tag*{(maximal)} \label{def:drop_maximal}\\ 
    & \left(\forall p \in \varrho \bullet least(\varrho, p) \implies p \in drop\_dup(\varrho)\right) \tag*{(minimal)} \label{def:drop_minimal}
\end{align*}
The first predicate \ref{def:drop_maximal} in the conjunction states that every element in the resultant $drop\_dup(\varrho)$ is $least$ in $\varrho$, and the second predicate \ref{def:drop_minimal} states that every least element in $\varrho$ must be in the resultant $drop\_dup(\varrho)$. 
For example, 
\begin{align*}
    drop\_dup\left( \langle (e_1 , e), (e_2 , e), (e_4 , eb) \rangle \right) = \langle (e_1 , e), (e_4 , eb) \rangle
\end{align*}
Here, the pair $(e_2, e)$ is dropped because it does not have the highest priority in terms of the target event $e$ (because $e$ appears early in $(e_1, e)$). We can define $\renamep{P}{\varrho}$ corecursively with the two functions above.

\begin{definition}[Renaming with priority]\label{def:rename_priority}
$ $\isalink{https://github.com/isabelle-utp/interaction-trees/blob/418b37554f808828610f10b40c051a562fe0c716/RoboChart/ITree_RoboChart.thy\#L56}
\begin{align*}
\begin{array}{l}
\renamep{\left(\Ret~x\right)}{\varrho} = \Ret~x \\[1pt] 
\renamep{\left(\Sil~P'\right)}{\varrho} = \Sil~\left( \renamep{P'}{\varrho} \right) \\[1pt]
\renamep{\left(\Vis~F\right)}{\varrho}
 = \left(
\begin{array}{l}
\LET 
G = F \circ mk\_functional\left( \left(\ZKey{$\ran$}\left(drop\_dup\left(\ZKey{$\dom$}(F) \dresl \varrho\right)\right)\right)^{\inv}\right) \\
@ \Vis \left(
\begin{array}{l}
\left\{ e_2 \mapsto \left(\renamep{P'}{\varrho}\right) | \left(e_2 \mapsto P' \right) \in G \right\}
\end{array}
\right) 
\end{array}
\right)
\end{array}
\end{align*}
\end{definition}

This definition is similar to Definition~\ref{def:rename} except that $\varrho$ here is a sequence of renaming pairs, and the relation in the inverse now has all its domain elements mapped to distinct values by $drop\_dup$. Because $drop\_dup$ defines a sequence of type $\nat \ffun (E_1 \cross E_2)$, we get the range of the sequence by the $\ZKey{$\ran$}$ function, which is a relation. 

The difference between renaming and renaming with priority is exemplified below. 
 \begin{align*}
 &\renamep{\left(
 e_1 \then P \extchoice 
 e_2 \then Q \extchoice 
 e_3 \then R
 \right)}{
     \langle (e_1 , e), (e_2 , e), (e_3 , ea), (e_4 , eb) \rangle
 } \\
 = &\left(
 \begin{array}[]{l}
 e \then \renamep{P}{ \langle (e_1 , e), (e_2 , e), (e_3 , ea), (e_4 , eb) \rangle }  \extchoice  \\
 ea \then \renamep{R}{ \langle (e_1 , e), (e_2 , e), (e_3 , ea), (e_4 , eb) \rangle }
 \end{array}
\right) \tag*{(renaming example 2)} \label{eqn:renaming_priority_example}
 \end{align*}

Compared to \changed[\C{49}]{the \ref{eqn:renaming_example} where renaming excludes nondeterminism}, the potential nondeterminism introduced by renaming both $e_1$ and $e_2$ to $e$ is resolved by giving priority to the renaming map $(e_1, e)$. If $(e_1 , e)$ and $(e_2 , e)$ are swapped, then the renaming will give priority to $e_2$ and its continuation is $Q$ (instead of $P$).

}


\section{RoboChart semantics in interaction trees}
\label{sec:rc_to_itrees}

In this section, we describe how we give semantics to RoboChart regarding ITrees in Isabelle/HOL. These include types, instantiations, functions, state machines, controllers, and modules. In implementing RoboChart's semantics, we also consider the practical details of the CSP semantics generation in RoboTool, such as naming and bounded primitive types. 

In particular, the nondeterministic choice between transitions, for example, in Fig.~\ref{fig:robochart_patrol}, is resolved using the prioritised renaming operator defined earlier through ordered transition event mappings. This is not covered in the previous work~\cite{Ye2022} and is the new contribution of this paper.

{
\subsection{Overview of RoboChart semantics}
\label{ssec:semantics_overview}

The RoboChart CSP semantics is sketched in Fig.~\ref{fig:robochart_semantics}. The semantics for modules, controllers, state machines, and (either composite or basic) states are CSP processes. However, the semantics for a robotic platform is different from controllers and state machines in that it does not have a particular behaviour. So, its semantics is not a CSP process because the platform is an abstraction of a physical robot through variables, events, and operations.
\begin{figure}[t]
  \centering%
  \includegraphics[width=1.00\textwidth]{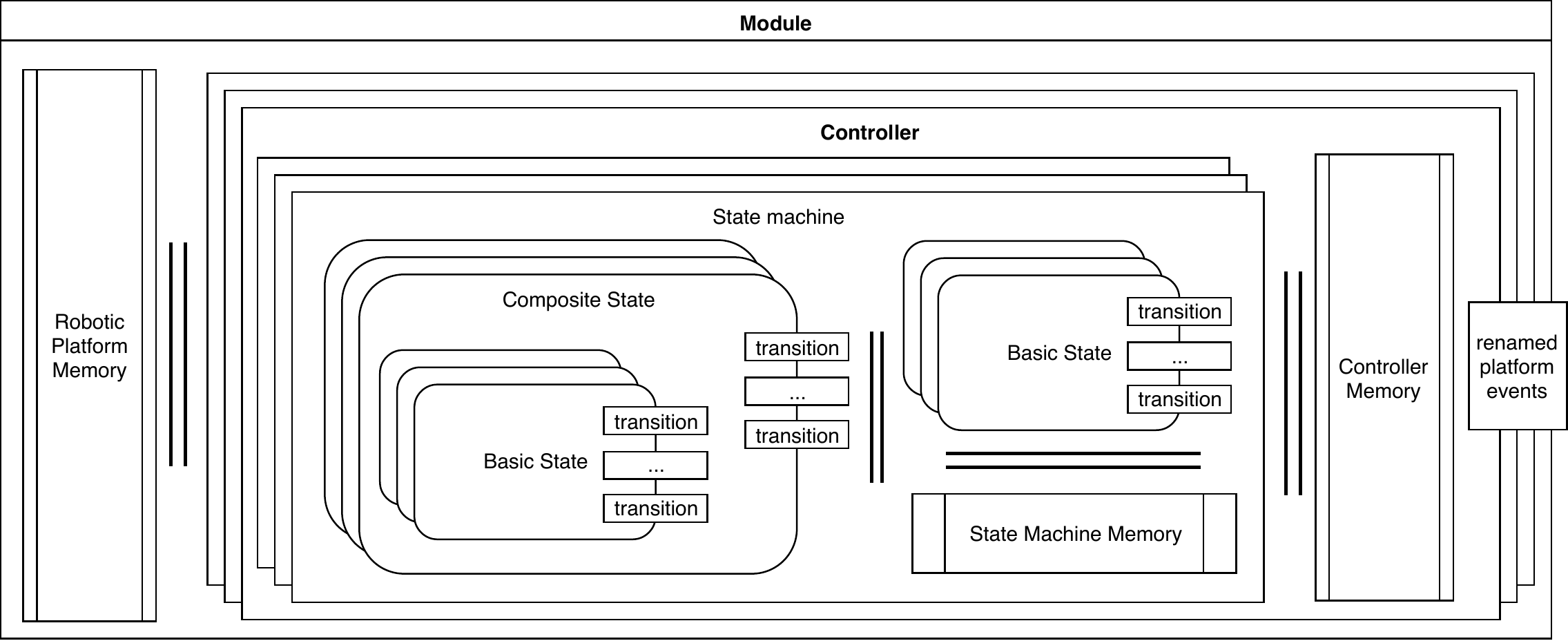}%
  \caption{From~\cite[Fig.~11]{Miyazawa2019}. Structure of the RoboChart semantics: stacked components and parallel lines indicate parallel composition; bordered boxes indicate points of interaction; the semantics of a container is composed of the semantics of its contained components.}%
  \label{fig:robochart_semantics}%
\end{figure}

RoboChart has a hierarchical memory model with memory for the robotic platform at the top, memories for controllers in the middle, and memories for state machines inside their container controllers. A memory process in scope records the reads and writes of variables for components (the platform, controllers, and state machines). The memory for a state machine caches the (both local and shared) variables it requires. So, the semantics of the machine is independent of the location where these variables are declared. However, the memory for a controller or the platform differs from that of a state machine in that it not only accepts updates to the variables in the memory but also propagates the updates down the hierarchy to the memories of state machines that require the updated variables.

The RoboChart semantics of the autonomous chemical detector model in Sect.~\ref{ssec:robochart_chemical} is shown in Fig.~\ref{fig:robochart_acd_semantics}. The module's semantics is a parallel composition of the two controllers with the robotic platform memory (RP memory) and a buffer process (Buffer). 
The buffer process models the asynchronous connection from \rcitem{MainController} to \rcitem{MicroController} on event \rcitem{turn} in Fig.~\ref{fig:robochart_acd_module}. For simplicity, this semantics for asynchronous connections is omitted in Fig.~\ref{fig:robochart_semantics}.

\begin{figure}[t]
  \centering%
  \includegraphics[width=1.00\textwidth]{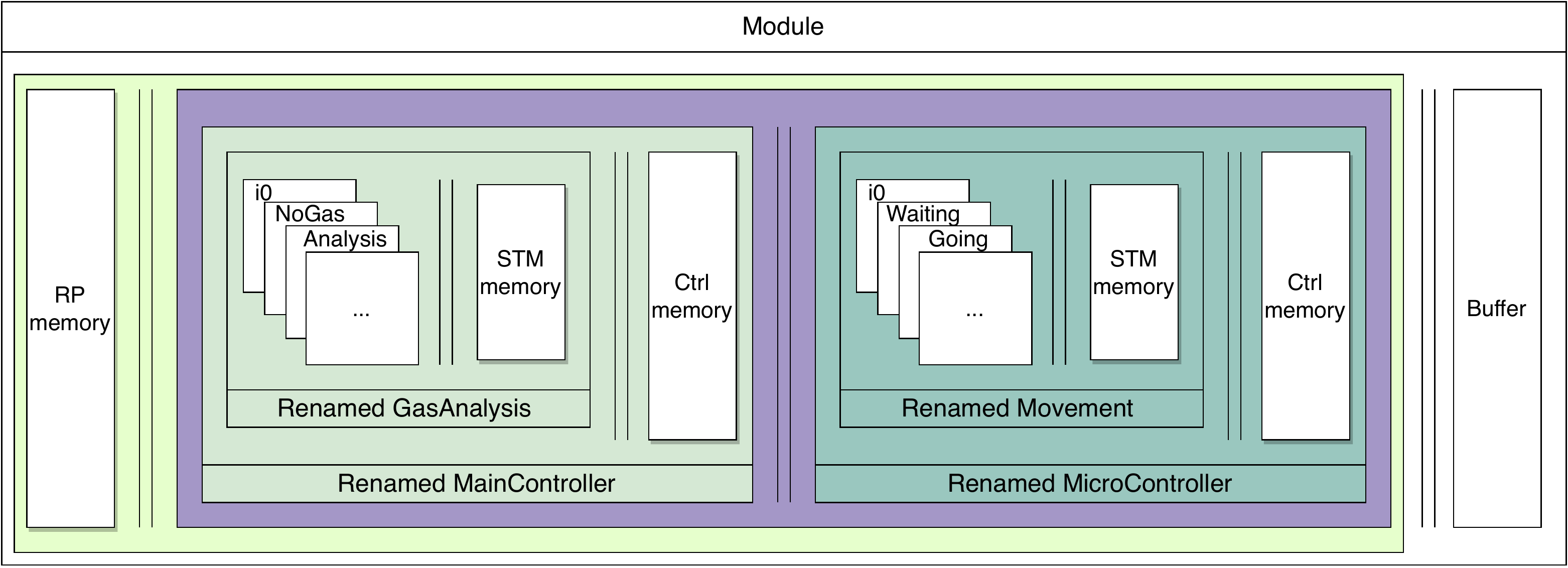}%
  \caption{RoboChart semantics of the autonomous chemical detector.}%
  \label{fig:robochart_acd_semantics}%
\end{figure}

The platform and controller memories for this chemical detector model are simply the CSP process \cspkey{skip} because the platform and the controllers do not provide any variables for sharing. The platform memory and the controller memory for the one-dimensional patrol robot in Sect.~\ref{ssec:robochart_patrol}, nevertheless, record the access and update of the shared variable $x$ and propagate its update to the controller, and finally to the state machines in the controller.

The semantics of \rcitem{MainController} (or \rcitem{MicroController}) is the composition of the CSP process for the semantics of its contained state machine \rcitem{GasAnalysis} (or \rcitem{Movement}) with the controller memory. We note the CSP processes for the controllers are renamed, such as {Renamed MainController}. This is because RoboChart uses directed connections for communication, but CSP's parallel composition requires composites to have the same channel names for communication. We, therefore, need to rename the exported channels of the controller CSP processes to ensure the channel names for the two controllers in a connection are the same so they can communicate with each other.

Similarly, the CSP processes for the semantics of the state machines (\rcitem{GasAnalysis} and \rcitem{Movement}) are also renamed for the same reason to ensure the state machines in a controller can communicate on directed connections. The renaming is not compulsory for the chemical detector model because each controller contains only one state machine. Still, it is mandatory for the patrol robot because the controller includes two state machines connected on event \rcitem{update}. So, in general, we always rename them.

The semantics of the state machine \rcitem{GasAnalysis} is a parallel composition of the node (including the junction \rcitem{i0} and the states \rcitem{NoGas}, etc.) processes with the machine memory recording the access and update of the local variables of the machine. The semantics of the machine \rcitem{Movement} has a similar structure. The memories of the state machines \rcitem{CalSTM} and \rcitem{MoveSTM} in Fig.~\ref{fig:robochart_patrol}, nonetheless, record not only their local variables \rcitem{l} but also the shared variable \rcitem{x}.
}

{
\paragraph{Practical consideration}
In this paper, our ITree-based semantics for RoboChart aims for consistency with the restricted CSP semantics generated in RoboTool.
Here, the restriction mainly refers to bounded types for basic RoboChart types and corresponding closed arithmetic operators. In the future, we could use unbounded basic types with usual operators, then enforce enumerations just in the animator, not in the semantics like this work.
One advantage of using this restricted semantics is the possibility of reusing the current CSP semantics generator in RoboTool to generate corresponding ITrees-based theories for Isabelle to generate Haskell code automatically.

RoboTool automatically generates multiple versions of CSP semantics for a RoboChart model, including a standard version (\textbf{S}), an optimised version (\textbf{D}) that reduces internal interaction to some extent, an optimised and visible version (\textbf{VS}) that further exposes internal interaction, and an optimised and compressed version (\textbf{O}) with compressions using strong bisimulation and diamond elimination~\cite{Roscoe2011}. All these versions use some compressions. Our ITree-based semantics for RoboChart in this paper is based on the \textbf{D} version because RoboTool uses this version for verification. Our semantics here does not support compressions in the \textbf{D} version. Instead, we implement a form of compression for internal events in the animator.  

}

\subsection{Types}
\label{ssec:semantics_type}
RoboChart has its type system based on the Z notation~\cite{Toyn2002}. It supports basic types: \rcitem{PrimitiveType}, \rcitem{Enumeration}, records (or schema types), and other additional types from the mathematical toolkits of Z.

The \rcitem{core} package of RoboTool provides five primitive types: \changed[\C{50}]{booleans, natural numbers, integer, real numbers, and strings}. 
We map integers, naturals, and strings onto the corresponding types in Isabelle/HOL, but with support for code generation to target language types. This improves the efficiency of evaluation and, thus, animation. We define the mappings below where we use the notation $\semL{\rcitem{rc}}{c}=\isacode{ic}$ to denote a mapping from a RoboChart entity \rcitem{rc} into an ITree-based CSP counterpart \isacode{ic} in the context $c$. In other words, the semantics of \rcitem{rc} in the context $c$ is $\isacode{ic}$ in our ITree-based CSP.
%
\begin{align*}
    & \meta{\semL{\rcitem{bool}}{t}} = \isacode{bool} \quad
    \meta{\semL{\rcitem{nat}}{t}} = \isacode{natural} \quad 
    \meta{\semL{\rcitem{int}}{t}} = \isacode{integer} 
\end{align*}
Here, the context $t$ means the type. 

For a record type \rcitem{T} (\rcitem{datatype T}) such as \rcitem{GasSensor} in Fig.~\ref{fig:robochart_acd_chemical} in RoboChart, we use \isakwmaj{record} in Isabelle. Its semantics is defined in the rule below.
\begin{align*}
    & \meta{\semL{\rcitem{datatype T}}{t}} = \quad
    \isacode{record }\meta{T}\isacode{ = } \meta{\left\{f : T.fields @ f.name \right.} \isacode{::''}\meta{\semL{f.type}{t}}\isacode{''}\meta{, \left.\right\}} 
\end{align*}
We also use the notation $\meta{a}$ to denote elements to be expanded, where we use the constructs from Z as a meta-notation and use $\isacode{a}$ to denote the resultant construct in our ITree-based CSP in Isabelle. 
The rule results in the definition of a record type $\meta{T}$ in Isabelle, which contains a set of fields corresponding to the fields ($\meta{T.fields}$) in \rcitem{T}. Here we use set comprehension $\meta{\left\{x : T_x @ expr(x)\right\}}$ from Z. For the sake of presentation, we introduce a comma $\meta{,}$ after the expression $\meta{expr(x)}$ to represent a character to separate each expression after the set is converted to concrete syntax in Isabelle. In the rule above, the character is empty, so there is a blank space to separate each expression. In the above rule, for each field $\meta{f}$, we use its name $\meta{f.name}$. The semantics of its type ($\meta{\semL{f.type}{t}}$) to construct each field in Isabelle by using Isabelle's syntax (\isacode{name::''type''}).

For an enumeration type such as \rcitem{Angle} and \rcitem{Status} in Fig.~\ref{fig:robochart_acd_chemical}, we use \isakwmaj{datatype} in Isabelle to define it.
\begin{align*}
    & \meta{\semL{\rcitem{enumeration T}}{t}} = \quad
    \isacode{datatype }\meta{T}\isacode{ = } 
        \meta{\left\{l : T.literals @ l.name, \right.} \isacode{|}\meta{\left.\right\}} 
\end{align*}
Similar to the rule for the record type, we use set comprehension to get a set of constructors: one for each literal $\meta{l}$ in the enumeration type \rcitem{T}. In Isabelle, only the name $\meta{l.name}$ of the literal matters and constructors are separated by \isacode{|}.

RoboChart models can also have abstract primitive types (\rcitem{type T}) with no explicit constructors, such as \rcitem{Chem} and \rcitem{Intensity} in the chemical detector model presented in Sect.~\ref{sec:robochart}. 
We map primitive types to finite enumerations for code generation. 
We define a finite type \isacode{PrimType} parametrised over two types: 
\isacode{\textquotesingle{t}} for specialisation and a numeral type \isacode{\textquotesingle{a}} for the number of elements.
\newcommand{\PrimType}{\isalink{https://github.com/isabelle-utp/interaction-trees/blob/418b37554f808828610f10b40c051a562fe0c716/RoboChart/ITree_RoboChart.thy\#L74}}
\begin{alltt}
\isakwmaj{datatype} (\textquotesingle{t}, \textquotesingle{a}::finite) PrimType = PrimTypeC \textquotesingle{a} \(\PrimType\)
\end{alltt}

We show the rule to generate \changed[\C{52}]{the semantics for primitive types} below.
\begin{align*}
  & \meta{\semL{\rcitem{type T}}{t}} = \metanobar{\left\{
      \begin{array}{@{}l}
          \isacodebl{typedef }\meta{T}\isacodebl{T = ''\{()\}'' by auto}\\
          \isacodebl{type\_synonym (\textquotesingle{a})} \meta{ T} \isacodebl{ = ''(}\meta{T}\isacodebl{T, \textquotesingle{a}) PrimType''}
      \end{array}
      \right.}
\end{align*}
This rule results in two statements in Isabelle: first to declare a new type $\meta{T}\isacode{T}$ using \isakwmaj{typedef}, and then the corresponding type $\meta{T}$ is just a synonym for \isacode{PrimType} instantiated to $\meta{T}\isacode{T}$. \changed[\C{52}]{We note that $\meta{T}\isacode{T}$ can be any type, and here we choose the type \isacode{\{()\}} containing only the unit \isacode{()}}. 

RoboChart additionally supports a large collection of data types from Z. We show the rules for set, product, and sequence types below and refer to the counterparts in the Z toolkit for other types.
\begin{align*}
    & \meta{\semL{\rcitem{Set(T)}}{t}} = \meta{\semL{T}{t}} \isacode{ set} \qquad 
    \meta{\semL{{T_1 * T_2}}{t}} = \meta{\semL{T_1}{t}} \isacode{ $\cross$ } \meta{\semL{T_2}{t}}\qquad
    \meta{\semL{\rcitem{Seq(T)}}{t}} = \meta{\semL{T}{t}} \isacode{ blist[\textquotesingle{n}]}
\end{align*}
The rules for set and product types are straightforward. \rcitem{Seq(T)} in RoboChart is a type representing an infinite set of all finite (any length) sequences of elements of type $T$. We need to bind the size of the set and also the length of sequences for code generation. 
For this reason, we define bounded lists or sequences \isacode{(\textquotesingle{a}, \textquotesingle{n}::finite) blist} 
 over two parametrised types: \isacode{\textquotesingle{a}} for the type of elements and a finite type \isacode{\textquotesingle{n}} for the maximum length of each list. We introduce a notation \isacode{\textquotesingle{a} blist[\textquotesingle{n}]} for this type. So the rule above for \rcitem{Seq(T)} results in a type with bounded sequences. We describe the definition of bounded sequences in Isabelle below.

\newcommand{\blist}{\isalink{https://github.com/isabelle-utp/Z_Toolkit/blob/90eefcdbc0dfd93cb1db6342a916add57a174381/Bounded_List.thy\#L10}}
\begin{alltt}
\isakwmaj{typedef} (\textquotesingle{a},\textquotesingle{n}::finite) blist = \{xs::\textquotesingle{a} list. length xs \(\leq\) CARD(\textquotesingle{n})\} \(\blist\)
\end{alltt}
\noindent 
\isacode{CARD} here retrieves the cardinality of \isacode{\textquotesingle{n}}. If \isacode{\textquotesingle{a}} is a finite type, then \isacode{\textquotesingle{a} blist[\textquotesingle{n}]} also defines a finite type. We define several functions: 
\begin{inparaenum}[(a)]
\item \isacode{blength} to get the length of a bounded sequence; 
\item \isacode{bnth} to get the nth element of a bounded sequence; 
\item \isacode{bappend} ($\text{@}_s$) to concatenate two bounded sequence; and 
\item \isacode{bmake} to construct a bounded sequence from a finite list. 
\end{inparaenum} 
\changed[\C{16}]{These functions are lifted from the corresponding functions for lists in Isabelle. Additionally, we instantiate the type to be comparable and enumerable, and so the equality of two bounded lists of the same type \isacode{\textquotesingle{a} blist[\textquotesingle{n}]} can be established, and all elements in such a bounded type can be enumerated. Equality and enumerability are essential for code generation in Isabelle.}

\begin{example}[types in the autonomous chemical detector]
    Using the rule $\semL{\varg}{t}$, we get \changed[\C{11}]{the corresponding definitions} in Isabelle for the RoboChart types in Fig.~\ref{fig:robochart_acd_chemical}. 
\newcommand{\ChemicalChem}{\isalink{https://github.com/isabelle-utp/interaction-trees/blob/81ca0b6ded0cfaf8e4ea1744520daed3d79fd441/RoboChart/examples/RoboChart_ChemicalDetector_autonomous/RoboChart_ChemicalDetector_autonomous_general.thy\#L121}}
\newcommand{\ChemicalStatus}{\isalink{https://github.com/isabelle-utp/interaction-trees/blob/418b37554f808828610f10b40c051a562fe0c716/RoboChart/examples/RoboChart_ChemicalDetector_autonomous/RoboChart_ChemicalDetector_autonomous_general.thy\#L190}}
\newcommand{\ChemicalGasSensor}{\isalink{https://github.com/isabelle-utp/interaction-trees/blob/418b37554f808828610f10b40c051a562fe0c716/RoboChart/examples/RoboChart_ChemicalDetector_autonomous/RoboChart_ChemicalDetector_autonomous_general.thy\#L266}}
\begin{alltt}
\isakwmaj{typedef} ChemT = "{()}"  \(\ChemicalChem\)
\isakwmaj{type_synonym} (\textquotesingle{a}) Chem = "(ChemT, \textquotesingle{a}) PrimType"
\isakwmaj{abbreviation} ChemC::"('a::finite \(\Rightarrow\) \textquotesingle{a} Chem)" \isakwmin{where} "ChemC \(\equiv\) PrimTypeC"
\isakwmaj{datatype} Status = Status_noGas | Status_gasD \(\ChemicalStatus\)
\isakwmaj{record} \textquotesingle{a} GasSensor = gs\_c :: "\textquotesingle{a} Chem"   gs\_i :: "\textquotesingle{a} Intensity" \(\ChemicalGasSensor\)
\end{alltt}

An example of a finite type \isacode{\textquotesingle{a}} is the numeral type in Isabelle, such as type \isacode{2}, which contains two elements: zero (\texttt{0::2}) and one (\texttt{1::2}). To construct an element of type \rcitem{Chem}, we define a dedicated constructor \isacode{ChemC}, which is simply a type cast of \isacode{PrimTypeC}. Finally, we use \isacode{(ChemC 0::2)} and \isacode{(ChemC 1::2)} to construct such two elements of type \isacode{(2 Chem)}.

An instantiation type \isacode{(2 GasSensor)} is a record containing two fields of finite types \isacode{(2 Chem)} and \isacode{(2 Intensity)}, both of which have two elements. We now can \changed[\C{53}]{use record brackets \isacode{\(\lparr\)\(\ldots\)\(\rparr\)} in Isabelle to} construct an element \isacode{\(\lparr\)gs\_c = Chem (1::2), gs\_i = IntensityC (0::2)\(\rparr\)} of this type: the chemical is 1, and the intensity is 0.  \qed

For the type \rcitem{Seq(GasSensor)} in RoboChart, its bounded type in Isabelle is \isacode{(2 GasSensor) blist[2]}, which denotes the length of sequences bounded to 2 and elements (in the sequences) of type \isacode{(2 GasSensor)}.
We now can use \isacode{bmake TYPE(2) [\(\lparr\)gs\_c = Chem (0::2), gs\_i = IntensityC (0::2)\(\rparr\), \(\lparr\)gs\_c = Chem (1::2), gs\_i = IntensityC (1::2)\(\rparr\)]} to construct a sequence containing two sensor readings. \qed

\end{example}

\subsection{Instantiations}
\label{ssec:semantics_inst}
The \cspcode{instantiation.csp} file of the CSP semantics contains common definitions used by all models for verification using FDR. These include the definitions of bounded core types such as \rcitem{core\_int} 
 and arithmetic operators under which these bounded types are closed. 
We use \isacode{\isakwmaj{locale}}~\cite{Ballarin2004} in Isabelle to define these for reuse in all models.
 Locales allow us to characterise abstract parameters (such as \isacode{min\_int} and \isacode{max\_int}, to define \changed[\C{54}]{the} bounded core type \rcitem{core\_int}) and assumptions in a local theory context.
 %
 \newcommand{\RoboChartConf}{\isalink{https://github.com/isabelle-utp/interaction-trees/blob/418b37554f808828610f10b40c051a562fe0c716/RoboChart/ITree_RoboChart.thy\#L115}}
 \begin{alltt}
\isakwmaj{locale} robochart_confs = 	\(\RoboChartConf\)
  \isakwmin{fixes} min_int::"integer" \isakwmin{and} max_int::"integer" \isakwmin{and} max_nat::"natural" \isakwmin{and} 
        min_real::"integer" \isakwmin{and} max_real::"integer"
\isakwmin{begin} ...  \isakwmin{end}
 \end{alltt}
%
\vspace{-2ex}
Here, we omit the definitions for simplicity.
%
%
In the theory of Isabelle for a RoboChart model, we instantiate this locale using \isakwmaj{interpretation} (\isaref{https://github.com/isabelle-utp/interaction-trees/blob/418b37554f808828610f10b40c051a562fe0c716/RoboChart/examples/RoboChart_ChemicalDetector_autonomous/RoboChart_ChemicalDetector_autonomous_general.thy\#L97}), which allows us to assign concrete values for the parameters. 

\begin{example}[instantiation]
    An example is shown below that instantiates the parameters (limits) of the locale to \changed[\C{24},\C{56}]{\texttt{-2, 2, ...}}  etc. 
\newcommand{\LocaleInter}{\isalink{https://github.com/isabelle-utp/interaction-trees/blob/418b37554f808828610f10b40c051a562fe0c716/RoboChart/examples/RoboChart_ChemicalDetector_autonomous/RoboChart_ChemicalDetector_autonomous_general.thy\#L97}}
 \begin{alltt} 
 \isakwmaj{interpretation} rc: robochart_confs "-2" "2" "2" "0" "1". \(\LocaleInter\)
 \end{alltt}
\vspace{-1em}
Then we can use \isacode{rc.core\_int\_set} and \isacode{rc.Plus} to access the instantiated definitions in the locale. \qed
\end{example}
 
\subsection{Functions}
\label{ssec:semantics_funcs}
Functions in RoboChart benefit from \changed[\C{56}]{the rich expressions and Z toolkit in Isabelle.} The expressions that are not supported in CSP-M, such as logical quantification, 
are naturally present in Isabelle. 
 Using the code generator, the preconditions and postconditions of a function definition can be solved effectively \changed[\C{30}]{(thanks to Isabelle's data refinement in the code generation)}, which is impossible in CSP-M and FDR. 

 We define the semantics of a function definition \rcitem{f} (of type \rcitem{Function} --- a class in RoboChart's metamodel) in RoboChart below.
\begin{align*}
    & \meta{\semLC{f : Function}{F}} = \\
    & \,\metanobar{\left\{
      \begin{array}{@{}l}
          \isacodebl{\isakwmaj{definition} ''pre\_}\meta{f.name~\langle p: f.parameters @ p.name, \rangle} \isacodebl{ = } \meta{\left\{c: f.preconditions @ \semL{c}{e}, \right.}\isacodebl{$\land$}\meta{\left.\right\}} \isacodebl{''}\\
          \isacodebl{\isakwmaj{definition} ''}\meta{f.name~\langle p: f.parameters @ p.name, \rangle} \isacodebl{ = (THE result.} \meta{\left\{c: f.postconditions @ \semL{c}{e}, \right.}\isacodebl{$\land$}\meta{\left.\right\}} \isacodebl{)''}
      \end{array}
      \right.}
\end{align*}
Here, the context $\mathcal{F}$ means functions. One function $\meta{f}$ corresponds to two definitions in Isabelle: one for its preconditions and one for its postconditions. This is due to the semantics of such a function \rcitem{f} in RoboChart: a \changed[\C{51}]{Boolean} guard \cspcode{(pre(f)$\guard$~P)} where \cspcode{pre(f)} is the preconditions of \rcitem{f} and \rcitem{f} is called in process $P$, and so if the preconditions are not satisfied, the semantics deadlocks. The name of the first definition has the name $\meta{f.name}$ of $\meta{f}$ with a prefix \isacode{pre\_}. Then, in the definition, the name is followed by a sequence of parameter names, constructed from sequence comprehension $\meta{\langle p: f.parameters @ p.name, \rangle}$, which is similar to set comprehension except that the result is a sequence instead of a set. We use the sequence here because the order of parameters matters. The body of the definition is a set of the semantics for preconditions, given by $\meta{\semL{c}{e}}$ in the context of expressions $\meta{e}$, combined using a conjunction operator \isacode{$\land$}. The second definition for the postconditions is similar except that a definite description ({\isacode{THE result}, denoting the unique \isacode{result} such that the predicate holds}) is used to return the \isacode{result} of the function.

\begin{example}[functions defined in the autonomous chemical detector]
 As mentioned in Sect.~\ref{ssec:robochart_chemical}, three (\rcitem{goreq}, \rcitem{angle}, and \rcitem{analysis}) among the five functions defined in the Chemical package of the autonomous chemical detector model in Fig.~\ref{fig:robochart_acd_chemical} are unspecified, and two (\rcitem{analysis} and \rcitem{intensity}) among them are specified in the original model. 
 Our model in Fig.~\ref{fig:robochart_acd_chemical} specifies all five functions.
 With the capability of solving preconditions and postconditions of functions introduced in our work, we detect two problems in the definitions of the two specified functions in the original model, which are corrected in our model. Next, we present the semantics of the two functions (\rcitem{intensity} and \rcitem{location}) by $\meta{\semLC{\varg}{F}}$ in our implementation.

 The \rcitem{intensity} function defined in Fig.~\ref{fig:robochart_acd_chemical} has a precondition (\includegraphics[align=c,height=8pt]{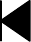}) that the length (\rcitem{size}) of the parameter \rcitem{gs} is more significant than 0, and two postconditions (\includegraphics[align=c,height=8pt]{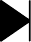}) involving universal and existential quantifications {where \rcitem{@} separates constraint and predicate parts, and \rcitem{goreq} is a $\geq$ relation on intensities.  The \rcitem{result} of the function is the largest intensity in \rcitem{gs}}.
For verification with FDR in RoboTool, an explicit implementation of this function must be supplied \changed[\C{30}]{through the \cspcode{instantiation.csp} file or the assertion language for RoboChart~\cite{RoboChartRef}}.
%
%
However, our definition of this function in Isabelle is directly \changed[\C{57}]{generated} from its specification and is shown below.
\newcommand{\pregs}{\isalink{https://github.com/isabelle-utp/interaction-trees/blob/418b37554f808828610f10b40c051a562fe0c716/RoboChart/examples/RoboChart_ChemicalDetector_autonomous/RoboChart_ChemicalDetector_autonomous_general.thy\#L341}}
\newcommand{\postgs}{\isalink{https://github.com/isabelle-utp/interaction-trees/blob/418b37554f808828610f10b40c051a562fe0c716/RoboChart/examples/RoboChart_ChemicalDetector_autonomous/RoboChart_ChemicalDetector_autonomous_general.thy\#L348}}
 \begin{alltt}
\isakwmaj{definition} "pre\_Chemical\_intensity\ gs = (blength gs > 0)" \(\pregs\)
\isakwmaj{definition} "Chemical\_intensity gs = (THE result. \(\postgs\)
    (\(\forall\)x::nat < blength gs. Chemical\_goreq(result, gs\_i (bnth gs x)))\(\land\)
    (\(\exists\)x::nat < blength gs. result = gs\_i (bnth gs x)))"
 \end{alltt}
\vspace{-3ex}
\noindent In the definitions, \isacode{blength gs} gives the length of a bounded sequence \isacode{gs}, \isacode{bnth gs n} gives the nth element in \isacode{gs}, and \isacode{gs\_i} returns the field value in a record of type \rcitem{GasSensor}. 

We note that there is an error in the definition of \rcitem{intensity} in the original model where $\leq$ (instead of $<$) is used for comparison between \rcitem{x} (and \rcitem{y}) and \rcitem{size(gs)}. {This is because sequences are zero-indexed.} Our animation detects this error 
and so we have fixed it. 

%

The \rcitem{location} function defined in Fig.~\ref{fig:robochart_acd_chemical} has one precondition and one postcondition; their corresponding definitions in Isabelle are shown below. The \rcitem{result} of this function is the location (on the right or in the front of the robot, according to the definition of the \rcitem{angle} function) where the largest intensity in \rcitem{gs} is detected.
\newcommand{\prelocation}{\isalink{https://github.com/isabelle-utp/interaction-trees/blob/418b37554f808828610f10b40c051a562fe0c716/RoboChart/examples/RoboChart_ChemicalDetector_autonomous/RoboChart_ChemicalDetector_autonomous_general.thy\#L386}}
\newcommand{\postlocation}{\isalink{https://github.com/isabelle-utp/interaction-trees/blob/418b37554f808828610f10b40c051a562fe0c716/RoboChart/examples/RoboChart_ChemicalDetector_autonomous/RoboChart_ChemicalDetector_autonomous_general.thy\#L389}}
 \begin{alltt}
\isakwmaj{definition} "pre\_Chemical\_location\ gs = (blength gs > 0)" \(\prelocation\)
\isakwmaj{definition} "Chemical\_location gs = (THE result. (\(\exists\)x::nat<blength gs. \(\postlocation\)
    (Chemical\_intensity(gs) = gs\_i (bnth gs x)) \(\land\)
    (\(\lnot\) (\(\exists\)y::nat < x. (Chemical\_intensity(gs) = gs\_i (bnth gs y)))) \(\land\)
    (\(\forall\)x::nat < blength gs. Chemical\_goreq(result, gs\_i (bnth gs x)))
    (result = Chemical\_angle(x))))"
 \end{alltt}
\vspace{-3ex}
Inside the definite description is an existential quantification over an index \isacode{x} of \isacode{gs} such that (a conjunction of three conjuncts)  
\begin{inparaenum}[(1)]
\item the intensity \isacode{gs\_i (bnth gs x)} of the sensor reading at index \isacode{x} is the largest intensity (\isacode{Chemical\_intensity(gs)}) in \isacode{gs}; 
\item there does not exist an index \isacode{y} such that \isacode{y} is less than \isacode{x} and the intensity at \isacode{y} is also the largest intensity in \isacode{gs} (in other \changed[\C{58}]{words}, \isacode{x} is the smallest index whose intensity is the largest); and   
\item the result of the function is just the angle \isacode{Chemical\_angle(x)} of \isacode{x}.
\end{inparaenum}

We also found another error in the postcondition of \rcitem{location} in the original model: the postcondition is not strong enough to identify a unique \rcitem{result} of the function for the same input (and so the result is a relation and not a function). Our function definition fixes this problem by specifying that \isacode{x} is the smallest index.

\changed[\C{59}]{While there are problems with the definitions of \rcitem{intensity} and \rcitem{location} in the original model\footnote{\url{robostar.cs.york.ac.uk/case_studies/autonomous-chemical-detector/autonomous-chemical-detector.html\#version4}}, their previous implementations\footnote{The implementations can be found in the \cspcode{instantiation.csp} file in the original model available at \url{robostar.cs.york.ac.uk/case_studies/autonomous-chemical-detector/autonomouschemicaldetector_v4.zip}.} (defined in CSP-M and shown below) in CSP for verification with FDR, however, are correct.} 
\begin{alltt}
intensity(gs) = let 
  aux(<>,max) = max
  aux(<g>^gs,max) = 
    if GasSensor_i(g) > max then aux(gs,GasSensor_i(g)) else aux(gs,max)
within
  aux(gs,0)

location(gs) = let
  aux(<>,max,n) = Front
  aux(<g>^gs,max,n) = if GasSensor_i(g) == max then angle(n) else aux(gs,max,n+1)
within
  aux(gs,intensity(gs),0)
\end{alltt}
\changed[\C{59}]{Both definitions make use of pattern matching (an empty sequence \cspcode{<>} or a non-empty sequence \cspcode{<g>\^{}gs}) to decompose values, initially starting (after \cspcode{within}) from the function application of \cspcode{aux} to \cspcode{gs}. The pattern matching, therefore, starts from the beginning of the sequence \cspcode{gs}. The definition of \cspcode{intensity} does not refer to the length of \cspcode{gs} and so avoids the comparison error introduced by using $\leq$ in its specification in the model. The definition of \cspcode{location} starts the comparison with the maximum intensity \cspcode{intensity(gs)} in \cspcode{gs} from the beginning of the sequence, and so the result is the \cspcode{angle} of the smallest index \cspcode{n} whose corresponding intensity is maximal. This is deterministic, and so avoids the problem in the specification of \cspcode{intensity} in the original model.} \qed
\end{example}

\changed[\C{59}]{The manual implementations, however, lead to an inconsistency between models and their semantic implementation for verification.} Our approach presented here translates the specification of functions from models to their semantic implementation \changed[\C{60}]{directly in Isabelle}, so the consistency is preserved. This is one benefit of our approach.
 
{
\subsection{Channels and alphabet transformation}
\label{ssec:semantics_channels}
%
A CSP process in our semantics is an interaction tree of type $(E,R) \cspkey{itree}$, parametrised over a type $E$ of events representing the event alphabet ($\Sigma$) of the CSP process and a type $R$ of return values. 
$E$ is declared through a \isakwmaj{chantype} command \changed[\C{25}]{, created in our ITree-based CSP,} and expressed as a finite set of channels declared in the command.  
We show an example to illustrate the creation of such an event alphabet \isacode{chan}.
\begin{alltt}
\isakwmaj{chantype} chan = 
  Input::integer \(\qquad\) Output::integer \(\qquad\)State::"integer list"
\end{alltt}
\noindent 
This \isacode{chan} declares three channels: \isacode{Input} and \isacode{Output} of type \isacode{integer}, and \isacode{State} of type \isacode{integer list}. 

External choice $P \extchoice Q$ requires that $P$ and $Q$ have the same type $(E, R) \cspkey{itree}$ and the same event type $E$. 
Parallel composition $P \parallel_A Q$ additionally requires that the type $R$ of return values is empty (\isacode{()} or \isacode{unit} in Isabelle) because CSP processes in parallel usually do not return data. So both $P$ and $Q$ should be the type of $(E, ()) \cspkey{itree}$. $A$ is a set of events and of type $\power E$\changed[\C{26}]{, the power set of $E$}.

The CSP processes with different types must be transformed into the same type before composition. The alphabet transformation of a process $P$ of type $(E_1,R) \cspkey{itree}$ is through renaming (or renaming with priority) according to a renaming relation $\rho$ of type $E_1 \rel E_2$ (or a finite sequence $\varrho$ of type $\seq\left(E_1 \cross E_2\right)$), and the resultant process $\rename{P}{\rho}$ (or $\renamep{P}{\varrho}$) is of type $(E_2,R) \cspkey{itree}$.

We also note that the renaming relation or the finite sequence for the renaming operators is ``total'' --- only events of type $E_1$ in the relation or the sequence are renamed. Others are blocked --- no matter whether this is a homogeneous ($E_1$ and $E_2$ are the same) or heterogeneous ($E_1$ and $E_2$ are different) renaming. This differs from the relational renaming operator in CSP-M, where the relation is partial. So, the renaming does not affect the events not in the relation. This difference is because ITrees or Isabelle's terms are strongly typed. 

In our ITree-based semantics for RoboChart, illustrated in Fig.~\ref{fig:robochart_acd_semantics} for the autonomous chemical detector model, each state machine, controller, or module has a different event alphabet by declaring an individual channel type. For example, the state machine \rcitem{GasAnalysis}, the controller \rcitem{MainController}, and the module \rcitem{ChemicalDetector} declare channel types \isacode{Chan\_\-GasAnalysis}, \isacode{Chan\_\-MainController}, and \isacode{Chan\_\-ChemicalDetector}. The renamed {GasAnalysis} transforms the event alphabet of \rcitem{GasAnalysis} from \isacode{Chan\_\-GasAnalysis} to \isacode{Chan\_\-MainController}, and so it can be composed in parallel with the controller memory (whose event type is \isacode{Chan\_\-MainController}). Similarly, the two controllers are renamed to the event alphabet \isacode{Chan\_\-ChemicalDetector} of the module to be composed in parallel with the robotic platform memory and buffer processes.


}

\subsection{State machines}
\label{ssec:semantics_stm}


The RoboChart semantics of a state machine is a parallel composition of memory processes for its variables (\cspcode{MemoryVar}) and transitions (\cspcode{MemoryTrans}), and a process (\cspcode{STM}) for its behaviour with internal events hidden and also catering for its termination using the exception operator. 

\cspcode{STM} is a parallel composition of the behaviour (\cspcode{STM\_I}) for its initial junction and the restricted behaviour (\cspcode{S\_R}) for each state \cspcode{S} synchronising on state entering and exiting events. A state's behaviour \cspcode{S} involves entering this state, the execution of its during-action, and the execution of one of its transitions. The execution of a transition exits the state, executes the action of the transition, and enters the target state of the transition. {Not all transitions are available for \cspcode{S}, such as the transitions from sibling states of \rcitem{S} and substates of \rcitem{S}. These transitions are excluded in the restricted behaviour \cspcode{S\_R}.}

\subsubsection{General definitions}
\label{ssec:semantics_stm:gen_def}
The state machine semantics uses a general type \cspcode{InOut} for the direction of an event in a connection.
\newcommand{\InOut}{\isalink{https://github.com/isabelle-utp/interaction-trees/blob/418b37554f808828610f10b40c051a562fe0c716/RoboChart/ITree_RoboChart.thy\#L162}}
 \begin{alltt} 
\isakwmaj{datatype} InOut = din | dout \(\InOut\)
\end{alltt}

Every state machine also has two data types for state and transition identifiers (\cspcode{SIDS} and \cspcode{TIDS}), and an event alphabet ($E$) for the process of this state machine. They are generated by three functions defined in the rules below. 
\begin{align*}
    & \meta{sidsOfSTM(stm:StateMachineDef) = }\\
    & \quad \isacode{\isakwmaj{datatype} SIDS}
    \isacode{ = SID\_}
    \isacode{ | } \meta{\left\{s : stm.nodes | s \in State @ \right.}\isacode{SID\_}
    \meta{s.name, }\isacode{|}\meta{\left.\right\}} \\
    & \meta{tidsOfSTM(stm:StateMachineDef) = }\\
    & \quad \isacode{\isakwmaj{datatype} TIDS}
    \isacode{ = }\meta{\left\{t : stm.transitions @ \right.}\isacode{TID\_}
    \meta{t.name, }\isacode{|}\meta{\left.\right\}} \\ 
    & \meta{channelsOfSTM(stm:StateMachineDef) = }\\
    & \quad \isacode{\isakwmaj{chantype} Chan}\isacode{ = internal::TIDS     terminate::unit } \\
    & \quad\ \  \isacode{enter::SIDS$\cross$SIDS }
           \isacode{entered::SIDS$\cross$SIDS }
           \isacode{exit::SIDS$\cross$SIDS }
           \isacode{exited::SIDS$\cross$SIDS } \\
    & \quad\ \ \meta{\bigcup \left\{v : allLocalVariables(stm) @ \left\{\right.\right.}
         \isacode{get\_}\meta{v.name}\isacode{::}\meta{\semL{v.type}{t},}
         \isacode{set\_}\meta{v.name}\isacode{::}\meta{\semL{v.type}{t}\left.\right\},\left.\right\}} \\
    & \quad\ \ \meta{\bigcup }\metanobar{\left\{\meta{v : requiredVariables(stm) @} \metanobar{\left\{
    \begin{array}[]{@{}l}
         \isacodebl{get\_}\meta{v.name}\isacode{::}\meta{\semL{v.type}{t},} \\
         \isacodebl{set\_}\meta{v.name}\isacode{::}\meta{\semL{v.type}{t},} \\
         \isacodebl{set\_EXT\_}\meta{v.name}\isacode{::}\meta{\semL{v.type}{t}}
    \end{array}\right\}}, \right\}} \\
    & \quad\ \ \meta{\bigcup }\metanobar{\left\{\meta{e : allEvents(stm) @} \metanobar{\left\{
    \begin{array}[]{@{}l}
        \meta{e.name}\isacodebl{\_::TIDS$\cross$InOut$\cross$}\meta{\semL{e.type}{t},} \\
        \meta{e.name}\isacodebl{::InOut$\cross$}\meta{\semL{e.type}{t}}
    \end{array}\right\}}, \right\}} \\
    & \quad\ \ \meta{\left\{op : requiredOperations(stm) @ \right.}
    \meta{op.name}\isacodebl{Call::}\meta{\langle p : op.parameters @ \semL{p.type}{t},}\isacodebl{$\cross$}\meta{\rangle} \meta{,\left.\right\}}
\end{align*}

The function $\meta{sidsOfSTM}$ declares an enumeration type \isacode{SIDS}\footnote{In the Isabelle code, we include suffixes to ensure that names do not collide, but omit them here} containing state identifiers representing the machine itself \isacode{SID\_} and other states of the machine, that is, the nodes $\meta{stm.nodes}$ of the machine that are states $\meta{s \in State}$. The function $\meta{tidsOfSTM}$ declares \isacode{TIDS} contains transition identifiers. Channels in the machine are declared by the statement generated by the function $\meta{channelsOfSTM}$. 
The channel type \isacode{Chan} includes four kinds of channels.
Firstly, flow control channels include 
\begin{inparaenum}[(a)]
\item \isacode{internal} for transitions without a trigger;
\item \isacode{enter}, \isacode{entered}, \isacode{exit}, and \isacode{exited} for the entering and exiting of a state; and
\item \isacode{terminate} for the termination of the machine.
\end{inparaenum}
Secondly,
variable channels contain a \isacode{set} and a \isacode{get} channel for each local variable ($\meta{v}$ of all local variables $\meta{allLocalVariables(stm)}$ of the machine) with an additional \isacode{set\_EXT} for each shared variable ($\meta{v}$ of the required variables $\meta{requiredVariables(stm)}$ of the machine) to accept an external update. We use generalised union $\meta{\bigcup}$ to combine all sets of channels for these variables into a large set.
Thirdly, event channels include two channels for each event ($\meta{e}$ of all events $\meta{allEvents(stm)}$) of the machine: one channel named $\meta{e.name}$ and another named $\meta{e.name}\isacode{\_}$. 
%
The distinction of two event channels (\isacode{ech} and \isacode{ech\_}) for each event (\rcitem{e}) is necessary because the guard of a transition is evaluated in \cspcode{MemoryTrans}, and so only the trigger event (not action event) of the transition is subject to the guard evaluation, and, therefore, has a new channel (\isacode{ech\_}) with a transition id of type \isacode{TIDS}. We note, however, that events \isacode{ech\_.tid} of this new channel are eventually renamed to the event channel \isacode{ech} in the process for the machine.
Fourthly, operation call channels include a channel named $\meta{op.name}\isacode{Call}$ for each required operation ($\meta{op}$ of all required operations $\meta{requiredOperations(stm)}$) of the machine. The type of the channel is the product \isacode{$\cross$} of the corresponding types ($\meta{\semL{p.type}{t}}$ for each parameter $\meta{p}$) of the operation parameters $\meta{op.parameters}$ in a sequence constructed by comprehension and so the order of the parameters is preserved.

\begin{example}[general types for Movement]
Below is an example of these data types for the \rcitem{Movement} machine in Fig.~\ref{fig:robochart_acd_movement}.
%
\newcommand{\SIDSMovement}{\isalink{https://github.com/isabelle-utp/interaction-trees/blob/418b37554f808828610f10b40c051a562fe0c716/RoboChart/examples/RoboChart_ChemicalDetector_autonomous/RoboChart_ChemicalDetector_autonomous_microcontroller.thy\#L77}}
\newcommand{\TIDSMovement}{\isalink{https://github.com/isabelle-utp/interaction-trees/blob/418b37554f808828610f10b40c051a562fe0c716/RoboChart/examples/RoboChart_ChemicalDetector_autonomous/RoboChart_ChemicalDetector_autonomous_microcontroller.thy\#L104}}
\newcommand{\ChanMovement}{\isalink{https://github.com/isabelle-utp/interaction-trees/blob/418b37554f808828610f10b40c051a562fe0c716/RoboChart/examples/RoboChart_ChemicalDetector_autonomous/RoboChart_ChemicalDetector_autonomous_microcontroller.thy\#L227}}
 \begin{alltt} 
\isakwmaj{datatype} SIDS_Movement = SID_Movement | SID_Movement_Waiting | ... \(\SIDSMovement\)
\isakwmaj{datatype} TIDS_Movement = TID_Movement_t1 | TID_Movement_t2 | ...\(\TIDSMovement\)
\isakwmaj{chantype} Chan_Movement =   \(\ChanMovement\)
  internal_Movement :: TIDS_Movement
  terminate_Movement:: unit
  enter_Movement    :: "SIDS_Movement\(\cross\)SIDS_Movement" ...
  get_l_Movement    :: "Location_Loc"  
  set_l_Movement    :: "Location_Loc"
  obstacle__Movement:: "TIDS_Movement\(\cross\)InOut\(\cross\)Location_Loc"
  obstacle_Movement :: "InOut\(\cross\)Location_Loc" ...
  moveCall_Movement :: "core_real\(\cross\)Chemical_Angle" ...
\end{alltt}
\end{example}
 
\subsubsection{Memory for variables and transitions}
\label{ssec:semantics_stm:memory}
The memory for a state machine is composed of the memory for its local variables, for its shared variables, and that for its transitions. Its semantics is defined below.
\begin{align*}
    & \meta{memSharedVar(s: Variable) = }\\
    & \quad \isacode{ \cspkey{loop} ($\lambda v$.} 
    \isacode{get\_}\meta{s.name}\isacode{!v $\to\subscriptIntt{\tick}{v} \extchoice$}
    \isacode{ set\_}\meta{s.name}\isacode{?x}\isacode{$\to\subscriptIntt{\tick}{x} \extchoice$}
    \isacode{ set\_EXT\_}\meta{s.name}\isacode{?x}\isacode{$\to\subscriptIntt{\tick}{x}$)} \\
    & \meta{memLocalVar(l: Variable) = } 
        \isacode{ \cspkey{loop} ($\lambda v$.} 
        \isacode{get\_}\meta{l.name}\isacode{!v $\to\subscriptIntt{\tick}{v} \extchoice$}
        \isacode{ set\_}\meta{l.name}\isacode{?x}\isacode{$\to\subscriptIntt{\tick}{x}$)} \\
        & \meta{guardExpr(e:Expression) = \left(\right.\IF\ e \neq null\ \THEN\ }\isacodebl{guard }\meta{\semL{e}{e}} \meta{\ \ELSE\ }\isacodebl{\cspkey{skip}}\meta{\left.\right)} \\
    & \meta{memTransition(t: Transition) = } \\
    & \metanobar{
        \circblockbegin
        \meta{\IF\ t.trigger=null\ \THEN\ }\\ 
        \quad \isacodebl{internal!TID\_}\meta{t.name}\isacodebl{$\to \cspkey{skip}$} \\
        \meta{\ELSE\ \IF\ t.trigger.type=CommunicationType.SIMPLE\ \THEN\ }\\ 
        \quad \meta{t.tigger.event.name}\isacodebl{\_!(TID\_}\meta{t.name}\isacodebl{,din)$\to \cspkey{skip}$} \\
        \meta{\ELSE\ \IF\ t.trigger.type=CommunicationType.INPUT\ \THEN\ }\\ 
        \quad \meta{\LET\ e==t.trigger.event; p==t.trigger.parameter @ }\\
        \quad \meta{e.name}\isacodebl{\_?}\meta{p.name : \left\{p.name : p.type | t.condition @ \right.}\isacodebl{(TID\_}\meta{t.name}\isacodebl{,din,}\meta{\semL{p.name}{e}}\isacodebl{)}\meta{\left.\right\}}\isacodebl{$\to \cspkey{skip}$} \\
        \meta{\ELSE\ \IF\ t.trigger.type=CommunicationType.OUTPUT\ \THEN\ }\\ 
        \quad \meta{guardExpr(t.condition)} \isacodebl{;} \meta{t.tigger.event.name}\isacodebl{\_!(TID\_}\meta{t.name}\isacodebl{,dout,}\meta{\semL{t.trigger.value}{e}}\isacodebl{)$\to \cspkey{skip}$} \\
        \meta{\ELSE\ }\\
        \quad \meta{guardExpr(t.condition)} \isacodebl{;} \meta{t.tigger.event.name}\isacodebl{\_!(TID\_}\meta{t.name}\isacodebl{,din,}\meta{\semL{t.trigger.value}{e}}\isacodebl{)$\to \cspkey{skip}$} \\
        \circblockend
} \\
    & \meta{memTransitions(stm:StateMachineDef) = }\\
    & \quad \isacode{ \cspkey{loop} ($\lambda id$. (} \isacode{$\Extchoice$}\meta{t:stm.transitions @ memTransition(t)}\isacode{))}  \\
    & \meta{\semLC{stm:StateMachineDef}{Mem} = }\\
    & \quad \metanobar{
        \circblockbegin
        \left(\isacodebl{$\Extchoice$}\meta{v : allLocalVariables(stm) @ memLocalVar(v)}\right) \isacodebl{$\extchoice$} \\
        \left(\isacodebl{$\Extchoice$}\meta{v : requiredVariables(stm) @ memSharedVar(v)}\right) \isacodebl{$\extchoice$} \\
        \meta{memTransitions(stm)}
        \circblockend
    }
\end{align*}

\noindent The memory $\meta{memSharedVar(s)}$ for a shared variable $\meta{s}$ is an infinite \isacode{\cspkey{loop}}. It provides three choices: output the value \isacode{v} on \isacode{get\_s} without updating the variable and
accept a local (or external) update of the variable through \isacode{set\_s} (or \isacode{set\_EXT\_s}). For a local variable $\meta{l}$, its memory process $\meta{memLocalVar(l)}$ does not provide an external update.

The memory $\meta{memTransition(tr)}$ for a transition $\meta{tr}$ depends on the trigger $\meta{tr.trigger}$ of $\meta{tr}$. Suppose $\meta{tr}$ has no trigger (such as the transition from the initial junction to \rcitem{Waiting} in the state machine \rcitem{Movement} in Fig.~\ref{fig:robochart_acd_movement}), $\meta{tr.trigger=null}$. In that case, its memory is an output of the corresponding transition identifier \isacode{TID\_}$\meta{tr.name}$ on channel \isacode{internal}. Otherwise, if the trigger type $\meta{tr.trigger.type}$ of $\meta{tr}$ is $\meta{SIMPLE}$ (with an event name but no value such as \rcitem{resume}
), its memory is an output of a tuple composed of the transition identifier and \isacode{din} on the corresponding channel $\meta{tr.tigger.event.name}$\isacode{\_}. 
Suppose the trigger type is $\meta{INPUT}$ (such as \rcitem{turn?a}). In that case, we use local definitions $\meta{\LET}$ to introduce $\meta{e}$ for the trigger event and $\meta{p}$ for the trigger input variable. The memory for the transition is an input of tuples, composed of the transition identifier, \isacode{din}, and the value $\meta{\semL{p.name}{e}}$ (restricted to the type $\meta{p.type}$ of the trigger event and satisfy the guard $\meta{tr.condition}$ of the transition), from the corresponding channel. 
Suppose the trigger type is $\meta{OUTPUT}$ (such as \rcitem{turn!v}). In that case, the memory for the transition evaluates its guard first by $\meta{guardExpr(tr.condition)}$ which uses the \isacode{guard} operator to test the expression if it is not $\meta{null}$, or is just \isacode{skip} otherwise, followed by an output of the corresponding value $\meta{tr.trigger.value}$ on the corresponding channel. 
If the trigger type is not from one of the presented types, it must be of type $\meta{SYNC}$ (such as $turn.v$). Then the memory for $\meta{tr}$ (the $\meta{\ELSE}$ branch) is similar to the output trigger except that the event direction is \isacode{din} now (because of RoboChart's semantics). 

The memory $\meta{memTransitions(stm)}$ for all transitions in a state machine $\meta{stm}$ is a loop infinitely offering the memory $\meta{memTransition(tr)}$ for each transition $\meta{tr}$ from the transitions $\meta{stm.transitions}$ of $\meta{stm}$. Here, we use replicated external choice \isacode{$\Extchoice$} to construct external choice from a set of processes. 

Then the memory $\meta{\semLC{stm}{Mem}}$ of a state machine offers an external choice for the memories of all the local variables $\meta{allLocalVariables(stm)}$, all the shared variables $\meta{requiredVariables(stm)}$, and all the transitions.

\begin{example}[memory of a shared variable]
The memory process \isacode{Memory\_x} for the shared variable \rcitem{x} in the state machine \rcitem{CalSTM} in Fig.~\ref{fig:robochart_patrol}, is shown below.
\newcommand{\MemorySharedV}{\isalink{https://github.com/isabelle-utp/interaction-trees/blob/418b37554f808828610f10b40c051a562fe0c716/RoboChart/ITree_RoboChart.thy\#L292}}
\begin{alltt}
  \cspkey{loop} (\(\lambda\)v. get_x!v \(\to\) \(\subscriptIntt{\tick}{v}\) \(\extchoice\) set_x?x \(\to\) \(\subscriptIntt{\tick}{x} \extchoice\) set_EXT_x?x \(\to\) \(\subscriptIntt{\tick}{x}\)) \(\MemorySharedV\)
\end{alltt}
\end{example}
%

\begin{example}[memory of transitions]
The memory process for transitions of the state machine \rcitem{Movement} in Fig.~\ref{fig:robochart_acd_movement} is partially (3 in 24 transitions) illustrated below.
\newcommand{\MovementMemoryTrans}{\isalink{https://github.com/isabelle-utp/interaction-trees/blob/418b37554f808828610f10b40c051a562fe0c716/RoboChart/examples/RoboChart_ChemicalDetector_autonomous/RoboChart_ChemicalDetector_autonomous_microcontroller.thy\#L747}}
\begin{alltt}
  Movement_MemoryTrans = \cspkey{loop} (\(\lambda\)id.  \(\MovementMemoryTrans\)
    internal!TID_t1 \(\to\) \(\subscriptIntt{\tick}{id}\) \(\extchoice\) 
    resume_!(TID_t0, din) \(\to\) \(\subscriptIntt{\tick}{id}\extchoice\) 
    turn_?(TID_t3, din, a\(\in\)Chemical_Angle) \(\to\) \(\subscriptIntt{\tick}{id}\) \(\extchoice\) 
    ...  \(\extchoice\) 
    get_d1?d1 \(\to\) get_d0?d0 \(\to\) 
      ((rc.Minus d1 d0 rc.core_int) > stuckDist)\(\guard\)(internal!TID_t12 \(\to\) \(\subscriptIntt{\tick}{id}\)) \(\extchoice\) 
    ...
\end{alltt}
\end{example}

\begin{example}[memory of transitions with an input trigger and a guard]
    The memory process for the transitions of the state machine \rcitem{MoveSTM} in Fig.~\ref{fig:robochart_patrol} has \changed[\C{64}]{the} choices below.
\newcommand{\MoveSTMMemoryTrans}{\isalink{https://github.com/isabelle-utp/interaction-trees/blob/418b37554f808828610f10b40c051a562fe0c716/RoboChart/examples/RoboChart_basic_v1/RoboChart_basic_v1_1.thy\#L725}}
\begin{alltt}
  MoveSTM_MemoryTrans = \cspkey{loop} (\(\lambda\)id. \(\MoveSTMMemoryTrans\)
    internal!(TID_t0, din) \(\to\) \(\subscriptIntt{\tick}{id}\extchoice\) 
    reset_!(TID_t2, din) \(\to\) \(\subscriptIntt{\tick}{id}\extchoice\) 
    inp update\_ \{(TID\_t1, din, l) | l \(\in\) rc.core\_int. l \(\geq\) (rc.Neg MAX rc.core\_int)\} \(\extchoice\) 
    inp update\_ \{(TID\_t3, din, l) | l \(\in\) rc.core\_int. l \(\leq\) MAX\})
\end{alltt}
\end{example}
%
\subsubsection{Transitions}
\label{ssec:semantics_stm:transitions}
This paper considers three kinds of nodes, initial junctions, basic states, and final states, used in the two RoboChart models. Semantics for other nodes, including normal junctions and composite states, are part of our future work. In the semantics for a transition defined below, we consider two kinds of source nodes because a final state is also a $\meta{State}$.
\begin{align*}
    & \meta{\semL{(t:Transition, stm:StateMachineDef)}{\mathscr{T}} =} \\
    & \quad \metanobar{
        \circblockbegin
        \meta{\IF\ t.src \in Initial \ \THEN\ }\\ 
        \quad \isacodebl{internal!TID\_}\meta{t.name}\isacodebl{ $\to$ }\meta{\semLC{t.action}{A}}\isacodebl{; enter!(SID\_}\meta{stm.name}\isacodebl{,SID\_} \meta{t.target.name}\isacodebl{)$\to$} \\
        \quad \isacodebl{entered!(SID\_}\meta{stm.name}\isacodebl{,SID\_} \meta{t.target.name}\isacodebl{) $\to  \cspkey{skip}$} \\
        \meta{\ELSE\ \IF\ t.src \in State \ \THEN\ }\\ 
        \quad \meta{\semLC{t.trigger}{TR}}\isacodebl{; exit!(SID\_}\meta{t.src.name}\isacodebl{,SID\_} \meta{t.src.name}\isacodebl{)$\to$} \meta{\semLC{t.src.exit}{A}} \isacodebl{;}\\
        \quad \isacodebl{exited!(SID\_}\meta{t.src.name}\isacodebl{,SID\_} \meta{t.src.name}\isacodebl{) $\to$} \meta{\semLC{t.action}{A}} \isacodebl{;} \\
        \quad\isacodebl{enter!(SID\_}\meta{t.src.name}\isacodebl{,SID\_} \meta{t.target.name}\isacodebl{)$\to$} \\
        \quad \isacodebl{entered!(SID\_}\meta{t.src.name}\isacodebl{,SID\_} \meta{t.target.name}\isacodebl{) $\to  \cspkey{skip}$} \\
        \circblockend
    }
\end{align*}
From the semantic point of view, a transition starts with the synchronising of its trigger, then initiates an exit from its source node, executes the exit action of the source node, makes its source node exited, executes its transitions action, and finally enters the target node of the transition. 
If the source node $\meta{t.src}$ of a transition $\meta{t}$ is an initial junction, $\meta{t.src \in Initial}$, then $\meta{t}$ has no trigger and so channel \isacode{internal} with \isacode{TID} for the transition is synchronised. Because the source node is an initial junction, $\meta{t}$ does not need to exit the junction explicitly. The next step after the trigger, therefore, is the execution of the transition action $\meta{\semLC{t.action}{A}}$ (whose definition in the context of actions $\meta{\mathscr{A}}$ is omitted here). After that, $\meta{t}$ \isacode{enter}s its target node, and the node is \isacode{entered}. 

If the source node $\meta{t.src}$ is a state, the transition trigger $\meta{\semLC{t.trigger}{TR}}$, defined below, is synchronised.  
\begin{align*}
    & \meta{\semLC{tr: Communication, t:Transition}{TR} = } \\
    & \quad \metanobar{
        \circblockbegin
        \meta{\IF\ tr = null \THEN\ }\\ 
        \quad \isacodebl{internal!TID\_}\meta{t.name}\isacodebl{$\to \cspkey{skip}$} \\
        \meta{\ELSE\ \IF\ tr.event.type = null \THEN\ }\\ 
        \quad \meta{tr.event.name}\isacodebl{\_!(TID\_}\meta{t.name}\isacodebl{,din)$\to \cspkey{skip}$} \\
        \meta{\ELSE\ }\\ 
        \quad \meta{\LET\ e==tr.event; p==tr.parameter @ }\\
        \quad \ \ \meta{e.name}\isacodebl{\_?}\meta{p.name : \left\{p.name : p.type @ \right.}\isacodebl{(TID\_}\meta{t.name}\isacodebl{,din,}\meta{\semL{p.name}{e}}\isacodebl{)}\meta{\left.\right\}}\\
        \qquad \ \ \isacodebl{$\to$ set\_} \meta{p.name}\isacodebl{!}\meta{p.name} \isacodebl{$\to \cspkey{skip}$} \\
        \circblockend
    }
\end{align*}
A trigger is of type $\meta{Communication}$. Depending on whether the trigger exists and the trigger type, the semantics of the trigger might synchronise on channel \isacodebl{internal}, the channel corresponding to the trigger event $\meta{tr.event.name}$, or accept input from the channel and then set the received value on $\meta{p.name}$ to the corresponding variable $\meta{p.name}$ in the memory using channel \isacodebl{set\_}$\meta{p.name}$.

After the trigger, the transition $\meta{t}$ starts to \isacodebl{exit} from its source node, executes the exit action $\meta{t.src.exit}$ of the node, and makes the node \isacodebl{exited}. The subsequent behaviour is similar to the $\meta{Initial}$ case.

\subsubsection{Nodes}
\label{ssec:semantics_stm:nodes}
The semantics of each node in a state machine is a CSP process. 

\paragraph{Initial junctions}
One state machine has only one initial junction and one outgoing transition. Its semantics is defined below.
\begin{align*}
    & \meta{\semL{(stm:StateMachineDef)}{\mathscr{NI}} =} \\
    & \quad \meta{\LET\ i==\left(\mu n : stm.nodes | n \in Initial\right); t==\left(\mu t : stm.transitions | t.source = i \right) @ \semL{t, stm}{\mathcal{T}}}
\end{align*}
The semantics of the only initial junction $\meta{i}$ (retrieved through the definite description operator $\mu$) in a state machine $\meta{stm}$ is just the semantics $\meta{\semLC{t, stm}{T}}$ of the only outgoing transition $\meta{t}$ (retrieved through $\mu$ too) from $\meta{i}$.

\begin{example}[initial junction]
 We show the semantics for the initial junction \rcitem{i0} in the state machine \rcitem{CalSTM} in Fig.~\ref{fig:robochart_patrol}, which has an outgoing transition \rcitem{t0} with an action \rcitem{x=0} to set the variable \rcitem{x} to 0. 
 \newcommand{\ICalSTMij}{\isalink{https://github.com/isabelle-utp/interaction-trees/blob/418b37554f808828610f10b40c051a562fe0c716/RoboChart/examples/RoboChart_basic_v1/RoboChart_basic_v1_1.thy\#L306}}
 \begin{alltt}
   I_i0 = internal!TID_t0 \(\to\) set_x!0 \(\to\) enter!(SID_CalSTM, SID_CalSTM_cal) \(\to\) 
     entered!(SID_CalSTM, SID_CalSTM_cal) \(\to\) \(\subscriptIntt{\tick}{}\) \(\ICalSTMij\) 
 \end{alltt}
\end{example}

\paragraph{States}
The semantics for a state node is given below.
\begin{align*}
    & \meta{\semLC{(s: State, stm:StateMachineDef)}{NS}} = \isacode{ \cspkey{loop} } \\
    & \ \ \circblockbegin
    \isacodebl{$\lambda id$.enter?sd:}\meta{\left\{sid : \right.}\isacodebl{SIDS} \meta{| sid \neq }\isacodebl{ SID\_}\meta{s.name} \meta{ @} \isacodebl{(}\meta{sid}\isacodebl{, SID\_} \meta{s.name} \isacodebl{)}\meta{\left.\right\}}\isacode{$\to$} \isacode{\Ret (True, fst sd);} 
    \\
    \circblockbegin
    \isacodebl{\cspkey{iterate} ($\lambda $r. fst r) } \\
    \circblockbegin
    \isacodebl{$\lambda $r.} \\
    \circblockbegin
    \meta{\semLC{s.entry}{A}} \isacodebl{;} \isacode{entered!(fst r, SID\_} \meta{s.name} \isacodebl{)}\isacode{$\to$ (} \meta{\semLC{s.during}{A}} \isacodebl{; stop) $\interrupt$}\\
    \circblockbegin
    \isacodebl{($\Extchoice$ }\meta{t : selfTransFromNode(s) @ } \isacodebl{(}\meta{\semLC{t, stm}{T}}\isacodebl{; \Ret (True, SID\_}\meta{s.name}\isacodebl{)))} \isacodebl{$\extchoice$} \\
    \isacodebl{($\Extchoice$ }\meta{t : nonSelfTransFromNode(s) @ } \isacodebl{(}\meta{\semLC{t, stm}{T}}\isacodebl{; \Ret (False, SID\_}\meta{s.name}\isacodebl{)))} \isacodebl{$\extchoice$} \\
    \circblockbegin
    \isacodebl{$\Extchoice$ }\meta{e : EvtChns @}\\
    \quad \meta{e.name} \isacodebl{\_?x:} \meta{\left\{t : \right.} \isacodebl{TIDS } \meta{ | t \in ITIDS(stm) \land t \notin ITIDS_s(s) @ } \isacodebl{(}\meta{t} \isacodebl{,\_,\_)} \meta{\left.\right\}} \isacodebl{$\to$}\\
    \quad \isacodebl{exit?sd:} \meta{\left\{sid: \right.}\isacodebl{SIDS }\meta{ | sid \neq } \isacodebl{SID\_}\meta{s.name @ }\isacodebl{(}\meta{sid}\isacodebl{,SID\_}\meta{s.name}\isacodebl{)} \meta{\left.\right\}} \isacodebl{$\to$} \meta{\semLC{s.exit}{A}} \isacodebl{;} \\
    \quad \isacodebl{exited!(fst sd, SID\_} \meta{s.name}\isacodebl{)$\to$ \Ret (False, SID\_}\meta{s.name}\isacodebl{)} 
    \circblockend\hspace*{-.5em}
    \circblockend\hspace*{-.5em}
    \circblockend\hspace*{-.5em}
    \circblockend\hspace*{-.5em}
    \circblockend\hspace*{-.5em} 
    \circblockend
\end{align*}
Basically, the semantics is an infinite loop, an ITree, whose state (a state of an ITree is its carried data or its return value) is an integer \isacode{id} (the parameter of the semantics process), with a nested conditional iteration by \isacode{iterate}. The state of the iteration is a pair such as \isacode{(True, fst sd)}, whose first element is a \changed[\C{52}]{Boolean} value to indicate if this iteration terminates or not, and whose second element is the RoboChart state (for example \isacode{fst r}) that initiates entering of the state $\meta{s}$. Initially, the state of the iteration is passed from its preceding \isacode{Ret} construct. 

Initially, the state $\meta{s}$ is waiting for the \isacode{enter}ing from other states ($\meta{sid \neq }\isacodebl{ SID\_}\meta{s.name}$). After that, the source node, the first \isacode{fst} element of the pair \isacode{sd}, is passed to the subsequent iteration. The first argument \isacode{($\lambda$r.fst r)} of the iteration is its condition, a boolean function from the state of the iteration (a pair and \isacode{fst r} is the termination condition), and the second argument is its body, an ITree process. The process starts with the execution of the entry action $\meta{\semLC{s.entry}{A}}$ of $\meta{s}$ and then signals that $\meta{s}$ is \isacode{entered} to the node, \isacode{fst sd}, that initiates the entering. Afterwards, the during action $\meta{\semLC{s.during}{A}}$ of $\meta{s}$ is being executed while offering the possibility of interruption \isacode{$\interrupt$} by a transition process on the right of \isacode{$\interrupt$}. The during action is composed sequentially with \isacode{stop}. So, the interrupt cannot be terminated here and can only be terminated by the transition process offered in an external choice of three groups:  
\begin{inparaenum}[(1)]
\item self-transitions $\meta{selfTransFromNode(s)}$ of $\meta{s}$; 
\item other transitions $\meta{nonSelfTransFromNode(s)}$ from $\meta{s}$ that are not self-transitions; and 
\item transitions that can interrupt $\meta{s}$.
\end{inparaenum}

The process for each self-transition $\meta{t}$ is terminated (\isacode{\Ret}), after $\meta{t}$ is taken ($\meta{\semLC{t.stm}{A}}$), and returned with a state \isacode{(True, SID\_)}$\meta{s.name}$. The iteration, therefore, is not terminated (because the condition is \isacode{True}), and so the self-transition doesn't need to \isacode{enter} $\meta{s}$ again. 

The process for each other transition $\meta{t}$ is terminated and returned with a state \isacode{(False, SID\_)}$\meta{s.name}$. The iteration, therefore, is terminated (because the condition is \isacode{False}), and so the semantics of $\meta{s}$ needs to re-\isacode{enter} it from the body of the loop again. 

The third group is for the transitions $\meta{ITIDS(stm)}$ of $\meta{stm}$ that are from any state (and so they could interrupt other states) of $\meta{stm}$, but do not include those transitions $\meta{ITIDS_s(s)}$ that are from or contained in $\meta{s}$. 
This group applies to composite states in RoboChart, which further contain nodes and transitions. No composite state is used for the two RoboChart examples in this paper, so the parent of $\meta{s}$ is just $\meta{stm}$. Therefore, no such interruptible transitions are in the examples. In general, for each event $\meta{e}$ of all trigger events $\meta{EvtChns}$ in $\meta{stm}$, its corresponding event channel is $\meta{e.name}$\isacode{\_}. Each trigger event can interrupt the during action if the corresponding transition is in this group. After the trigger event, the semantics of $\meta{s}$ accepts \isacode{exit} from any other state $\meta{sid}$ ($\meta{sid\neq }$\isacode{SID\_}$\meta{s.name}$), executes its exit action, signals that $\meta{s}$ is \isacode{exited}, and then terminates.

\begin{example}[state]
The process for state \rcitem{Waiting} in \rcitem{Movement} in Fig.~\ref{fig:robochart_acd_movement} is shown below.
\newcommand{\StateWaiting}{\isalink{https://github.com/isabelle-utp/interaction-trees/blob/418b37554f808828610f10b40c051a562fe0c716/RoboChart/examples/RoboChart_ChemicalDetector_autonomous/RoboChart_ChemicalDetector_autonomous_microcontroller.thy\#L988}}
\begin{alltt}
1 State_Waiting = \cspkey{loop} (\(\lambda\)id::integer. \(\StateWaiting\)
2   sd \(\gets\) inp enter \{(s, SID_Waiting) | s. \(\lnot\) s \(\in\) SIDS_Movement_no_Waiting\} ;
3   ret \(\gets\) \Ret (True, fst sd);  
4   (\cspkey{iterate} (\(\lambda\)r. fst r) (\(\lambda\)r. 
5     entered!(snd (snd r), SID_Waiting); (
6     ({CALL__randomWalk(id); stop}) \(\interrupt\) (
7       (resume_!(TID_t0, din) \(\to\) exit!(SID_Waiting, SID_Waiting) \(\to\) 
8        exited!(SID_Waiting, SID_Waiting) \(\to\) enter!(SID_Waiting, SID_Waiting); 
9        \Ret (True, SID_Waiting)) \(\extchoice\)
10      (turn_?(TID_t2, din, a\(\in\)Chemical_Angle) \(\to\) set_a!a \(\to\) 
11       exit!(SID_Waiting, SID_Waiting) \(\to\) exited!(SID_Waiting, SID_Waiting) \(\to\) 
12       enter!(SID_Waiting, SID_Going) \(\to\) entered!(SID_Waiting, SID_Going) ; 
13       \Ret (False, SID_Waiting)) \(\extchoice\) 
14       ...
15  ))) (ret)); \Ret (id)) \qed
\end{alltt}
\end{example}

The semantics of a state $\meta{s}$ also captures the transitions from its sibling states, which have the same parent as $\meta{s}$. These transitions shall be blocked because they are part of the behaviours of its sibling states. This is implemented as the restricted behaviour of \rcitem{s} shown below.
\begin{align*}
    & \meta{restrictedState(s: State, stm:StateMachineDef) = }\\
    & \quad \isacode{$\lambda$id.} \meta{\semLC{(s: State, stm:StateMachineDef)}{NS}}~\isacodebl{$\parallel_{\meta{eventsAllTransFromSiblings(s, stm)}}$ \cspkey{skip}}
\end{align*}
The synchronisation of the semantics of $\meta{s}$ with \cspkey{skip} over a set of events $\meta{eventsAllTransFromSiblings(s, stm)}$ on the transitions that are from the sibling states of $\meta{s}$. These events are blocked in the restricted semantics because of the synchronisation with \cspkey{skip}.

\newcommand{\StateWaitingR}{\isalink{https://github.com/isabelle-utp/interaction-trees/blob/418b37554f808828610f10b40c051a562fe0c716/RoboChart/examples/RoboChart_ChemicalDetector_autonomous/RoboChart_ChemicalDetector_autonomous_microcontroller.thy\#L}}

The behaviour of multiple states is composed of their restricted behaviours.
\begin{align*}
    &\meta{composeStates(ss: \seq State, stm: StateMachineDef) =} \\    
    & \quad \metanobar{
        \circblockbegin
        \meta{\IF \# ss = 1 \THEN\ }\\ 
        \quad \meta{restrictedState(head~ss, stm)}\\
        \meta{\ELSE\ }\\ 
        \quad \meta{restrictedState(head~ss, stm)} \isacodebl{~$\parallel_{\meta{flowEvents(ss, stm)}}$~} \meta{composeStates(tail~ss, stm)}
        \circblockend
    }
\end{align*}
Here, $\meta{ss}$ is a sequence of states. If there is only one state in the sequence (that is, the length $\meta{\# ss}$ of $\meta{ss}$ is 1), the semantics is just that of the only state (in the $\meta{head}$ of $\meta{ss}$). Otherwise, the semantics is the parallel composition of the behaviour of the $\meta{head}$ state with the composed behaviour $\meta{composeStates(tail~ss, stm)}$ of the $\meta{tail}$ states ($\meta{tail~ss}$, $\meta{ss}$ after its head is removed), synchronised on a set of events $\meta{flowEvents(ss, stm)}$ that are flow channel (\isacodebl{enter}, \isacodebl{entered}, \isacodebl{exit}, and \isacodebl{exited}) events used in the restricted behaviours of all states in $\meta{tail~ss}$ to enter the head state or exit from the head state.

The behaviour of nodes in a state machine is then the composition of the initial junction and that of states.
\begin{align*}
    &\meta{\semLC{stm: StateMachineDef}{N} = }\\ 
    & \quad \meta{\semLC{stm: StateMachineDef}{NI}} \isacodebl{~$\parallel_{\meta{initFlowEvents(stm)}}$~} \meta{composeStates(\langle n:stm.nodes | n \in State \rangle, stm)}
\end{align*}
Here, we use sequence comprehension to construct all states in $\meta{stm}$ in a sequence. The composition synchronises on a set of events $\meta{initFlowEvents(stm)}$ that are flow channel events used to enter the states of $\meta{stm}$ from the initial junction and not from other states.

\begin{example}[nodes]
    The semantics of nodes in \rcitem{GasAnalysis} in Fig.~\ref{fig:robochart_acd_gasanalysis} can be found online \isaref{https://github.com/isabelle-utp/interaction-trees/blob/418b37554f808828610f10b40c051a562fe0c716/RoboChart/examples/RoboChart_ChemicalDetector_autonomous/RoboChart_ChemicalDetector_autonomous_maincontroller.thy\#L659}.
    
\end{example}

\subsubsection{Composition of STM and memory processes}
\label{ssec:semantics_stm:stm_memory}
The semantics of nodes in $\meta{stm}$ is composed with the memory of $\meta{stm}$.
\begin{align*}
    & \meta{MemorySTM(stm:StateMachineDef) =} \\
    & \quad \metanobar{ \left(\left(\meta{\semLC{stm}{N}} \isacodebl{$\hidep$} \meta{allFlowEvents(stm)} \right) \isacodebl{$\parallel_{\meta{\left(varChannelEvents(stm)\cup\, trigEvents(stm)\right)}}$} \meta{\semLC{stm}{Mem}} \right) \isacodebl{$\hidep$} \meta{localVarEvents(stm)}}
\end{align*}
In the semantics, all flow channel events $\meta{allFlowEvents(stm)}$ are hidden from the behaviour of nodes. Then the nodes synchronise with the memory on the union of two sets of events: $\meta{varChannelEvents(stm)}$ for all \isacode{get\_},  \isacode{set\_}, and \isacode{set\_EXT\_} variable channel events, and $\meta{trigEvents(stm)}$ for all trigger events. Subsequently, all local variable channel events $\meta{localVarEvents(stm)}$ are hidden.

As discussed previously \changed[\C{65}]{in Sect.~\ref{ssec:semantics_stm:gen_def}}, each event \rcitem{e} of a state machine has two corresponding event channels \isacode{e\_} (for triggers) and \isacode{e} (for actions) in the semantics of the machines, such as \isacode{obstacle\_} and \isacode{obstacle} for the event \rcitem{obstacle} of the machine \rcitem{Movement}. Trigger event channels \isacode{e\_} in $\meta{MemorySTM(stm)}$ are renamed to \isacode{e} by forgetting the first element (a transition identifier) of the value carried on \isacode{e\_}. 
\begin{align*}
    & \meta{MemorySTM\_renamedp(stm:StateMachineDef) =~}\isacode{$\renamep{\meta{MemorySTM(stm)}}{\meta{triggerMap(stm)}}$}
\end{align*}
Here, we use the renaming operator with priority to avoid nondeterminism. The renaming mapping list $\meta{triggerMap(stm)}$ includes the mappings for renaming and also the mappings whose names are not changed.

\begin{example}[rename]
We show the renaming of the state machine \rcitem{MoveSTM} in Fig.~\ref{fig:robochart_patrol}.
\newcommand{\RenamedpMemorySTMMoveSTM}{\isalink{https://github.com/isabelle-utp/interaction-trees/blob/98a6268ea827d6da98db843d2525520afead0d0c/RoboChart/examples/Patrol_Robot/Patrol_Robot.thy\#L1028}}
\begin{alltt}
MemorySTM\_renamedp = \(\renamep{\isacode{MemorySTM\_MoveSTM }}{\isacode{event\_map\_list}}\) \(\RenamedpMemorySTMMoveSTM\)
\end{alltt}

The \isacode{event\_map\_list} contains the mappings like \isacode{(update\_.(TID\_t1, din, v), update.(din, v))} and \isacode{(update\_.(TID\_t3, din, v), update.(din, v))} where \isacode{v} is an integer number. This renaming gives priority to the event at the front of the list. If \isacode{(update\_.(TID\_t3, din, v), update.(din, v))} is after \isacode{(update\_.(TID\_t1, din, v), update.(din, v))}, then \rcitem{t1} has a priority and so the nondeterminism is resolved. This corresponds to moving towards only one direction (to the left) when the robot is in Section S2 in Fig.~\ref{fig:robochart_patrol_sections} instead of the nondeterministic choice of two directions. If we change the order of the two mappings, then \rcitem{t3} will have a priority, and the direction of the movement to the right will be chosen. In our example, \isacode{event\_map\_list} is arranged based on the order in which the transitions are created. \qed
\end{example}

The semantics of a state machine also needs to take its termination (for example, the final state is reached) into consideration. 
\begin{align*}
    & \meta{\semLC{stm:StateMachineDef}{STM}} = \\ 
    &\quad \metanobar{\left(
        {{{\meta{MemorySTM\_renamedp(stm)}}
        ~\isacodebl{$\lbrack\mkern-3mu\lbrack$}
        {\isacodebl{\{terminate.()\}}}
        \isacodebl{$\rres$} 
        \isacodebl{\cspkey{skip}}}}
    \right)} \isacodebl{$\hidep$} \meta{internalEvents(stm)} 
\end{align*}
Based on the definition of the exception operator, if $\meta{MemorySTM\_renamedp(stm)}$ ever performs a \isacode{terminate} event, then \cspkey{skip} will take over, and so the process terminates. The process also hides all \isacode{internal} events and flow control events, defined in $\meta{internalEvents(stm)}$.

\subsection{Operations}
\label{ssec:semantics_operation}

Operations in RoboChart can be provided by robotic platforms such as \rcitem{move} and \rcitem{randomWalk} in Fig.~\ref{fig:robochart_acd_module} or defined by state machines such as \rcitem{changeDirection} in Fig.~\ref{fig:robochart_acd_location}.

The semantics of a call (an action) to an operation provided by a robotic platform is an event to record the call with appropriate arguments. 
\begin{align*}
    & \meta{\semLC{s: Call}{A} = s.operation.name}\isacodebl{Call!(}\meta{\langle a:s.args @ \semL{a}{e}, \rangle} \isacodebl{) $\to \cspkey{skip}$ }
\end{align*}
Here, $\meta{s.operation.name}$\isacodebl{Call} is the corresponding channel name to the operation $\meta{s.operation.name}$ which $s$ calls. The call arguments $\meta{s.args}$ become a tuple composed of corresponding arguments $\meta{\semL{a}{e}}$ for each argument $\meta{a}$.


The semantics of a state machine-defined operation differs from that of a state machine because an operation is not an independent execution element like a state machine. Its behaviour is within the scope of the state machine that calls the operation. 
For this reason, the semantics of the operation does not include a separate channel type (or an event alphabet) and does not have a separate memory. Instead, all channels required for the operation are declared along with the channels for the caller state machine in a channel-type declaration. However, the semantics of an operation's nodes, transitions, and memory are the same as those in a state machine.
 We omit the semantics of operations for simplicity. Instead, we illustrate it with an example below.

 \begin{example}[operation]
The channel type \isacode{Chan\_Movement} of \rcitem{Movement} has the following additional channels for the operation \rcitem{changeDirection}.
\newcommand{\ChanMovement}{\isalink{https://github.com/isabelle-utp/interaction-trees/blob/418b37554f808828610f10b40c051a562fe0c716/RoboChart/examples/RoboChart_ChemicalDetector_autonomous/RoboChart_ChemicalDetector_autonomous_microcontroller.thy\#L227}}
 \begin{alltt} 
\isakwmaj{chantype} Chan_Movement =   \(\ChanMovement\)
  ...
  internal_changeDirection :: TIDS_changeDirection
  enter_changeDirection    :: "SIDS_changeDirection\(\cross\)SIDS_changeDirection"
  entered_changeDirection  :: "SIDS_changeDirection\(\cross\)SIDS_changeDirection"
  exit_changeDirection     :: "SIDS_changeDirection\(\cross\)SIDS_changeDirection"
  exited_changeDirection   :: "SIDS_changeDirection\(\cross\)SIDS_changeDirection"
  terminate_changeDirection:: unit
  get_l_changeDirection    :: "Location_Loc"
  set_l_changeDirection    :: "Location_Loc"
\end{alltt}
The \isacode{get\_l} and \isacode{set\_l} channels are for the parameter \rcitem{l} of the operation and not for a local variable \rcitem{l} (indeed, there is no such local variable). Different from local variables, \rcitem{l} is only set once (\isacode{set\_l}) by the caller of the operation for passing its value, not inside the operation like local variables. The call \rcitem{changeDirection(l)} to the operation in the entry action of the state \rcitem{Avoiding} has its semantics in CSP as follows.

\newcommand{\CallChangeDirectionMovement}{\isalink{https://github.com/isabelle-utp/interaction-trees/blob/418b37554f808828610f10b40c051a562fe0c716/RoboChart/examples/RoboChart_ChemicalDetector_autonomous/RoboChart_ChemicalDetector_autonomous_microcontroller.thy\#L670}}
\begin{alltt}
CALL__changeDirection_Movement = \(\CallChangeDirectionMovement\)
  {get_l_Movement?l \(\to\) set_l_changeDirection!l \(\to\) \(\subscriptIntt{\tick}{}\); D_changeDirection \(\changed[\C{66}]{}\)} 
\end{alltt}
The first input event gets the value of the local variable \rcitem{l} of \rcitem{Movement}, and the value is recorded in \isacode{l}. The second event updates the value of the parameter \rcitem{l} (in the memory) of \rcitem{changeDirection} to \isacode{l}. Finally, the process for this call behaves as the process \isacode{D\_changeDirection} (for \rcitem{changeDirection}). 
\qed
 \end{example}

\subsection{Controllers}
\label{ssec:semantics_ctrl}

The event alphabet of the process for a controller  
contains a termination, shared variable, event, and operation call channels. The event channels include not only the events of the controller but also those in connections between its state machines. 
\begin{align*}
    & \meta{channelsOfCTRL(c:ControllerDef) = }\\
    & \quad \isacode{\isakwmaj{chantype} Chan}\isacode{ = terminate::unit } \\
    & \quad\ \ \meta{\bigcup }\metanobar{\left\{
    \begin{array}[]{@{}l}
            \meta{v : allLocalVariables(c) @} \\
            \metanobar{\meta{\bigcup}\left\{
    \begin{array}[]{@{}l}
        \meta{\left\{\right.}\isacodebl{get\_}\meta{v.name}\isacode{::}\meta{\semL{v.type}{t},\ } \isacodebl{set\_}\meta{v.name}\isacode{::}\meta{\semL{v.type}{t}\ } \meta{\left.\right\},}\\
         \meta{\left\{s:c.machines|v \in requiredVariables(s) @\right.}\isacodebl{set\_EXT\_}\meta{s.name}\isacodebl{\_}\meta{v.name}\isacode{::}\meta{\semL{v.type}{t},\left.\right\}}
    \end{array}\right\}}, 
    \end{array}\right\}} \\
    & \quad\ \ \meta{\bigcup }\metanobar{\left\{
    \begin{array}[]{@{}l}
            \meta{v : requiredVariables(c) @} \\
            \metanobar{\meta{\bigcup}\left\{
    \begin{array}[]{@{}l}
        \meta{\left\{\right.}\isacodebl{get\_}\meta{v.name}\isacode{::}\meta{\semL{v.type}{t},\ } \isacodebl{set\_}\meta{v.name}\isacode{::}\meta{\semL{v.type}{t},\ } 
         \isacodebl{set\_EXT\_}\meta{v.name}\isacode{::}\meta{\semL{v.type}{t}\left.\right\},}\\
         \meta{\left\{s:c.machines|v \in requiredVariables(s) @\right.}\isacodebl{set\_EXT\_}\meta{s.name}\isacodebl{\_}\meta{v.name}\isacode{::}\meta{\semL{v.type}{t},\left.\right\}}
    \end{array}\right\}}, 
    \end{array}\right\}} \\
    & \quad\ \ \metanobar{\left\{\meta{e : allEvents(c) @ e.name}\isacodebl{::InOut$\cross$}\meta{\semL{e.type}{t}} , \right\}} \\
    & \quad\ \ \metanobar{\left\{\meta{e : internalEvents(c) @ e.name}\isacodebl{::InOut$\cross$}\meta{\semL{e.type}{t}} , \right\}} \\
    & \quad\ \ \meta{\left\{op : requiredOperations(c) @ \right.}
    \meta{op.name}\isacodebl{Call::}\meta{\langle p : op.parameters @ \semL{p.type}{t},}\isacodebl{$\cross$}\meta{\rangle} \meta{,\left.\right\}}
\end{align*}
The channel type for a controller includes channels for termination, local variables, shared variables, events, and operations. Different from those for a state machine, for each local or shared variable $\meta{v}$ a controller $\meta{c}$, additionally, declares a \isacodebl{set\_EXT\_}$\meta{s.name}$\isacode{\_}$\meta{v.name}$ channel for each state machine $\meta{s}$ of $\meta{c}$ that requires $\meta{v}$. These channels propagate the update of $\meta{v}$ to all state machines that require $\meta{v}$. 

In addition to event channels for all events, $\meta{allEvents{c}}$ of $\meta{c}$,  an event channel for each event $\meta{e}$ of all internal events (that is, the events that are used in connections between state machines) is also declared. These channels are used for communication between state machines.

\begin{example}[controller channel type]
The channel type of the controller \rcitem{Ctrl} in Fig.~\ref{fig:robochart_patrol} is defined below. 
\newcommand{\ChanCtrl}{\isalink{https://github.com/isabelle-utp/interaction-trees/blob/98a6268ea827d6da98db843d2525520afead0d0c/RoboChart/examples/Patrol_Robot/Patrol_Robot.thy\#L1071}}
\begin{alltt}
\isakwmaj{chantype} Chan_Ctrl = \(\ChanCtrl\)
  terminate_Ctrl         :: unit 
  set_x_Ctrl             :: core_int
  get_x_Ctrl             :: core_int
  set_EXT_x_Ctrl         :: core_int
  set_EXT_x_Ctrl_CalSTM  :: core_int
  set_EXT_x_Ctrl_MoveSTM :: core_int
  rec_Ctrl               :: "InOut\(\cross\)core_int"
  reset_Ctrl             :: "InOut"
  update_Ctrl            :: "InOut\(\cross\)core_int" \(\hfill\)\qed
\end{alltt}
\end{example}

The memory of a controller deals with the update of its local and shared variables, but not transitions like that of a state machine.
\begin{align*}
    & \meta{memSharedVarCtrl(s: Variable, c: ControllerDef) = }\\
    & \quad \isacode{\cspkey{loop}} 
        \circblockbegin
    \isacodebl{$\lambda v$.} \isacode{set\_EXT\_}\meta{s.name}\isacode{?x} \isacode{$\to$}  \\
    \quad \metanobar{\left(\isacodebl{$\fatsemi$ }\meta{m: \left\{m: c.machines | s\in requiredVariables(m)\right\} @ }\isacodebl{set\_EXT\_}\meta{m.name}\isacodebl{\_}\meta{s.name}\isacodebl{!x$\to$\cspkey{skip}}\right)}
        \circblockend \\
    & \meta{memLocalVarCtrl(l: Variable, c: ControllerDef) = } \isacode{ \cspkey{loop}} \\ 
    & \ \ \left(\begin{array}{@{}l}
    \isacodebl{$\lambda v$.} \\
    \ \metanobar{
        \left(\begin{array}{@{}l}
        \isacode{(set\_}\meta{l.name}\isacodebl{?x} \isacodebl{$\to$} \\
        \left(\isacodebl{$\fatsemi$ }\meta{m: \left\{m: c.machines | l\in requiredVariables(m)\right\} @ }\isacodebl{set\_EXT\_}\meta{m.name}\isacodebl{\_}\meta{l.name}\isacodebl{!x$\to$\cspkey{skip}}\right)
        \isacodebl{;$\subscriptIntt{\tick}{x}$)} 
        \end{array}\hspace*{-.5em}\right)
        } \\
    \ \extchoice \isacode{get\_}\meta{l.name}\isacode{!v $\to\subscriptIntt{\tick}{v}$}
    \end{array}\hspace*{-.5em}\right)
        \\
    & \meta{\semLC{c: ControllerDef}{Mem} = }\\
    & \quad \metanobar{
        \circblockbegin
        \left(\isacodebl{$\Extchoice$}\meta{v : allLocalVariables(c) @ memLocalVarCtrl(v)}\right) \isacodebl{$\extchoice$} \\
        \left(\isacodebl{$\Extchoice$}\meta{v : requiredVariables(c) @ memSharedVarCtrl(v,c)}\right) \\
        \circblockend
    }
\end{align*}
The memory $\meta{memSharedVarCtrl(s, c)}$ of a shared variable $\meta{s}$ in a controller $\meta{c}$ is an infinite loop. It accepts an update of $\meta{s}$ on channel \isacodebl{set\_EXT\_}$\meta{s.name}$ and then propagates the updated value $x$ to the state machines that require $\meta{s}$ through the corresponding channels, using a replicated sequential composition \isacode{$\fatsemi$}. The memory $\meta{memLocalVarCtrl(l, c)}$ of a local variable $\meta{l}$ also offers an update of $\meta{l}$ but on a different channel \isacodebl{set\_}$\meta{l.name}$, then propagates the update. Additionally, it offers access to the value of $\meta{l}$ through channel \isacodebl{get\_}$\meta{l.name}$. We also note that the loop state $v$ of $\meta{memSharedVarCtrl(s,c)}$ is dummy (that is, not used and updated) because $\meta{c}$ does not store the value of a shared variable, but the loop state $v$ of $\meta{memLocalVarCtrl(l,c)}$ is usual (for update and access).


The memory $\meta{\semLC{c}{Mem}}$ of a controller $\meta{c}$ offers an external choice for the memories of all the local variables $\meta{allLocalVariables(c)}$, and all the shared variables $\meta{requiredVariables(c)}$.

\begin{example}[controller memory]
The memory of \rcitem{Ctrl} in Fig.~\ref{fig:robochart_patrol} is shown below. 
\newcommand{\CtrlMemory}{\isalink{https://github.com/isabelle-utp/interaction-trees/blob/98a6268ea827d6da98db843d2525520afead0d0c/RoboChart/examples/Patrol_Robot/Patrol_Robot.thy\#L1104}}
\begin{alltt}
Memory_Ctrl = \cspkey{loop} (\(\lambda\)id. \(\CtrlMemory\)
    set_EXT_x_Ctrl?x \(\to\) set_EXT_x_Ctrl_CalSTM!x \(\to\) set_EXT_x_Ctrl_MoveSTM!x \(\to\) \(\subscriptIntt{\tick}{id}\)) \(\hfill\) \qed
\end{alltt}
\end{example}

Each state machine in a controller has a different event alphabet or channel type. To compose state machines, their event alphabets must be transformed into the same channel type for the controller. We use renaming defined in Sect.~\ref{sec:itree} to achieve it.
\begin{align*}
    & \meta{STM\_renamedp(stm:StateMachineDef, c: ControllerDef) =~}\isacode{$\renamep{\meta{\left(\semLC{stm}{STM}\right)}}{\meta{stm2CtrlMap(stm,c)}}$}
\end{align*}
We use renaming with priority to resolve potential nondeterminism introduced from the renaming mapping list $\meta{stm2CtrlMap(stm,c)}$. This list contains event mappings between $\meta{stm}$ and $\meta{c}$ for the \isacode{terminate} channel, (local and shared) variable channels, event channels, and operation call channels. Additionally, it includes event channels that are used for internal communication between state machines, such as the \rcitem{update} event between \rcitem{CalSTM} and \rcitem{MoveSTM} in Fig.~\ref{fig:robochart_patrol}.

\begin{example}[state machine renaming]
    The renaming list for \rcitem{CalSTM} in Fig.~\ref{fig:robochart_patrol} can be found online \isaref{https://github.com/isabelle-utp/interaction-trees/blob/98a6268ea827d6da98db843d2525520afead0d0c/RoboChart/examples/Patrol_Robot/Patrol_Robot.thy\#L1118}.
    \qed
\end{example}

Renamed state machines are composed together using parallel composition.
\begin{align*}
    &\meta{composeMachines(ms: \seq StateMachineDef, c: Controllers) =} \\    
    & \quad \metanobar{
        \circblockbegin
        \meta{\IF \# ss = 1 \THEN\ }\\ 
        \quad \meta{STM\_renamedp(head~ms, c)}\\
        \meta{\ELSE\ }\\ 
        \quad \meta{STM\_renamedp(head~ms, c)} \isacodebl{~$\parallel_{\meta{connEvents(ms, c)}}$~} \meta{composeMachines(tail~ms, c)}
        \circblockend
    }
\end{align*}
Here, $\meta{ms}$ is a sequence of state machines. If there is only one machine in the sequence, the semantics is just that of the only machine (in the $\meta{head}$ of $\meta{ms}$). Otherwise, the semantics is the parallel composition of the behaviour of the $\meta{head}$ machine with the composed behaviour $\meta{composeMachines(tail~ms, c)}$ of the $\meta{tail}$ machines, synchronised on a set of events $\meta{connEvents(ms, c)}$ which includes the \isacode{terminate} channel and event channels used in all machines in $\meta{tail~ms}$ to communicate with the head machine. 

The events for internal communication between state machines are hidden.
\begin{align*}
    \meta{hiddenMachines(c) = composeMachines(\langle m: c.machine @ m\rangle, c)} \isacode{$\hidep$ } \meta{intConnEvents(c)}
\end{align*}
Here, $\meta{intConnEvents(c)}$ defines a set of events used for internal communication between state machines in $\meta{c}$.

The semantics of a controller is the parallel composition of that of its composed machines with its memory and also deals with its termination. 
\begin{align*}
    & \meta{\semLC{c:ControllerDef}{C}} = \\ 
    &\quad \metanobar{
        {\left(\left(\meta{hiddenMachines(c)} \isacodebl{ $\parallel_{\meta{ctrlMemEvents(c)}}$ } \meta{\semLC{c}{Mem}}\right) \isacodebl{$\hidep$ } ctrlMemEvents(c)\right)}
        ~\isacodebl{$\lbrack\mkern-3mu\lbrack$}
        {\isacodebl{\{terminate.()\}}}
        \isacodebl{$\rres$} 
        \isacodebl{\cspkey{skip}}
    }
\end{align*}
The $\meta{ctrlMemEvents(c)}$ is a set of variable channel events in $\meta{c}$ used to access, update, and propagate variables.

\begin{example}[controller]
    The semantics of \rcitem{Ctrl} in Fig.~\ref{fig:robochart_patrol} can be found online \isaref{https://github.com/isabelle-utp/interaction-trees/blob/98a6268ea827d6da98db843d2525520afead0d0c/RoboChart/examples/Patrol_Robot/Patrol_Robot.thy\#L1190}. \qed
\end{example}

\subsection{Modules}
\label{ssec:semantics_module}

Similar to the event alphabet of the process for a controller, the process for a module also contains a termination channel, shared variable channels, event channels, and operation call channels. The event channels include the events of its platform and the events in connections between its controllers for the same reason. 

The process for a module is a parallel composition of the renamed processes for its controllers, memory processes, and buffer processes for asynchronous connections between its controllers, such as the connection on event \rcitem{turn} from \rcitem{MainController} to \rcitem{MicroController} in Fig.~\ref{fig:robochart_acd_module}. 
The semantics for an asynchronous connection is a one-place buffer.
\begin{align*}
    & \meta{singleBuffer(efrom: Event, eto: Event) = } \\ 
    & \quad \isacode{\cspkey{loop}} 
        \circblockbegin
            \isacodebl{$\lambda lv$.} 
            \circblockbegin
                \circblockbegin
                    \isacodebl{guard (length lv $\geq$ 0 $\land$ length lv $\leq$ 1); } \\
                    \meta{efrom.name}\isacodebl{?x:}\meta{\left\{v:efrom.type @ \right.}\isacodebl{(dout,}\meta{v}\isacodebl{)}\meta{\left.\right\}} \isacodebl{$\to\subscriptIntt{\tick}{[snd~x]}$}
                \circblockend \isacodebl{$\extchoice$} \\
                \circblockbegin
                    \isacodebl{guard (length lv > 0); } \meta{eto.name}\isacodebl{!(din, hd lv)}\isacodebl{$\to\subscriptIntt{\tick}{[]}$}
                \circblockend \\ 
            \circblockend \\
        \circblockend \\
\end{align*}
The connection is from a $\meta{efrom}$ event to a $\meta{eto}$ event. The semantics is an infinite loop whose state is a list \isacodebl{lv} in Isabelle. It offers two choices: either accepting a write operation if the list is empty or contains one element, using channel $\meta{efrom.name}$  and then updating the state to a list containing the write value \isacodebl{snd x} in the second part of \isacodebl{x}, or accepting a read operation, if the list is not empty, using channel $\meta{eto.name}$ with current element \isacodebl{hd lv} and then updating the list to be empty \isacodebl{[]}.

The memory of a module deals with the update of shared variables and the propagation of the update to its controllers, which is similar to that of a controller.   
The definition of the process for a module is omitted for simplicity. It can be found online (\isacode{D\_ChemicalDetector} \isaref{https://github.com/isabelle-utp/interaction-trees/blob/418b37554f808828610f10b40c051a562fe0c716/RoboChart/examples/RoboChart_ChemicalDetector_autonomous/RoboChart_ChemicalDetector_autonomous.thy\#L110} for the autonomous chemical detector and \isacode{D\_PatrolMod} \isaref{https://github.com/isabelle-utp/interaction-trees/blob/98a6268ea827d6da98db843d2525520afead0d0c/RoboChart/examples/Patrol_Robot/Patrol_Robot.thy\#L1300} for the patrol robot). 

\section{Code generation, animation, and case studies}
\label{sec:animation}

As discussed previously in \cite[Sect.~5]{Foster2021}, the animation of ITrees is achieved through code generation~\cite{Haftmann2010} in Isabelle. 
Infinite corecursive definitions over ITrees are implemented using lazy evaluation in Haskell. 
Associative lists are used to implement partial functions in ITrees, and a simple animator in Haskell is presented. 
Using the same approach for animation, we can animate the two RoboChart models shown in Sect.~\ref{sec:robochart}. 

\subsection{Autonomous chemical detector}
\label{ssec:animation_chemical}
We illustrate two scenarios \changed[\C{2}]{\textbf{SCE-ACD-1} and \textbf{SCE-ACD-2}} of the animation of the autonomous chemical detector in Figs.~\ref{fig:animate_stop} and \ref{fig:animate_resume}. 
Here, we instantiate \isacode{Chem} and \isacode{Intensity} to be a numeral type \isacode{2} and the sequence of \isacode{GasSensor} is bounded to 2, which is the same as the instantiations for the verification with FDR4.
An animation scenario represents the interaction of the model with its environment: the lines starting with \lstinline[language=Animation]{Events} are produced by the model and describe all enabled events, and the lines beginning with \lstinline[language=Animation]{[Choose: 1-n]} represents a user's choice of enabled events from number 1 to n. In Fig.~\ref{fig:animate_resume}, we omit the lines for enabled events and append the chosen event to the selected number for simplicity. 

\begin{figure}[t]
\begin{lstlisting}[language=Animation, caption={}, label={lst:animation_acd_1}]
Starting ITree animation...
Events: (1) RandomWalkCall (); (2) Gas (Din, []); ...;
[Choose: 1-22]: 1
Events: (1) Gas []; (2) Gas [(0,0)]; (3) Gas [(0,1)]; (4) Gas [(1,0)]; 
  (5) Gas [(1,1)]; (6) Gas [(0,0),(0,0)]; (7) Gas [(0,0),(0,1)]; (8) Gas 
  [(0,0),(1,0)]; (9) Gas [(0,0),(1,1)]; ...; (21) Gas [(1,1),(1,1)];
[Choose: 1-21]: 9
Events: (1) MoveCall (0,Chemical_Angle_Front);
[Choose: 1-1]: 1
Events: (1) Flag Dout;
[Choose: 1-1]: 1
Terminated: ()
\end{lstlisting}
\caption{\label{fig:animate_stop} \changed[\C{2}]{\textbf{SCE-ACD-1}:} animation of the example when dangerous chemical detected.}
\vspace{-3ex}
\end{figure}



\paragraph{\textbf{\changed[\C{2}]{SCE-ACD-1}}}
Figure~\ref{fig:animate_stop} illustrates the behaviour of the model when detecting a dangerous chemical: 
\begin{inparaenum}[(1)]
\item initially the controller calls the platform to perform a random walk: the number \lstinline[language=Animation]{1} event is chosen on line \verb+#3+, which corresponds to the call of the during action \rcitem{randomWalk()} 
    of state \rcitem{Waiting} in Fig.~\ref{fig:robochart_acd_movement};
    \item then a sequence of gas sensor readings is received through the \rcitem{gas} event, and we choose number \lstinline[language=Animation]{9} (among 21 enabled \rcitem{gas} events shown on lines \verb+#4-6+ where the first element \lstinline[language=Animation]{Din} of each event is omitted) on line \verb+#7+: \lstinline[language=Animation]{Gas [(0,0),(1,1)]},  representing a chemical being detected and its intensity is high in the second pair of the sequence;
    \item the controllers call the \rcitem{move} operation with speed 0 (on line \verb+#9+), provided by the platform, to stop the robot; 
    \item the controllers indicate the platform to drop a flag (on line \verb+#11+); and finally 
    \item the controllers terminate (on line \verb+#12+).
\end{inparaenum}

\begin{figure}[t]
\begin{lstlisting}[language=Animation, caption={}, label={lst:animation_acd_2}]
/*Starting ITree animation...
*/[Choose: 1-22]: 1   RandomWalkCall ()
[Choose: 1-21]: 4   Gas (Din,[(1, 0)])
[Choose: 1-22]: 1   MoveCall (1,Chemical_Angle_Front)
[Choose: 1-24]: 2   Obstacle (Din,Location_Loc_right)
[Choose: 1-23]: 1   Odometer (Din,0)
[Choose: 1-22]: 1   MoveCall (1,Chemical_Angle_Left)
[Choose: 1-21]: 8   Gas (Din,[(0, 0),(1, 0)])
[Choose: 1-22]: 1   MoveCall (1,Chemical_Angle_Front)
[Choose: 1-24]: 1   Obstacle (Din, Location_Loc_left)
[Choose: 1-23]: 2   Odometer (Din,1)
[Choose: 1-23]: 1   Odometer (Din,0)
/*[Choose: 1-22]: 1   MoveCall (1,Chemical_Angle_Right)
[Choose: 1-21]: 4   Gas (Din,[(1, 0)])
[Choose: 1-22]: 1   MoveCall (1,Chemical_Angle_Front)
[Choose: 1-24]: 2   Obstacle (Din,Location_Loc_right)
[Choose: 1-23]: 1   Odometer (Din,0)
[Choose: 1-22]: 1   Stuck_timeout Din
[Choose: 1-22]: 1   ShortRandomWalkCall ()
*/[Choose: 1-22]: ...
\end{lstlisting}
\caption{\label{fig:animate_resume} \changed[\C{2}]{\textbf{SCE-ACD-2}:} animation of the example when chemical detected with low intensity.}
\vspace{-3ex}
\end{figure}

\paragraph{\textbf{\changed[\C{2}]{SCE-ACD-2}}}
In Fig.~\ref{fig:animate_resume}, we illustrate another scenario: a chemical is detected, but its intensity is low for the two readings on lines \verb+#2+ and \verb+#7+. The model behaves as follows: 
\begin{inparaenum}[(1)]
    \item the initial behaviour is the same: calling the platform to request a random walk;
    \item a sequence of gas sensor readings is received (online \verb+#2+);
    \item the controllers call the \rcitem{move} operation (the entry action of the state \rcitem{Going} in Fig.~\ref{fig:robochart_acd_movement}) to request the robot to move forward at speed 1 (on line \verb+#3+);
    \item an obstacle on its right is encountered (on line \verb+#4+);
    \item the odometer reading is 0 (on line \verb+#5+);
    \item the controllers call \rcitem{move} (the action of a transition in the defined operation \rcitem{changeDirection}) to request the robot to move towards its opposite direction (left here) to the obstacle at speed 1 (online \verb+#6+);
    \item another reading of the gas sensor shows there is still a chemical detected with low intensity (on line \verb+#7+);
    \item the controllers call \rcitem{move} (the entry action of state \rcitem{TryingAgain} in machine \rcitem{Movement}) to request the robot to move towards its front at speed 1 (on line \verb+#8+);
    \item an obstacle on its left is encountered (on line \verb+#9+);
    \item the odometer reading (the action of the transition from state \rcitem{TryingAgain} to state \rcitem{AvoidingAgain}) is 1 (on line \verb+#10+);
    \item there is another odometer reading (0) on line \verb+#11+, which corresponds to the entry action of state \rcitem{Avoiding} (the entering of this state has resulted from the transition taken from state \rcitem{AvoidingAgain} to state \rcitem{Avoiding} due to its guard \rcitem{d1-d0>stuckDist} is true where the values of \rcitem{d0} and \rcitem{d1} are the previous two odometer readings 0 and 1, and the value of \rcitem{stuckDist} is set 0 in this animation);
    \item we omit further interactions.
\end{inparaenum}

Based on the animation, we also observe that if no chemical is detected, the model returns to its initial state. 
If the low-intensity chemical is detected, even without progress of \rcitem{MicroController}, the model can continuously read through the \rcitem{gas} event without blocking. This is due to the connection between the controllers on event \rcitem{turn} being asynchronous, and so \rcitem{MainController} can continuously send a \rcitem{turn} event without waiting for the synchronisation of \rcitem{MicroController}. 
In our implementation in ITrees, the buffer process defined previously for the connection reflects this behaviour: overwriting the buffer is always allowed.

\subsection{The patrol robot}
\label{ssec:animation_patrol}
In this example, we instantiate \isacode{MAX\_INT} to 3 and \isacode{MAX} to 2 and illustrate three scenarios \changed[\C{2}]{\textbf{SCE-PR-1}, \textbf{SCE-PR-2}, and \textbf{SCE-PR-3}} of the animation of the patrol robot corresponding to the three sections (S1, S2, and S3) of the corridor in Fig.~\ref{fig:robochart_patrol_sections}.

\begin{figure}[t]
\begin{lstlisting}[language=Animation, caption={}, label={lst:animation_pr_1}]
Starting ITree Simulation...
Events: (1) Reset_PatrolMod Din; (2) Cal_PatrolMod (Din,-3); (3) Cal_PatrolMod (Din,-2); 
         (4) Cal_PatrolMod (Din,-1); (5) Cal_PatrolMod (Din,0); (6) Cal_PatrolMod (Din,1); 
         (7) Cal_PatrolMod (Din,2); (8) Cal_PatrolMod (Din,3);
[Choose: 1-8]: 2 Cal_PatrolMod (Din,-3)
[Choose: 1-1]: Right_PatrolMod (Dout,-2)
[Choose: 1-1]: Right_PatrolMod (Dout,-2)
[Choose: 1-1]: Right_PatrolMod (Dout,-1)
[Choose: 1-1]: Right_PatrolMod (Dout,-1)
[Choose: 1-1]: Right_PatrolMod (Dout,0)
Events: (1) Right_PatrolMod (Dout,0); (2) Cal_PatrolMod (Din,-3); (3) Cal_PatrolMod (Din,-2); 
         (4) Cal_PatrolMod (Din,-1); (5) Cal_PatrolMod (Din,0); (6) Cal_PatrolMod (Din,1); 
         (7) Cal_PatrolMod (Din,2); (8) Cal_PatrolMod (Din,3);
[Choose: 1-8]: 1 Right_PatrolMod (Dout,0)
Events: (1) Reset_PatrolMod Din; (2) Cal_PatrolMod (Din,-3); (3) Cal_PatrolMod (Din,-2); 
         (4) Cal_PatrolMod (Din,-1); (5) Cal_PatrolMod (Din,0); (6) Cal_PatrolMod (Din,1); 
         (7) Cal_PatrolMod (Din,2); (8) Cal_PatrolMod (Din,3);
[Choose: 1-8]:
\end{lstlisting}
\caption{\label{fig:animate_patrol_s1} \changed[\C{2}]{\textbf{SCE-PR-1:}} animation of the patrol robot when the calibrated position is in S1.}
\end{figure}

\paragraph{\textbf{\changed[\C{2}]{SCE-PR-1}}}
We show the first scenario in Fig.~\ref{fig:animate_patrol_s1}, related to the calibrated position in S1. The model behaves as follows: 
\begin{inparaenum}[(1)]
\item initially, the controller provides eight events on lines \verb+#2-4+ for users to choose: one to reset the position and another seven to set the calibrated position to an integer value between -3 and 3;
\item the second event (\lstinline[language=Animation]{2}) is chosen on line \verb+#5+, denoting the calibrated position is \lstinline[language=Animation]{-3} and so the robot is in S1;
\item the only available event on lines \verb+#6-10+ is \lstinline[language=Animation]{Right_PatrolMod}\footnote{The change of the name from \isacode{right\_PatrolMod} to \lstinline[language=Animation]{Right_PatrolMod} is due to the code generation in Isabelle to generate Haskell. In Haskell, it is conventional to use capitalised names for data types.} which corresponds to the \rcitem{right} event in the RoboChart model, denoting the movement of the robot towards the right side of the corridor at the new positions (\lstinline[language=Animation]{-2}, \lstinline[language=Animation]{-1}, and \lstinline[language=Animation]{0} respectively);
\item since the new position on line \verb+#10+ is 0 now, the controller could accept a \rcitem{right} event and calibration events on lines \verb+#11-13+;
\item the \rcitem{right} event is chosen on line \verb+14+, and after that, 
\item the controller returns to its initial state: having the same available events on lines \verb+#15-17+ as initially available events on lines \verb+2-4+.
\end{inparaenum}
When \rcitem{x} is equal to -3 and -2 (in section S1), the robot moves towards the right side, and \rcitem{x} is increased by 1, as illustrated on line \verb+#6-9+. This behaviour is consistent with the semantics of the model. After \rcitem{x} becomes -1 (in section S2), the model nondeterministically chooses to move towards the left or right side. However, our animation on lines \verb+#6-10+, and \verb+#14+ shows only the right side is chosen. This is because of the use of renaming with priority in \isacode{Renamedp\_MemorySTM\_MoveSTM} and the higher priority of the \rcitem{update} event on \rcitem{t1} than \rcitem{t3} to resolve the nondeterminism (and so the priority is given to the movement towards the right side).

\begin{figure}[t]
  \centering%
  \includegraphics[width=0.70\textwidth]{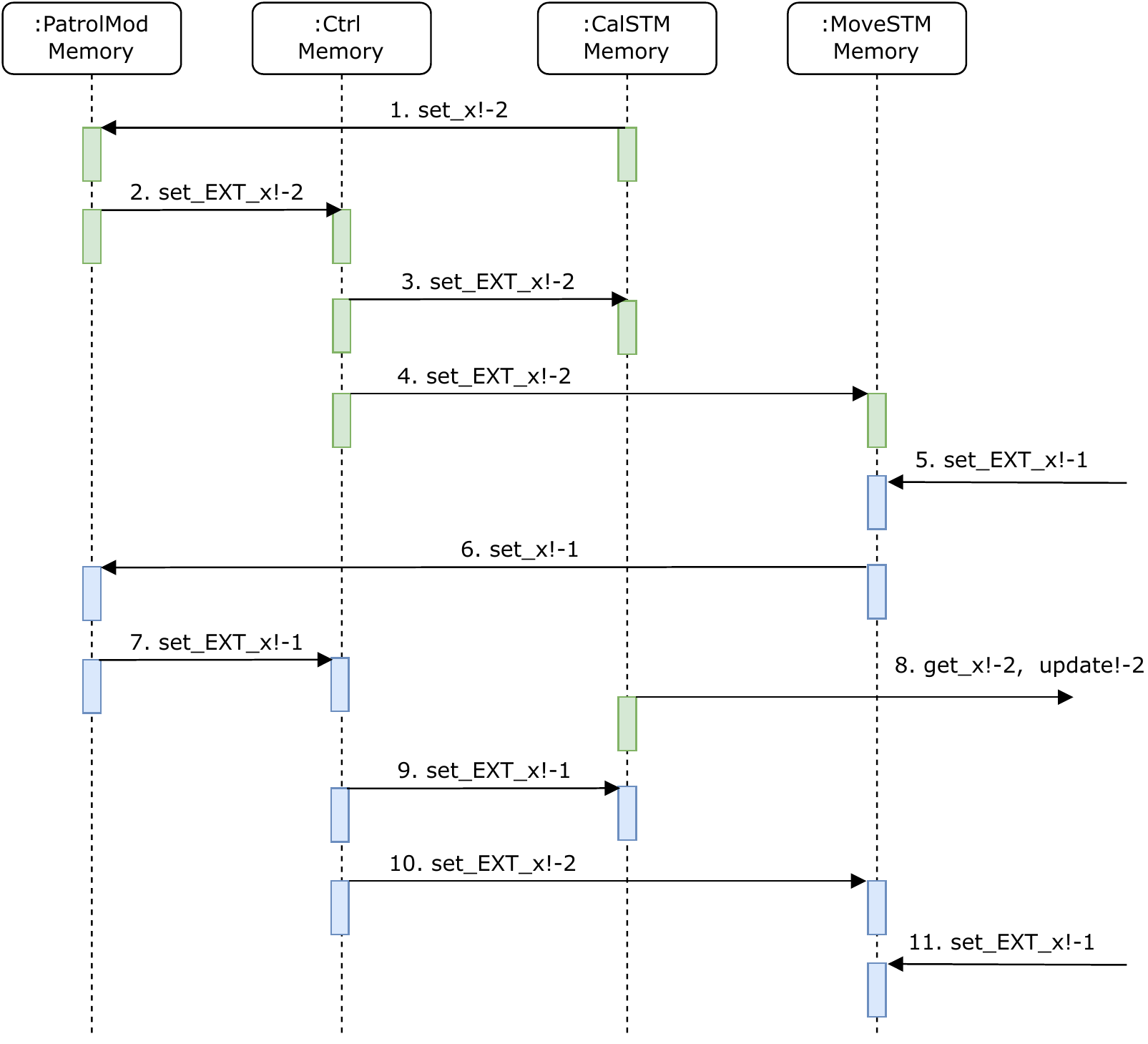}%
  \caption{The update of the shared variable \rcitem{x} and its propagation in the patrol robot model.}%
  \label{fig:basic_shared_variables}%
\end{figure}

In the RoboChart model, we expect each \rcitem{left} or \rcitem{right} event to correspond to decrease or increase \rcitem{x} by 1. The animation, however, shows that the new positions (\lstinline[language=Animation]{-2}, \lstinline[language=Animation]{-1}, and \lstinline[language=Animation]{0}) on the \rcitem{right} event on lines \verb+#7+, \verb+#9+, and \verb+#14+ stay the same as their previous positions on lines \verb+#5+, \verb+#8+, and \verb+#10+. This is actually due to the semantics of shared variables in RoboChart, specifically, the mechanism used to update shared variables and propagate the updates, which is subtle. We illustrate the implemented mechanism in our semantics in Fig.~\ref{fig:basic_shared_variables} where the exchange of the value of \rcitem{x} in the module, the controller, and the two machines is through communication (labelled with an identifier, a channel, and a message over the channel).

The communications 1 to 4 show that the change of \rcitem{x} to -2 in \rcitem{CalSTM} is updated to the module \rcitem{PatrolMod}, and then this update is propagated down the memory hierarchy to the controller \rcitem{Ctrl}, subsequently to the state machines \rcitem{CalSTM} and \rcitem{MoveSTM}. The update and propagation, however, are not atomic. The memories can be accessed and evaluated between these communications using the outdated value. This is further demonstrated by communications 5 to 11. Consider a new update of \rcitem{x} to -1 (by communication 5) in the memory of \rcitem{MoveSTM}. The update and propagation are similar to the previous update (to -2). We here, however, consider the value of \rcitem{x} in \rcitem{CalSTM} is accessed (by the event \isacode{get\_x} on communication 8) and evaluated in the guard \rcitem{[x!=0]} of the transition in the machine before the new value -1 is propagated to the machine (on communication 9). This means the action of the transition still outputs -2 on the event \rcitem{update!-2}, not the -1, because the new value has not been seen in this machine. As a consequence, the input trigger \rcitem{update?l} in the machine \rcitem{MoveSTM} will receive a value -2, and so \rcitem{x} is set to -1 (\rcitem{x=l+1}) again, as indicated on communication 11. The updates of \rcitem{x} to -1 on communications 5 and 11 correspond to the action \rcitem{x=l+1} (\isacode{rc.Plus l 1 rc.core\_int\_set} in our semantics) in the self transition of \rcitem{MoveSTM}, which is followed by an output event \rcitem{right!x}. The animation, therefore, shows two \lstinline[language=Animation]{Right_PatrolMod} events on lines \verb+#8+ and \verb+#9+.

It is worth mentioning that the RoboChart semantics in this model with the shared variable \rcitem{x} has a high degree of nondeterminism because of the interleaving of events between the module, the controller, and the state machines, and eventually nondeterminism due to the hiding of these interleaving events. Our implementation of the semantics reduces nondeterminism in a particular way: the maximal progress assumption (internal events $\tau$ have a higher priority)~\cite{Foster2021}. We also note that the animated behaviour of two \isacode{Right\_PatrolMod} events for a position is one behaviour of the RoboChart's standard semantics. This has been verified using FDR that this scenario is a trace refinement of the standard semantics (generated in RoboTool). For the verification, we encode the scenario in Fig.~\ref{fig:animate_patrol_s1} in a CSP process \lstinline[language=CSP]{Scenario1} below and then use FDR to check the assertion satisfied.
\begin{lstlisting}[language=CSP, caption={}]
Scenario1 = PatrolMod::cal.in.-3    -> PatrolMod::right.out.-2 -> PatrolMod::right.out.-2 ->
	           PatrolMod::right.out.-1 -> PatrolMod::right.out.-1 -> PatrolMod::right.out.0  -> 
              PatrolMod::right.out.0  -> Scenario1
assert PatrolMod [T= Scenario1  
\end{lstlisting}
In the assertion, \lstinline[language=CSP]{PatrolMod} is the CSP process for the module \rcitem{PatrolMod} in the generated CSP semantics in RoboTool.

Though semantically allowed, the model does not reflect the optimal way to use shared variables \changed[\C{28}]{in terms of the unnecessary interleaving behaviour by updating the shared variables from multiple state machines}. We could, for example, design models to allow only one state machine to update a shared variable and other state machines to access its value or add additional events (such as start\_update and end\_update) to enforce a synchronisation of updates to shared variables. Our patrol robot model here is presented to reveal the subtle semantics of using shared variables.

\begin{figure}[t]
\begin{lstlisting}[language=Animation, caption={}, label={lst:animation_pr_2}]
Starting ITree Simulation...
Events: (1) Reset_PatrolMod Din; (2) Cal_PatrolMod (Din,-3); (3) Cal_PatrolMod (Din,-2); 
         (4) Cal_PatrolMod (Din,-1); (5) Cal_PatrolMod (Din,0); (6) Cal_PatrolMod (Din,1); 
         (7) Cal_PatrolMod (Din,2); (8) Cal_PatrolMod (Din,3);
[Choose: 1-8]: 6 Cal_PatrolMod (Din,1)
[Choose: 1-1]: 1 Right_PatrolMod (Dout,2)
[Choose: 1-1]: 1 Right_PatrolMod (Dout,2)
[Choose: 1-1]: 1 Left_PatrolMod (Dout,1)
[Choose: 1-1]: 1 Left_PatrolMod (Dout,1)
[Choose: 1-1]: 1 Right_PatrolMod (Dout,2)
[Choose: 1-1]: 1 Right_PatrolMod (Dout,2)
[Choose: 1-1]: 1 Left_PatrolMod (Dout,1)
[Choose: 1-1]: 1 Left_PatrolMod (Dout,1)
[Choose: 1-1]: ... 
\end{lstlisting}
\caption{\label{fig:animate_patrol_s2} \changed[\C{2}]{\textbf{SCE-PR-2:}} animation of the patrol robot when the calibrated position is in S2.}
\end{figure}
\paragraph{\textbf{\changed[\C{2}]{SCE-PR-2}}}
In Fig.~\ref{fig:animate_patrol_s2}, we consider the second scenario where the calibrated position is 1 (in S2). The model behaves as follows:
\begin{inparaenum}[(1)]
\item the sixth event (\lstinline[language=Animation]{6}) is chosen on line \verb+#5+, denoting the calibrated position is \lstinline[language=Animation]{1} and so the robot is in S2;
\item then the robot moves towards the right side at position 2 (lines \verb+#6+ and \verb+#7+); 
\item subsequently the robot moves towards the left direction to position 1 (lines \verb+#8+ and \verb+#9+); and 
\item finally, the robot repeats steps (2) and (3) to patrol between position 1 and position 2.
\end{inparaenum}
We also verified this scenario is a trace refinement of the standard semantics, shown below. 
\begin{lstlisting}[language=CSP, caption={}]
Repeat = PatrolMod::right.out.2 -> PatrolMod::right.out.2 ->
          PatrolMod::left.out.1 -> PatrolMod::left.out.1 -> Repeat
Scenario2 = PatrolMod::cal.in.1 -> Repeat
assert PatrolMod [T= Scenario2
\end{lstlisting}

\begin{figure}[t]
\begin{lstlisting}[language=Animation, caption={}, label={lst:animation_pr_3}]
Starting ITree Simulation...
Events: (1) Reset_PatrolMod Din; (2) Cal_PatrolMod (Din,-3); (3) Cal_PatrolMod (Din,-2); 
         (4) Cal_PatrolMod (Din,-1); (5) Cal_PatrolMod (Din,0); (6) Cal_PatrolMod (Din,1); 
         (7) Cal_PatrolMod (Din,2); (8) Cal_PatrolMod (Din,3);
[Choose: 1-8]: 8 Cal_PatrolMod (Din,3)
[Choose: 1-1]: 1 Left_PatrolMod (Dout,2)
[Choose: 1-1]: 1 Left_PatrolMod (Dout,2)
[Choose: 1-1]: 1 Left_PatrolMod (Dout,1)
[Choose: 1-1]: 1 Left_PatrolMod (Dout,1)
[Choose: 1-1]: 1 Right_PatrolMod (Dout,2)
[Choose: 1-1]: 1 Right_PatrolMod (Dout,2)
[Choose: 1-1]: ... 
\end{lstlisting}
\caption{\label{fig:animate_patrol_s3} \changed[\C{2}]{\textbf{SCE-PR-3:}} animation of the patrol robot when the calibrated position is in S3.}
\end{figure}

\paragraph{\textbf{\changed[\C{2}]{SCE-PR-3}}}
The third scenario we consider is shown in Fig.~\ref{fig:animate_patrol_s3} where the calibrated position is initially 3 (in S3) on line \verb+#5+. The robot starts to move towards the left side to position 2 (on lines \verb+#6+ and \verb+#7+) and position 1 (on lines \verb+#8+ and \verb+#9+), and then towards the right side back to position 2 (on lines \verb+#10+ and \verb+#11+). After that, the behaviour is the same as in the second scenario in Fig.~\ref{fig:animate_patrol_s2}. Similarly, we verified this scenario is a trace refinement of the standard semantics.
\begin{lstlisting}[language=CSP, caption={}]
Scenario2 = PatrolMod::cal.in.3 -> PatrolMod::left.out.2 -> PatrolMod::left.out.2 -> 
	           PatrolMod::left.out.1 -> PatrolMod::left.out.1 -> Repeat 
assert PatrolMod [T= Scenario3
\end{lstlisting}

\paragraph{Summary}
From the three scenarios, we have seen that the event \isacode{Reset\_PatrolMod} (\rcitem{reset} in the model) is only enabled when the current position (the value of \rcitem{x} on the event \rcitem{left} or \rcitem{right}) is 0. The events enabled on lines \verb+#2+ and \verb+#15+ in Fig.~\ref{fig:animate_patrol_s1} and on line \verb+#2+ in Figs.~\ref{fig:animate_patrol_s2} and \ref{fig:animate_patrol_s3} are such examples. The standard CSP semantics, however, allows \rcitem{reset} when the current position is other than 0. The following analysis using FDR illustrates it clearly.

\begin{lstlisting}[language=CSP, caption={}]
Reset = PatrolMod::cal.in.-2 -> PatrolMod::left.out.-1 -> 
		     PatrolMod::reset.in -> PatrolMod::right.out.0 -> 
		     PatrolMod::right.out.1 -> PatrolMod::reset.in -> Reset
assert PatrolMod [T= Reset
\end{lstlisting}
In this example, \rcitem{reset} is enabled when \rcitem{x} is -1 (on line \verb+#2+) and 1 (on line \verb+#3+).
This difference is due to the maximal progress assumption in the definition of external choice and hiding in our approach: internal events have priority over external events. In this patrol robot model, the event \rcitem{update} is internal and \rcitem{reset} is external, so \rcitem{update} has priority over \rcitem{reset}. When \rcitem{x} is not 0, the guard \rcitem{[x!=0]} in the self transition of \rcitem{Cal} in the machine \rcitem{CalSTM} is true, which enables \rcitem{CalSTM} to communicate with the machine \rcitem{MoveSTM} on \rcitem{update}, and then the transition with the trigger \rcitem{reset} cannot be taken due to its lower priority than \rcitem{update}. Therefore, the \rcitem{reset} is only enabled when \rcitem{x} is 0.

\section{Related work}
\label{sec:related}
%

{Animation is a lightweight formal method. Kazmierczak et al.~\cite{Kazmierczak1998} describe the advantages of using animation to \changed[\C{29}]{test} models. It is highly automated and cheap to perform. It provides an insight into the specification and its implicit assumptions and is very suitable for demonstrating the system. It is a form of interactive testing of the model and its properties. It requires little expertise: less than model checking and much less than theorem proving. However, its biggest drawback is that it cannot verify consistency, correctness, or completeness.

Animation can be tailored to specific application domains. For example, Boichut et al.~\cite{Boichut2007} report on using animation to improve the formal specifications of security protocols. They animate these specifications to draw diagrams of typical executions of the protocols. They use this to visualise protocol termination and understand interleaved execution. They experiment with the animation to detect unwanted side effects. Finally, they use visualisation to simulate intruders to find attacks not detected by other protocol analysis tools.

We use ITrees to implement a framework for the animation of formal specifications. The ProB animator and model checker provide a different framework~\cite{Leuschel2003}. ProB contains model and constraint-based checkers that can detect errors in B specifications. It implements a back-end in a framework for various specification languages, including the B language, Event-B, CSP-M, TLA+, and Z.

De Souza~\cite{Souza2011} provides another framework: Joker. This is a tool for producing animators for formal languages. The application is based on general labelled transition systems and provides graphical animation, supporting B, CSP, and Z.}

{Rosu et al.~\cite{Rosu2010} develop K,\footnote{\url{https://kframework.org/}} a rewriting-based executable semantic framework. The operational semantics of programming languages such as C~\cite{Ellison2012} and Java~\cite{Bogdanas2015} are proposed based on K. Our ITree-based approach is also an executable semantic framework enabling the definition of operational semantics but for both abstract specification languages and concrete refinements. Thus, program development by refinement is supported in our framework. Higher-order logic and nondeterminism are some features of interaction trees, but not \changed[\C{67}]{of} K.}

Stateflow is a graphical language integrated into Matlab's Simulink to model and simulate decision logic using state machines and flow charts. During simulation, transitions in state machines are evaluated, by default, based on the order in which they are created~\cite{MathWorks2022}. This is the same as our approach to resolving nondeterministic choices between transitions using the prioritised renaming operator.

\changed[\C{1}]{The automatically generated CSP semantics of a RoboChart model in RoboTool targets at verification with FDR4 and so uses CSP-M with modules. This naturally makes Probe in FDR the first choice for the animation of RoboChart. 
The benefits of Probe include 
\begin{inparaenum}[(1)]
    \item the consistency of the standard and timed semantics of RoboChart between verification and animation because they use the same generated CSP code; and
    \item various compression methods in FDR are also used for animation, which could potentially reduce the size of state space and improve the efficiency of animation.
\end{inparaenum}
 Probe presents all internal behaviours to users to let them choose each nondeterministic choice, making the animation challenging even for experts. Our animation, presented in this paper, simplifies the process by eliminating internal behaviours so only observable events are visible to users, which makes our animation accessible to normal users. 

ProB~\cite{prob} also supports CSP-M for model checking and animation without modules. It also has some limitations \footnote{\url{prob.hhu.de/w/index.php?title=CSP-M}} in terms of supported CSP constructors. ProB cannot directly model check and animate the generated CSP code (in the new CSP-M version) in RoboTool because ProB only supports early versions of CSP-M. To use the user-friendly (GUI-based) animator in ProB, we need to encode the CSP semantics of RoboChart in the CSP-M supported by ProB.  
It is possible to encode the CSP semantics of RoboChart in the CSP-M supported by ProB, but we have not investigated it yet. The GUI-based animation in ProB is straightforward and easily used even by normal users. This is an advantage of ProB in terms of animation. 
}

\section{Conclusions \changed[\C{6}]{ and future work}}
\label{sec:concl}

{This work gives RoboChart an ITree-based operational semantics and enables the animation of RoboChart using code generation in Isabelle/HOL.}
To provide animation support, we extend ITree-based CSP with extra operators
 and present their definitions. We describe how the semantics of RoboChart 
 is implemented in ITree-based CSP and illustrate it with an autonomous chemical detector model and a patrol robot model.
With the semantics of a RoboChart model in Isabelle, we generate Haskell code and animate it using a simple simulator. Using animation, we show two concrete scenarios of the chemical detector example and three concrete scenarios of the patrol robot model.
The FDR analysis shows these scenarios are trace refinements of the standard RoboChart CSP semantics, so our approach gives and animates a refinement of the original models.

This work targets deterministic RoboChart and nondeterministic RoboChart models (but nondeterminism is resolved in a priority way in the semantics and so deterministic semantics eventually). Our work covers many RoboChart features (but not all). 
Our immediate future work is to investigate the support of nondeterminism in semantics and give semantics to more features, such as hierarchical state machines and timed semantics.


In this paper, we manually translate the RoboChart semantics to Isabelle. 
Mainly, we take RoboChart's CSP semantics generated in RoboTool into account to define consistent and restricted semantics based on an optimised version of the CSP semantics. This practical consideration could entitle us to reuse the current CSP semantics generator in RoboTool to generate ITree-based CSP semantics automatically. 
Then, the workflow from RoboChart models to Haskell code can be fully automated, and our work brings insights into it. This is part of our future work. 

With the RoboChart semantics in ITrees, we can also conduct verification in Isabelle/HOL and animation in this paper. We will investigate using temporal logic as a property language for verifying ITrees. We note that verification can also capitalise on the contributions of this work. 

ITrees can also be extended to other semantic domains. Further work would be of great help in extending ITrees with probability and linking them to discrete-time Markov chains (DTMCs)~\cite{Kemeny1976c,Kemeny1983}, which will allow us to give an ITree-based probabilistic semantics to RoboChart.

Our work has many potential applications in robotics. Further research could investigate the development of verified ROS nodes using code generation here for a concrete implementation of RoboChart controllers. We could also use this approach to automatically generate a sound runtime monitor from RoboChart models to observe the behaviour of systems derived from the models.

We use a basic textual animation in this work. This will be improved to allow the visualisation of RoboChart models in RoboTool for animation. Eventually, RoboTool users can animate a state machine, a controller, or a whole model by clicking transitions or events.

\section*{Acknowledgements}

This work is funded by the EPSRC projects CyPhyAssure%
\footnote{%
  CyPhyAssure Project: \url{www.cs.york.ac.uk/circus/CyPhyAssure/}.%
} (Grant EP/S001190/1), RoboCalc (Grant EP/M025756/1), and RoboTest (Grant EP/R025479/1).
The icons used in RoboChart have been made by Sarfraz Shoukat, Freepik, Google, Icomoon and Madebyoliver from \url{www.flaticon.com} and are licensed under CC 3.0 BY.



\bibliographystyle{elsarticle-num} 
\bibliography{main}

\begin{thebibliography}{10}
\expandafter\ifx\csname url\endcsname\relax
  \def\url#1{\texttt{#1}}\fi
\expandafter\ifx\csname urlprefix\endcsname\relax\def\urlprefix{URL }\fi
\expandafter\ifx\csname href\endcsname\relax
  \def\href#1#2{#2} \def\path#1{#1}\fi

\bibitem{Cavalcanti2021}
A.~Cavalcanti, W.~Barnett, J.~Baxter, G.~Carvalho, M.~C. Filho, A.~Miyazawa,
  P.~Ribeiro, A.~Sampaio,
  \href{https://doi.org/10.1007/978-3-030-66494-7_9}{{RoboStar Technology: A
  Roboticist's Toolbox for Combined Proof, Simulation, and Testing}}, Springer
  International Publishing, Cham, 2021, pp. 249--293.
\newblock \href {https://doi.org/10.1007/978-3-030-66494-7_9}
  {\path{doi:10.1007/978-3-030-66494-7_9}}.
\newline\urlprefix\url{https://doi.org/10.1007/978-3-030-66494-7_9}

\bibitem{Hoare1998}
C.~A.~R. Hoare, J.~He, {Unifying Theories of Programming}, Prentice-Hall, 1998.

\bibitem{Miyazawa2019}
A.~Miyazawa, P.~Ribeiro, W.~Li, A.~Cavalcanti, J.~Timmis, J.~Woodcock,
  {RoboChart: modelling and verification of the functional behaviour of robotic
  applications}, Softw. Syst. Model. 18~(5) (2019) 3097--3149.
\newblock \href {https://doi.org/10.1007/s10270-018-00710-z}
  {\path{doi:10.1007/s10270-018-00710-z}}.

\bibitem{Ye2021}
K.~Ye, A.~Cavalcanti, S.~Foster, A.~Miyazawa, J.~Woodcock,
  \href{https://doi.org/10.1007/s10270-021-00916-8}{{Probabilistic modelling
  and verification using RoboChart and PRISM}}, Software and Systems Modeling
  (Oct 2021).
\newblock \href {https://doi.org/10.1007/s10270-021-00916-8}
  {\path{doi:10.1007/s10270-021-00916-8}}.
\newline\urlprefix\url{https://doi.org/10.1007/s10270-021-00916-8}

\bibitem{Woodcock2019}
J.~Woodcock, A.~Cavalcanti, S.~Foster, A.~Mota, K.~Ye, {Probabilistic Semantics
  for RoboChart}, in: P.~Ribeiro, A.~Sampaio (Eds.), Unifying Theories of
  Programming, Springer International Publishing, Cham, 2019, pp. 80--105.

\bibitem{Ye2021a}
K.~Ye, S.~Foster, J.~Woodcock, {Automated Reasoning for Probabilistic
  Sequential Programs with Theorem Proving}, in: U.~Fahrenberg, M.~Gehrke,
  L.~Santocanale, M.~Winter (Eds.), Relational and Algebraic Methods in
  Computer Science, Springer International Publishing, Cham, 2021, pp.
  465--482.

\bibitem{Cavalcanti2019a}
A.~L.~C. Cavalcanti, A.~C.~A. Sampaio, A.~Miyazawa, P.~Ribeiro, M.~S.~C. Filho,
  A.~Didier, W.~Li, J.~Timmis,
  \href{https://www-users.cs.york.ac.uk/%7Ealcc/publications/papers/CSMRCD19.pdf}{Verified
  simulation for robotics}, Science of Computer Programming 174 (2019) 1--37.
\newblock \href {https://doi.org/10.1016/j.scico.2019.01.004}
  {\path{doi:10.1016/j.scico.2019.01.004}}.
\newline\urlprefix\url{https://www-users.cs.york.ac.uk/%7Ealcc/publications/papers/CSMRCD19.pdf}

\bibitem{Foster2020a}
S.~Foster, J.~J. Huerta~y Munive, G.~Struth, Differential hoare logics and
  refinement calculi for hybrid systems with isabelle/hol, in: U.~Fahrenberg,
  P.~Jipsen, M.~Winter (Eds.), Relational and Algebraic Methods in Computer
  Science, Springer International Publishing, Cham, 2020, pp. 169--186.

\bibitem{Foster2021a}
S.~Foster, J.~J. Huerta~y Munive, M.~Gleirscher, G.~Struth, Hybrid systems
  verification with isabelle/hol: Simpler syntax, better models, faster proofs,
  in: M.~Huisman, C.~P{\u{a}}s{\u{a}}reanu, N.~Zhan (Eds.), Formal Methods,
  Springer International Publishing, Cham, 2021, pp. 367--386.

\bibitem{Murray2022}
Y.~Murray, M.~Sirevåg, P.~Ribeiro, D.~A. Anisi, M.~Mossige,
  \href{https://www.sciencedirect.com/science/article/pii/S0167642321001593}{Safety
  assurance of an industrial robotic control system using hardware/software
  co-verification}, Science of Computer Programming 216 (2022) 102766.
\newblock \href {https://doi.org/https://doi.org/10.1016/j.scico.2021.102766}
  {\path{doi:https://doi.org/10.1016/j.scico.2021.102766}}.
\newline\urlprefix\url{https://www.sciencedirect.com/science/article/pii/S0167642321001593}

\bibitem{Hoare1985}
C.~A.~R. Hoare, {Communicating Sequential Processes}, Prentice-Hall Int., 1985.

\bibitem{Roscoe2011}
A.~W. Roscoe, {Understanding Concurrent Systems}, Texts in Computer Science,
  Springer, 2011.

\bibitem{Xia2019}
L.-y. Xia, Y.~Zakowski, P.~He, C.-K. Hur, G.~Malecha, B.~C. Pierce,
  S.~Zdancewic, \href{https://doi.org/10.1145/3371119}{{Interaction Trees:
  Representing Recursive and Impure Programs in Coq}}, Proc. ACM Program. Lang.
  4~(POPL) (Dec. 2019).
\newblock \href {https://doi.org/10.1145/3371119} {\path{doi:10.1145/3371119}}.
\newline\urlprefix\url{https://doi.org/10.1145/3371119}

\bibitem{Foster2021}
S.~Foster, C.-K. Hur, J.~Woodcock,
  \href{https://drops.dagstuhl.de/opus/volltexte/2021/14397}{{Formally Verified
  Simulations of State-Rich Processes Using Interaction Trees in
  Isabelle/HOL}}, in: S.~Haddad, D.~Varacca (Eds.), 32nd International
  Conference on Concurrency Theory (CONCUR 2021), Vol. 203 of Leibniz
  International Proceedings in Informatics (LIPIcs), Schloss Dagstuhl --
  Leibniz-Zentrum f{\"u}r Informatik, Dagstuhl, Germany, 2021, pp. 20:1--20:18.
\newblock \href {https://doi.org/10.4230/LIPIcs.CONCUR.2021.20}
  {\path{doi:10.4230/LIPIcs.CONCUR.2021.20}}.
\newline\urlprefix\url{https://drops.dagstuhl.de/opus/volltexte/2021/14397}

\bibitem{Brookes1984}
S.~D. Brookes, C.~A.~R. Hoare, A.~W. Roscoe,
  \href{https://doi.org/10.1145/828.833}{{A Theory of Communicating Sequential
  Processes}}  560--599\href {https://doi.org/10.1145/828.833}
  {\path{doi:10.1145/828.833}}.
\newline\urlprefix\url{https://doi.org/10.1145/828.833}

\bibitem{GABR14}
T.~Gibson-Robinson, P.~Armstrong, A.~Boulgakov, A.~W. Roscoe, {FDR3 - A Modern
  Refinement Checker for CSP}, in: Tools and Algorithms for the Construction
  and Analysis of Systems, 2014, pp. 187--201.

\bibitem{Toyn2002}
I.~Toyn (Ed.), {Information Technology --- {Z} Formal Specification Notation
  --- Syntax, Type System and Semantics}, ISO, 2002, iSO/IEC 13568:2002(E).

\bibitem{Spivey1992}
J.~M. Spivey, {The Z Notation: A Reference Manual}, 2nd, Prentice-Hall, 1992.

\bibitem{Haftmann2010}
F.~Haftmann, T.~Nipkow,
  \href{https://doi.org/10.1007/978-3-642-12251-4\_9}{{Code Generation via
  Higher-Order Rewrite Systems}}, in: M.~Blume, N.~Kobayashi, G.~Vidal (Eds.),
  Functional and Logic Programming, 10th International Symposium, {FLOPS} 2010,
  Sendai, Japan, April 19-21, 2010. Proceedings, Vol. 6009 of Lecture Notes in
  Computer Science, Springer, 2010, pp. 103--117.
\newblock \href {https://doi.org/10.1007/978-3-642-12251-4\_9}
  {\path{doi:10.1007/978-3-642-12251-4\_9}}.
\newline\urlprefix\url{https://doi.org/10.1007/978-3-642-12251-4\_9}

\bibitem{Mayr1998}
R.~Mayr, T.~Nipkow, Higher-order rewrite systems and their confluence,
  Theoretical computer science 192~(1) (1998) 3--29.

\bibitem{Leuschel2003}
M.~Leuschel, M.~J. Butler,
  \href{https://doi.org/10.1007/978-3-540-45236-2\_46}{{ProB: {A} Model Checker
  for {B}}}, in: K.~Araki, S.~Gnesi, D.~Mandrioli (Eds.), {FME} 2003: Formal
  Methods, International Symposium of Formal Methods Europe, Pisa, Italy,
  September 8-14, 2003, Proceedings, Vol. 2805 of Lecture Notes in Computer
  Science, Springer, 2003, pp. 855--874.
\newblock \href {https://doi.org/10.1007/978-3-540-45236-2\_46}
  {\path{doi:10.1007/978-3-540-45236-2\_46}}.
\newline\urlprefix\url{https://doi.org/10.1007/978-3-540-45236-2\_46}

\bibitem{Ye2022}
K.~Ye, S.~Foster, J.~Woodcock, {Formally Verified Animation for RoboChart Using
  Interaction Trees}, in: A.~Riesco, M.~Zhang (Eds.), Formal Methods and
  Software Engineering, Springer International Publishing, Cham, 2022, pp.
  404--420.

\bibitem{Hoare1987}
C.~A.~R. Hoare, J.~He, {The weakest prespecification}, Information Processing
  Letters 24~(2) (1987) 127--132.

\bibitem{Baxter2021}
J.~Baxter, P.~Ribeiro, A.~Cavalcanti, {Sound reasoning in tock-CSP}, Acta
  Informatica (04 2021).
\newblock \href {https://doi.org/10.1007/s00236-020-00394-3}
  {\path{doi:10.1007/s00236-020-00394-3}}.

\bibitem{Schneider1999}
S.~Schneider, Concurrent and Real Time Systems: The CSP Approach, 1st Edition,
  John Wiley \& Sons, Inc., USA, 1999.

\bibitem{Woodcock2002}
J.~Woodcock, A.~Cavalcanti, The semantics of circus, in: D.~Bert, J.~P. Bowen,
  M.~C. Henson, K.~Robinson (Eds.), ZB 2002:Formal Specification and
  Development in Z and B, Springer Berlin Heidelberg, Berlin, Heidelberg, 2002,
  pp. 184--203.

\bibitem{Woodcock2012}
J.~C.~P. Woodcock, A.~L.~C. Cavalcanti, J.~Fitzgerald, P.~G. Larsen,
  A.~Miyazawa, S.~Perry, Features of cml: a formal modelling language for
  systems of systems, in: 7th International Conference on Systems of Systems
  Engineering, Vol.~6 of IEEE Systems Journal, IEEE, 2012.

\bibitem{Hilder2012}
J.~A. Hilder, N.~D.~L. Owens, M.~J. Neal, P.~J. Hickey, S.~N. Cairns, D.~P.~A.
  Kilgour, J.~Timmis, A.~M. Tyrrell, {Chemical Detection Using the Receptor
  Density Algorithm}, {IEEE} Trans. Syst. Man Cybern. Part {C} 42~(6) (2012)
  1730--1741.
\newblock \href {https://doi.org/10.1109/TSMCC.2012.2218236}
  {\path{doi:10.1109/TSMCC.2012.2218236}}.

\bibitem{RoboChartRef}
A.~Miyazawa, A.~Cavalcanti, P.~Ribeiro, K.~Ye, W.~Li, J.~Woodcock, J.~Timmis,
  {RoboChart Reference Manual}, Tech. rep., University of York,
  \url{www.cs.york.ac.uk/circus/publications/techreports/reports/robochart-reference.pdf}
  (2020).

\bibitem{Blanchette2014}
J.~C. Blanchette, J.~H{\"{o}}lzl, A.~Lochbihler, L.~Panny, A.~Popescu,
  D.~Traytel, \href{https://doi.org/10.1007/978-3-319-08970-6\_7}{{Truly
  Modular (Co)datatypes for Isabelle/HOL}}, in: G.~Klein, R.~Gamboa (Eds.),
  Interactive Theorem Proving - 5th International Conference, {ITP} 2014, Held
  as Part of the Vienna Summer of Logic, {VSL} 2014, Vienna, Austria, July
  14-17, 2014. Proceedings, Vol. 8558 of Lecture Notes in Computer Science,
  Springer, 2014, pp. 93--110.
\newblock \href {https://doi.org/10.1007/978-3-319-08970-6\_7}
  {\path{doi:10.1007/978-3-319-08970-6\_7}}.
\newline\urlprefix\url{https://doi.org/10.1007/978-3-319-08970-6\_7}

\bibitem{Ballarin2004}
C.~Ballarin, {Locales and Locale Expressions in Isabelle/Isar}, in: S.~Berardi,
  M.~Coppo, F.~Damiani (Eds.), Types for Proofs and Programs, Springer Berlin
  Heidelberg, Berlin, Heidelberg, 2004, pp. 34--50.

\bibitem{Kazmierczak1998}
E.~Kazmierczak, M.~Winikoff, P.~W. Dart, {Verifying Model Oriented
  Specifications through Animation}, in: 5th Asia-Pacific Software Engineering
  Conference {(APSEC} '98), 2-4 December 1998, Taipei, Taiwan, {ROC}, {IEEE}
  Computer Society, pp. 254--261.
\newblock \href {https://doi.org/10.1109/APSEC.1998.733727}
  {\path{doi:10.1109/APSEC.1998.733727}}.

\bibitem{Boichut2007}
Y.~Boichut, T.~Genet, Y.~Glouche, O.~Heen, {Using Animation to Improve Formal
  Specifications of Security Protocols}, in: 2nd Conference on Security in
  Network Architectures and Information Systems (SARSSI 2007, 2007, pp.
  169--182.

\bibitem{Souza2011}
D.~H.~O. de~Souza, {Joker: An Animator for Formal Languages}, Ph.D. thesis,
  Departamento de Inform\'{a}tica e Matem\'{a}tica Aplicada, Universidade
  Federal do Rio Grande do Norte (2011).

\bibitem{Rosu2010}
G.~Rosu, T.~Serbanuta, An overview of the {K} semantic framework, J. Log.
  Algebraic Methods Program. 79~(6) (2010) 397--434.
\newblock \href {https://doi.org/10.1016/j.jlap.2010.03.012}
  {\path{doi:10.1016/j.jlap.2010.03.012}}.

\bibitem{Ellison2012}
C.~Ellison, G.~Rosu, An executable formal semantics of {C} with applications,
  in: J.~Field, M.~Hicks (Eds.), Proceedings of the 39th {ACM} {SIGPLAN-SIGACT}
  Symposium on Principles of Programming Languages, {POPL} 2012, Philadelphia,
  Pennsylvania, USA, January 22-28, 2012, {ACM}, 2012, pp. 533--544.
\newblock \href {https://doi.org/10.1145/2103656.2103719}
  {\path{doi:10.1145/2103656.2103719}}.

\bibitem{Bogdanas2015}
D.~Bogdanas, G.~Rosu, K-java: {A} complete semantics of java, in: S.~K.
  Rajamani, D.~Walker (Eds.), Proceedings of the 42nd Annual {ACM}
  {SIGPLAN-SIGACT} Symposium on Principles of Programming Languages, {POPL}
  2015, Mumbai, India, January 15-17, 2015, {ACM}, 2015, pp. 445--456.
\newblock \href {https://doi.org/10.1145/2676726.2676982}
  {\path{doi:10.1145/2676726.2676982}}.

\bibitem{MathWorks2022}
MathWorks,
  \href{www.mathworks.com/help/pdf_doc/stateflow/stateflow_ug.pdf}{{Stateflow:
  User's Guide (R2022b)}}, The MathWorks Inc., 2022.
\newline\urlprefix\url{www.mathworks.com/help/pdf_doc/stateflow/stateflow_ug.pdf}

\bibitem{prob}
{The ProB Animator and Model Checker},
  \url{https://prob.hhu.de/w/index.php?title=Main_Page}, accessed: 2021-03-25.

\bibitem{Kemeny1976c}
J.~G. Kemeny, J.~L. Snell, A.~W. Knapp, {Denumerable Markov Chains}, 1976.
\newblock \href {https://doi.org/10.1007/978-1-4684-9455-6}
  {\path{doi:10.1007/978-1-4684-9455-6}}.

\bibitem{Kemeny1983}
J.~G. Kemeny, J.~L. Snell, {Finite Markov Chains: With a New Appendix
  "Generalization of a Fundamental Matrix" (Undergraduate Texts in
  Mathematics)}, Springer, 1983.

\end{thebibliography}

\ifdefined \CHANGES \indexprologue{%
  This index lists for each comment the pages where the text has been modified to address the comment. Since the same page may contain multiple changes, the page number contains the index of the change in superscript to identify different changes. Finally, the page number contains a hyperlink that takes the reader to corresponding change.%
}%
\printindex[changes] \fi

\end{document}